\documentclass[aps,prc,twocolumn,groupedaddress,showpacs,preprintnumbers,superscriptaddress,floatfix,amsmath,amssymb]{revtex4}
\usepackage{graphicx}
\usepackage{multirow}
\usepackage{xspace}
\usepackage{dcolumn}
\usepackage{array}
\usepackage{bm}
\usepackage{hhline}
\usepackage{hyperref}
\usepackage{color}
\usepackage{upgreek}

\newcommand{\sqrtsNN}{\mbox{$\sqrt{s_{_{\mathrm{NN}}}}$}}
\newcommand{\axi}{$\overline{\Xi}^+$}
\newcommand{\xim}{$\Xi^-$}
\newcommand{\alam}{$\overline{\Lambda}$}
\newcommand{\lam}{$\Lambda$}
\newcommand{\ks}{$\mathrm{K}^{0}_{\mathrm S}$}
\newcommand{\omm}{$\Omega^-$}
\newcommand{\aom}{$\overline{\Omega}^+$}

\newcommand{\ppt}{$p_{\rm T}$}

\newcommand{\GeVc}{\mbox{$\mathrm{GeV} / c$}}

\newcolumntype{d}{D{$\pm$}{$\pm$}{-1}}

\begin{document}
%\linenumbers

\title{Strange hadron production in Au+Au collisions at \sqrtsNN\ = 7.7, 11.5, 19.6, 27, and 39 GeV}             
\affiliation{Abilene Christian University, Abilene, Texas   79699}
\affiliation{AGH University of Science and Technology, FPACS, Cracow 30-059, Poland}
\affiliation{Alikhanov Institute for Theoretical and Experimental Physics, Moscow 117218, Russia}
\affiliation{Argonne National Laboratory, Argonne, Illinois 60439}
\affiliation{American Univerisity of Cairo, Cairo, Egypt}
\affiliation{Brookhaven National Laboratory, Upton, New York 11973}
\affiliation{University of California, Berkeley, California 94720}
\affiliation{University of California, Davis, California 95616}
\affiliation{University of California, Los Angeles, California 90095}
\affiliation{University of California, Riverside, California 92521}
\affiliation{Central China Normal University, Wuhan, Hubei 430079 }
\affiliation{University of Illinois at Chicago, Chicago, Illinois 60607}
\affiliation{Creighton University, Omaha, Nebraska 68178}
\affiliation{Czech Technical University in Prague, FNSPE, Prague 115 19, Czech Republic}
\affiliation{Technische Universit\"at Darmstadt, Darmstadt 64289, Germany}
\affiliation{E\"otv\"os Lor\'and University, Budapest, Hungary H-1117}
\affiliation{Frankfurt Institute for Advanced Studies FIAS, Frankfurt 60438, Germany}
\affiliation{Fudan University, Shanghai, 200433 }
\affiliation{University of Heidelberg, Heidelberg 69120, Germany }
\affiliation{University of Houston, Houston, Texas 77204}
\affiliation{Huzhou University, Huzhou, Zhejiang  313000}
\affiliation{Indian Institute of Science Education and Research (IISER), Berhampur 760010 , India}
\affiliation{Indian Institute of Science Education and Research, Tirupati 517507, India}
\affiliation{Indian Institute Technology, Patna, Bihar, India}
\affiliation{Indiana University, Bloomington, Indiana 47408}
\affiliation{Institute of Physics, Bhubaneswar 751005, India}
\affiliation{University of Jammu, Jammu 180001, India}
\affiliation{Joint Institute for Nuclear Research, Dubna 141 980, Russia}
\affiliation{Kent State University, Kent, Ohio 44242}
\affiliation{University of Kentucky, Lexington, Kentucky 40506-0055}
\affiliation{Lawrence Berkeley National Laboratory, Berkeley, California 94720}
\affiliation{Lehigh University, Bethlehem, Pennsylvania 18015}
\affiliation{Max-Planck-Institut f\"ur Physik, Munich 80805, Germany}
\affiliation{Michigan State University, East Lansing, Michigan 48824}
\affiliation{National Research Nuclear University MEPhI, Moscow 115409, Russia}
\affiliation{National Institute of Science Education and Research, HBNI, Jatni 752050, India}
\affiliation{National Cheng Kung University, Tainan 70101 }
\affiliation{Nuclear Physics Institute of the CAS, Rez 250 68, Czech Republic}
\affiliation{Ohio State University, Columbus, Ohio 43210}
\affiliation{Panjab University, Chandigarh 160014, India}
\affiliation{Pennsylvania State University, University Park, Pennsylvania 16802}
\affiliation{NRC "Kurchatov Institute", Institute of High Energy Physics, Protvino 142281, Russia}
\affiliation{Purdue University, West Lafayette, Indiana 47907}
\affiliation{Pusan National University, Pusan 46241, Korea}
\affiliation{Rice University, Houston, Texas 77251}
\affiliation{Rutgers University, Piscataway, New Jersey 08854}
\affiliation{Universidade de S\~ao Paulo, S\~ao Paulo, Brazil 05314-970}
\affiliation{University of Science and Technology of China, Hefei, Anhui 230026}
\affiliation{Shandong University, Qingdao, Shandong 266237}
\affiliation{Shanghai Institute of Applied Physics, Chinese Academy of Sciences, Shanghai 201800}
\affiliation{Southern Connecticut State University, New Haven, Connecticut 06515}
\affiliation{State University of New York, Stony Brook, New York 11794}
\affiliation{Temple University, Philadelphia, Pennsylvania 19122}
\affiliation{Texas A\&M University, College Station, Texas 77843}
\affiliation{University of Texas, Austin, Texas 78712}
\affiliation{Tsinghua University, Beijing 100084}
\affiliation{University of Tsukuba, Tsukuba, Ibaraki 305-8571, Japan}
\affiliation{United States Naval Academy, Annapolis, Maryland 21402}
\affiliation{Valparaiso University, Valparaiso, Indiana 46383}
\affiliation{Variable Energy Cyclotron Centre, Kolkata 700064, India}
\affiliation{Warsaw University of Technology, Warsaw 00-661, Poland}
\affiliation{Wayne State University, Detroit, Michigan 48201}
\affiliation{Yale University, New Haven, Connecticut 06520}

\author{J.~Adam}\affiliation{Brookhaven National Laboratory, Upton, New York 11973}
\author{L.~Adamczyk}\affiliation{AGH University of Science and Technology, FPACS, Cracow 30-059, Poland}
\author{J.~R.~Adams}\affiliation{Ohio State University, Columbus, Ohio 43210}
\author{J.~K.~Adkins}\affiliation{University of Kentucky, Lexington, Kentucky 40506-0055}
\author{G.~Agakishiev}\affiliation{Joint Institute for Nuclear Research, Dubna 141 980, Russia}
\author{M.~M.~Aggarwal}\affiliation{Panjab University, Chandigarh 160014, India}
\author{Z.~Ahammed}\affiliation{Variable Energy Cyclotron Centre, Kolkata 700064, India}
\author{I.~Alekseev}\affiliation{Alikhanov Institute for Theoretical and Experimental Physics, Moscow 117218, Russia}\affiliation{National Research Nuclear University MEPhI, Moscow 115409, Russia}
\author{D.~M.~Anderson}\affiliation{Texas A\&M University, College Station, Texas 77843}
\author{R.~Aoyama}\affiliation{University of Tsukuba, Tsukuba, Ibaraki 305-8571, Japan}
% ...

\collaboration{STAR Collaboration}\noaffiliation

\begin{abstract}
We present STAR measurements of strange hadron (\ks, $\Lambda$, \alam, \xim, \axi, \omm, \aom, and $\phi$) production at mid-rapidity ($|y| < 0.5$) in Au+Au collisions at
\sqrtsNN\ =\ 7.7 -- 39 GeV from the Beam Energy Scan Program at the Relativistic Heavy Ion Collider (RHIC). Transverse momentum spectra, averaged transverse mass, and the overall integrated yields of these strange hadrons are presented versus the centrality and collision energy. Antibaryon-to-baryon ratios (\alam/\lam, \axi/\xim, \aom/\omm) are presented as well, and used to test a thermal statistical model and to extract the temperature normalized strangeness and baryon chemical potentials at hadronic freeze-out ($\mu_{B}/T_{\rm ch}$ and $\mu_{S}/T_{\rm ch}$) in central collisions. Strange baryon-to-pion ratios are compared to various model predictions in central collisions for all energies. The nuclear modification factors ($R_{\textrm{\tiny{CP}}}$) and antibaryon-to-meson ratios as a function of transverse momentum are presented for all collision energies. The \ks\ $R_{\textrm{\tiny{CP}}}$ shows no suppression for \ppt\ up to 3.5 \GeVc\ at energies of 7.7 and 11.5 GeV. The \alam/\ks\ ratio also shows baryon-to-meson enhancement at intermediate \ppt\ ($\approx$2.5 \GeVc) in central collisions at energies above 19.6 GeV. Both observations suggest that there is likely a change of the underlying strange quark dynamics at collision energies below 19.6 GeV.
\end{abstract}

\pacs{25.75.-q, 25.75.Dw, 25.75.Nq}

\maketitle

\section{Introduction}\label{section-introduction}

The main motivation of the RHIC Beam Energy Scan (BES) Program is to study the quantum chromodynamics (QCD) phase diagram \cite{Stephanov:2004wx,Mohanty:2009vb, Aggarwal:2010cw}. Systematic analysis of Au+Au collisions from \sqrtsNN\ = 39 GeV down to 7.7 GeV in the RHIC BES Program could help to achieve the following goals: 1) to find the QCD critical point where the first order phase transition at finite baryon chemical potential ($\mu_B$) ends and to identify the phase boundary of the first order phase transition \cite{Adamczyk:2013dal,Adamczyk:2014ipa,Adamczyk:2014fia,bes_pid,Adamczyk:2017wsl}; 2) to locate the collision energy where deconfinement begins \cite{Adamczyk:2013gv,Adamczyk:2013gw,Adamczyk:2015lvo,Adamczyk:2017nof}.

Strange hadrons are an excellent probe for identifying the phase boundary and onset of deconfinement. Strangeness enhancement in {\it A}+{\it A} with respect to {\it p}+{\it p} collisions has long been suggested as a signature of quark-gluon plasma (QGP) \cite{raf82}. Strangeness has been extensively measured in many experiments at different accelerator facilities \cite{e896,e802,e895,e917,e891,e891_2,Antinori:2004ee,Antinori:2006ij,Afanasiev:2002mx,Alt:2007aa,Anticic:2003ux,na49prl,na49prc,Anticic:2009ie,ceres,Adler:2002uv,starpid_130,Adams:2003fy,Adams:2006ke,Abelev:2007xp,Abelev:2008zk,starprc83,Agakishiev:2011ar,Abelev:2010rv,phenix_lambda,lhc,lhc2,Chen:2018tnh}. 
Generally, the yields of strange hadrons in nuclear collisions are close to those expected from statistical models \cite{becattini_prc,pbmnpa,pbm_shm,Redlich:2001kb}. The precise measurement of these yields in heavy ion collisions in the BES may lead to a better understanding of strangeness production mechanisms in nuclear collisions and a more constrained extraction of the chemical freeze-out parameters.

The measurement of strange hadrons at high \ppt\ can probe hard parton scatterings in the QGP medium, through the central-to-peripheral nuclear modification factor $R_{\textrm{\tiny{CP}}}=(\textrm{yield}/N_\textrm{\scriptsize{coll}})_\textrm{central}/(\textrm{yield}/N_\textrm{\scriptsize{coll}})_\textrm{peripheral}$, where $N_\textrm{\scriptsize{coll}}$ is the averge number of binary nucleon-nucleon collisions. 
It has been observed in Au+Au collisions at \sqrtsNN\,\,=~200 GeV at RHIC that, at high \ppt,  $R_{\textrm{\tiny{CP}}}$ of various particles is much less than unity \cite{Adams:2003kv,Adams:2006ke}, indicating a significant energy loss of the scattered partons in the dense matter. $R_{\textrm{\tiny{CP}}}$ of strange hadrons in the BES, together with other non-strange hadron suppression results, can potentially pin down the beam energy at which energy loss to the medium begins to dominate hard parton interactions \cite{Aggarwal:2010cw}. 

At intermediate \ppt\ (2--5 \GeVc), as first discovered in central Au+Au events at RHIC \cite{Adcox:2003nr,starb2m,Abelev:2006jr,Agakishiev:2011ar} and later observed at the Large Hadron Collider \cite{lhc,Abelev:2014laa}, the $p/\pi$ and $\Lambda/$\ks\ ratios are larger than unity and much higher than those observed in peripheral {\it A}+{\it A} and in {\it p}+{\it p} collisions. These results may indicate different hadronization mechanisms in this \ppt\ range in {\it A}+{\it A} collisions. There are recombination/coalescence models which allow soft partons to coalesce into hadrons, or soft and hard partons to recombine into hadrons \cite{coal,coal2,reconbination0,reconbination1,reconbination2,reconbination21,reconbination3,reconbination31}. They naturally reproduce enhanced baryon-to-meson ratios for any quickly falling distribution of parton \ppt. Such models rely on recombination or coalescence of constituent quarks, thus existence of a partonic medium. Hence observation of such behavior at hadronization is a fundamental piece of evidence for the formation of the partonic QGP medium. It is also interesting to investigate at which collision energies these phenomena are prevalent \cite{Aggarwal:2010cw}, in order to locate the energy range over which the onset of the deconfinement happens.

We present strangeness data obtained from Au+Au collisions at \sqrtsNN\,\,=~7.7, 11.5, 19.6, 27, and 39~GeV, collected by the STAR experiment during the first phase of the RHIC BES Program in 2010 and 2011.

This paper is organized as follows. Section~\ref{section-experiment} briefly describes
the experimental setup, the event selection, and the centrality determination. Section~\ref{section-analysis} discusses the reconstruction methods of various strange hadrons, the signal extraction methods, the acceptance and efficiency correction factors, the feed-down corrections for $\Lambda$ hyperons, the extrapolations to low \ppt, and the systematic uncertainties. Section~\ref{section-results} presents transverse momentum spectra, averaged transverse mass, integrated yields, and various particle ratios of those strange hadrons and comparisons to theory for different centralities and collision energies. Finally, Sec.~\ref{Conclusions} is the summary.

\section{Experimental setup}\label{section-experiment}

The Solenoidal Tracker At RHIC (STAR) is a versatile particle detector at the RHIC collider at Brookhaven National Laboratory. A detailed description of its solenoidal magnet and various sub-detectors for tracking, particle identification, and triggering can be found in Ref.~\cite{STARNIM}.

The Time Projection Chamber (TPC) is the main detector at STAR which provides tracking capability covering $2\pi$ azimuthal angle in the transverse direction and $-1$ to 1 in pseudo-rapidity, $\eta$~\cite{STARTPC}. The TPC is immersed in a constant 0.5 Tesla magnetic field parallel to the beam direction, which is generated by the surrounding solenoidal magnet. The track of a charged particle can be reconstructed with a maximum of 45 hit points within the TPC fiducial radius of $0.5 < r < 2$ m. The location of the primary vertex of a collision event is determined using the reconstructed charged particle tracks. A primary vertex resolution in the transverse plane of 350 ${\rm \upmu m}$ can be achieved with $\approx$1000 tracks. The fitted primary vertex can be included in the track fitting of the charged particles to improve their momentum resolution. With this procedure, a relative momentum resolution for pions of $\approx$2\% at \ppt\,=\,1 \GeVc\ can be achieved. The TPC also measures the energy loss of charged particles, which allows separation of $\pi$ and K up to \ppt\,$\simeq$\,0.7 \GeVc, and statistical proton identification up to \ppt\,$\simeq$\,1.1 \GeVc\ \cite{STARTPC}. 

In 2010, the STAR experiment recorded Au+Au collisions at the nucleon-nucleon center-of-mass energy (\sqrtsNN) of 7.7, 11.5, and 39 GeV. The data of Au+Au collisions at \sqrtsNN\ = 19.6 and 27 GeV were further collected in 2011. The minimum bias trigger at all five energies was defined by the coincidence of the zero-degree calorimeters (ZDC), vertex position detectors (VPD) \cite{Llope:2003ti}, and/or beam-beam counters (BBC) signals \cite{STARTRG}. However, at the lowest beam energies, most of the triggered events are from Au beam nuclei with large emittance that hit the nuclei at rest in the beam pipe. This background can be removed by requiring the primary vertex of an event to be within a radius $r$ of less than 2 cm of the geometrical center of the detector system, which is much less than that of the beam pipe (3.95 cm).
The primary vertex position in the beam direction ($z$-direction) was limited to the values listed in Table~\ref{table_1}. These values were selected according to the offline $z$-vertex trigger conditions which were different for different energies. It was further required that at least two tracks from the primary vertex were matched to the cells of the barrel time-of-flight detector (BTOF) \cite{Bonner:2003bv} in order to remove the pile-up events from different bunch-crossings. Finally, events from bad runs were removed according to an extensive quality assurance analysis of the events (see Ref.~\cite{Adamczyk:2013gw}). The accepted number of minimum bias events for each of the five energies are also listed in Table~\ref{table_1}.

\begin{table}[hbt]
\caption{The $z$-vertex acceptance, and the total number of minimum-bias (MB) events used, for different energies.
\label{table_1}}
\begin{center}
\begin{tabular}{ >{\centering\arraybackslash}m{0.75in}  | >{\centering\arraybackslash}m{1.3in} | >{\centering\arraybackslash}m{1.1in} } 
\hline
\sqrtsNN\ (GeV) & $z$-vertex range (cm) & MB events ($10^{6}$)\\
\hline
7.7&[-70, 70]&4.4\\
11.5&[-50, 50]&12.0\\
19.6&[-70, 70]&36.3\\
27&[-70, 70]&72.8\\
39&[-40, 40]&134.3\\
\hline
\end{tabular}
\end{center}
\end{table}

The centrality selection of the events was chosen to be 0--80\% of the total reaction cross section due to trigger inefficiencies for the most peripheral events. The centrality definition was based on an uncorrected multiplicity distribution and a Glauber Monte Carlo simulation \cite{Miller:2007ri}, and details can be found in Ref.~\cite{Adamczyk:2013gw}. By comparison of the Glauber simulation to the measured multiplicity distribution at each energy, it is possible to determine, for each centrality class, the average number of participant nucleons $\langle N_{\rm part} \rangle$. The values of $\langle N_{\rm part} \rangle$ at different centralities and collision energies are listed in Table~\ref{tab:npart}.

\begin{table*}[hbt]
\begin{center}
\caption{The average number of participating nucleons $\langle N_{\rm part} \rangle $ for various collision centralities in Au+Au collisions at $7.7$--$39$ GeV, determined using the charged particle multiplicity distributions and the Glauber Monte Carlo simulation \cite{Adamczyk:2013gw}. The errors represent systematic uncertainties. The inelastic $p+p$ cross-sections used in the simulations are 30.8, 31.2, 32, 33, and 34 mb for $\sqrt{s}$\,=\,7.7, 11.5, 19.6, 27, and 39 GeV, respectively \cite{Miller:2007ri}.
\label{tab:npart}}
\begin{tabular}{c|ccccccccccccc}
\hline
 \sqrtsNN\ (GeV)  & 0--5\% & \hspace{0.25cm} & 5--10\% & \hspace{0.25cm} & 10--20\% & \hspace{0.25cm} & 20--30\% & \hspace{0.25cm} & 30--40\% & \hspace{0.25cm} & 40--60\% & \hspace{0.25cm} & 60--80\% \\
\hline
7.7 & 337.4 $\pm$ 2.1 & & 290.4 $\pm$ 6.0 & & 226.2 $\pm$ 7.9 & & 160.2 $\pm$ 10.2 & & 109.9 $\pm$ 11.0 & & 58.4  $\pm$ 9.8 & & 20.2  $\pm$ 5.3  \\
11.5 & 338.2 $\pm$ 2.0 & & 290.6 $\pm$ 6.2 & & 226.0 $\pm$ 8.2 & & 159.6 $\pm$ 9.5\,\,\, & & 110.0 $\pm$ 10.3 & & 58.5 $\pm$ 9.4 & & 20.1 $\pm$ 6.7 \\
19.6 & 338.0 $\pm$ 2.3 & & 289.2 $\pm$ 6.0 & & 224.9 $\pm$ 8.6 & & 158.1 $\pm$ 10.5 & & 108.0 $\pm$ 10.6 & & 57.7 $\pm$ 9.1 & & 19.9 $\pm$ 5.9 \\
27 & 343.3 $\pm$ 2.0 & & 299.3 $\pm$ 6.2 & & 233.6 $\pm$ 9.0 & & 165.5 $\pm$ 10.7 & & 114.0 $\pm$ 11.3 & & \,\,\,61.2 $\pm$ 10.4 & & 20.5 $\pm$ 7.1 \\
39 & 341.7 $\pm$ 2.2 & & 293.9 $\pm$ 6.4 & & 229.8 $\pm$ 8.7 & & 162.4 $\pm$ 10.2 & & 111.4 $\pm$ 10.8 & & 59.2 $\pm$ 9.7 & & 20.0 $\pm$ 6.4 \\
\hline
\end{tabular}
\end{center}
\end{table*}

\section{Analysis details}\label{section-analysis}

\subsection{Strange particle reconstruction}

The strange hadrons, \ks, $\Lambda$(\alam), \xim(\axi), \omm(\aom), and $\phi$, have short lifetimes, and decay into a pair of charged particles or into one charged particle plus a $\Lambda$(\alam). All of them can be reconstructed using the invariant mass technique.
The corresponding decay channels and branching ratios are \cite{Olive:2016xmw} 
\begin{center}
\begin{tabular}{ll}
$\mathrm{K}^{0}_{\mathrm S} \rightarrow \pi^{+} + \pi^{-}$, & 69.20\%; \\
$\Lambda(\overline{\Lambda})  \rightarrow p(\overline{p}) + \pi^{-}(\pi^{+})$, & 63.9\%; \\
$\Xi^-(\overline{\Xi}^{+}) \rightarrow \Lambda(\overline{\Lambda}) + \pi^-(\pi^{+})$, & 99.887\%; \\
$\Omega^-(\overline{\Omega}^{+})  \rightarrow \Lambda(\overline{\Lambda}) + \mbox{K}^-(\mbox{K}^+)$, & 67.8\%; \\
$\phi  \rightarrow \mbox{K}^+ + \mbox{K}^-$, & 49.1\%. \\
\end{tabular}
\end{center}

The truncated mean of the ionization energy loss, $\langle dE/dx \rangle$, measured by the TPC, was used for identification of the charged daughter particles, $\pi^\pm$, $\mbox{K}^\pm$, and $p(\bar{p})$ \cite{Shao:2005iu}. Despite the finite statistical precision of the measured $\langle dE/dx \rangle$ for a certain track arising from a limited number of hit points measured by the TPC, the central value of the measured $\langle dE/dx \rangle$ as a function of momentum is well described by the Bichsel function for each particle species \cite{bichsel}. Hence a normalized $\langle dE/dx \rangle$, $n\sigma_{\rm particle}$, was used in particle identification. It is defined as
\begin{eqnarray}
n\sigma_{\rm particle} = \frac{1}{\sigma_{\rm particle}}\log \frac{\left< dE/dx\right>_{\rm measured}}{\left< dE/dx \right>^{\rm Bichsel}_{\rm particle}},
\label{form_dEdx} 
\end{eqnarray}
where $\left< dE/dx \right>^{\rm Bichsel}_{\rm particle}$ is the expected $\langle dE/dx \rangle$ from the Bichsel function for a certain particle species at a given momentum, and $\sigma_{\rm particle}$ is the $\langle dE/dx \rangle$ resolution of the TPC for the same particle species at the same momentum. The $n\sigma_{\rm particle}$ distribution at a given momentum is nearly Gaussian and is calibrated to be centered at zero with a width of unity for each particle species. By default, a loose cut of $|n\sigma_{\rm particle}|<4.0$ was used to select all the corresponding charged daughter particles for the reconstruction of \ks, $\Lambda$(\alam), and \xim(\axi). In order to reduce the combinatorial background, a tighter $|n\sigma_{p}|<3.0$ was used for selecting the protons in $\Omega$ reconstruction, and $|n\sigma_{\rm K}|<2.0$ was used for the kaons in $\phi$ meson reconstruction. In order to improve the average momentum and energy-loss resolution, the charged daughter particle tracks were required to consist of at least 16 TPC hit points for the reconstruction of \ks, $\Lambda$(\alam), \xim(\axi), and \omm(\aom), while at least 16 hit points (including the primary vertex) were required for the kaons in $\phi$ meson reconstruction. The \ppt\ of daughter particles was required to be larger than 0.10 \GeVc\ for \ks, $\Lambda$, and $\Xi$ reconstruction, and larger than 0.15 \GeVc\ for $\Omega$ and $\phi$ reconstruction.

Due to the large number of final state particles in Au+Au collisions, there is a significant amount of combinatorial background in the invariant mass distributions of all strange hadrons.
The weakly decaying strange hadrons, \ks, $\Lambda$(\alam), \xim(\axi), and \omm(\aom), have a typical decay length of $c\tau \approx 2$--$7$ cm. Their decay topology can be reconstructed well with their daughter particle tracks measured by the TPC with a precision of $\approx$1 mm. Therefore, a certain set of cuts can be applied to the topological variables in order to significantly reduce the combinatorial background. Such variables include the distance of closest approach (DCA) between the two daughter tracks, the DCA of the daughter tracks to the primary vertex, the DCA of the projected strange hadron path to the primary vertex, the decay length of strange hadrons, and the angles between the spatial vector pointing from the production vertex to the decay vertex and the momentum vector of strange hadrons. These cuts were optimized as a compromise between background reduction and signal efficiency. Table \ref{tab:cuts_V0} shows the default topological cuts used for $V^0$ particle (\ks, $\Lambda$, and \alam) reconstruction in this analysis. For the reconstruction of multi-strange hyperons, \xim(\axi) and \omm(\aom), the $\Lambda$ candidates reconstructed with $p$ and $\pi$ daughter tracks are further combined with the ``bachelor" tracks --- the identified $\pi^{\pm}$ for $\Xi$ reconstruction or the identified $\mbox{K}^{\pm}$ for $\Omega$ reconstruction. In order to reduce the combinatorial background, the $\Lambda$ candidates were required to be inside the invariant mass window of $[M_{\Lambda}-0.012$ GeV$/c^2$, $M_{\Lambda}+0.012$ GeV$/c^2]$ and $[M_{\Lambda}-0.006$ GeV$/c^2$, $M_{\Lambda}+0.006$ GeV$/c^2]$ for $\Xi$ and $\Omega$ reconstruction, respectively, with the known $\Lambda$ mass $M_{\Lambda}=1.115683$ GeV$/c^2$ \cite{Olive:2016xmw}. The decay topology of multi-strange hyperons is more complicated compared to those of $V^0$ particles, and hence more topological cuts were used in these hyperon reconstructions. Tables \ref{tab:cuts_Xi} and \ref{tab:cuts_Om} show the default topological cuts for $\Xi$ and $\Omega$ reconstruction, respectively.

The $\phi$ meson decays strongly at the primary collision vertex and has a short lifetime. Hence its two daughter kaons also appear to originate from the primary vertex. Therefore the primary tracks, which have the primary vertex included in their fit, were used for $\phi$ meson reconstruction. The DCA of their associated TPC tracks, which exclude primary vertex in their fit, to the primary vertex were required to be less than 3 cm. In order to avoid split tracks, the ratio of the number of hits on a track to the maximum possible number of hits that track may possess was required to be larger than $0.52$. Due to the electron/positron contamination in the selected kaon candidates, photon conversions ($\gamma^* \rightarrow e^+e^-$) contribute significantly to the residual background in $\mbox{K}^+\mbox{K}^-$ invariant mass distributions. This contribution can be removed effectively by a cut on the dip angle $\delta$ \cite{Abelev:2007rw,Abelev:2008aa,Adams:2004ux,Abelev:2008zk}, which is defined as
\begin{eqnarray}
\delta = \cos^{-1}\left[\frac{p_{\rm T1}p_{\rm T2}+p_{\rm z1}p_{\rm z2}}{p_{\rm 1}p_{\rm 2}}\right],
\end{eqnarray}
where $p_{\rm 1}$, $p_{\rm 2}$, $p_{\rm T1}$, $p_{\rm T2}$, $p_{\rm z1}$, $p_{\rm z2}$ are total, transverse, and longitudinal momenta of the two candidiate tracks. By default, the $\delta$ was required to be greater than 0.04 radians in this analysis.

% Table III
\begin{table*}[hbt]
\begin{center}
\caption{Topological cuts used for $V^0$ particle (\ks, $\Lambda$, and \alam) reconstruction. In this table, $\vec{r}_{V^0}$ and $\vec{r}_{\rm PV}$ denote the $V^0$ decay vertex position vector and the primary vertex position vector in the STAR coordinate system, respectively. $\vec{p}_{V^0}$ is the reconstructed $V^0$ momentum vector. Slightly tighter topological cuts, together with a tighter particle identification cut of $|n\sigma_{p(\pi)}|<3.8$, were used for \alam\ reconstruction at \sqrtsNN\,=\,7.7 GeV to reduce the combinatorial background.
\label{tab:cuts_V0}}
\begin{tabular}{c|c|c|c}
\hline
\begin{minipage}[c][0.5cm][c]{0.2\columnwidth} Cut \end{minipage} & \begin{minipage}[c][0.5cm][c]{0.4\columnwidth} \ks\ \end{minipage} & \begin{minipage}[c][0.5cm][c]{0.46\columnwidth} \alam\ ($\ge11.5$ GeV), $\Lambda$ \end{minipage} & \begin{minipage}[c][0.5cm][c]{0.46\columnwidth} \alam\ (7.7 GeV) \end{minipage}\\
\hline
DCA of $V^0$ to primary vertex & $ < 0.8$ cm & $< 0.8$ cm & $< 0.8$ cm \\
DCA of daughters to primary vertex & $> 0.7$ cm & $> 0.3$ cm ($p$), $>1.0$ cm ($\pi$)  & $> 0.5$ cm ($p$), $>1.5$ cm ($\pi$) \\
DCA between daughters & $< 0.8$ cm & $< 0.8$ cm & $< 0.8$ cm \\
$V^0$ decay length & $> 2.5$ cm & $>3$ cm & $>4$ cm \\
$(\vec{r}_{V^0}-\vec{r}_{\rm PV})\cdot\vec{p}_{V^0}$ & $>0$ & $>0$ & $>0$\\
\hline
\end{tabular}
\end{center}
\end{table*}

% Table IV
\begin{table*}[hbt]
\begin{center}
\caption{Topological cuts used for \xim\ and \axi\ reconstruction. In this table, $\vec{r}_{\Xi}$, $\vec{r}_{\Lambda}$ and $\vec{r}_{\rm PV}$ denote the $\Xi$ and $\Lambda$ decay vertex position vectors and the primary vertex position vector in the STAR coordinate system, respectively. $\vec{p}_{\Lambda}$ and $\vec{p}_{\Xi}$ are the reconstructed $\Lambda$ and $\Xi$ momentum vectors. Slightly tighter topological cuts, together with a tighter particle identification cut of $|n\sigma_{p(\pi)}|<3.6$ and a narrower \lam\ invariant mass window of $[M_{\Lambda}-0.010$ GeV$/c^2$, $M_{\Lambda}+0.010$ GeV$/c^2]$, were used for $\Xi$ reconstruction at \sqrtsNN\,=\,7.7 and 11.5 GeV to reduce the combinatorial background.
\label{tab:cuts_Xi}}
\begin{tabular}{c|c|c}
\hline
 \begin{minipage}[c][0.5cm][c]{0.33\columnwidth} Cut \end{minipage} & \begin{minipage}[c][0.5cm][c]{0.55\columnwidth} \xim\ and \axi\ ($\ge19.6$ GeV)\end{minipage} & \begin{minipage}[c][0.5cm][c]{0.55\columnwidth} \xim\ and \axi\ ($\le11.5$ GeV)\end{minipage}\\
\hline
DCA of $\Xi$ to primary vertex & $ < 0.8$ cm & $< 0.8$ cm \\
DCA of bachelor $\pi$ to primary vertex & $ > 0.8$ cm & $>0.8$ cm \\
DCA of $\Lambda$ to primary vertex & $[0.2, 5.0]$ cm & $[0.2, 5.0]$ cm \\
DCA of $\Lambda$-daughter $p$ to primary vertex & $> 0.5$ cm & $>0.5$ cm \\
DCA of $\Lambda$-daughter $\pi$ to primary vertex & $> 1.0$ cm & $>1.5$ cm \\
DCA between $\Lambda$ and bachelor $\pi$ & $< 0.8$ cm & $< 0.8$ cm \\
DCA between $\Lambda$-daughters & $< 0.8$ cm & $< 0.8$ cm \\
$\Xi$ decay length & $> 3.4$ cm & $> 4.0$ cm \\
$\Lambda$ decay length & $> 5.0$ cm & $> 5.0$ cm \\
$(\vec{r}_{\Lambda}-\vec{r}_{\rm PV})\cdot\vec{p}_{\Lambda}$ & $>0$ & $>0$\\
$(\vec{r}_{\Lambda}-\vec{r}_{\Xi})\cdot\vec{p}_{\Lambda}$ & $>0$ & $>0$ \\
$(\vec{r}_{\Xi}-\vec{r}_{\rm PV})\cdot\vec{p}_{\Xi}$ & $>0$ & $>0$  \\
$(\vec{r}_{\Xi}-\vec{r}_{\rm PV})\times\vec{p}_{\Xi}/|\vec{r}_{\Xi}-\vec{r}_{\rm PV}|/|\vec{p}_{\Xi}|$ & $<0.2$ & $<0.12$ \\ 
\hline
\end{tabular}
\end{center}
\end{table*}

% Table V
\begin{table*}[hbt]
\begin{center}
\caption{Topological cuts used for \omm\ and \aom\ reconstruction. In this table, $\vec{r}_{\Omega}$, $\vec{r}_{\Lambda}$ and $\vec{r}_{\rm PV}$ denote the $\Omega$ and $\Lambda$ decay vertex position vectors and the primary vertex position vector in the STAR coordinate system, respectively. $\vec{p}_{\Lambda}$ and $\vec{p}_{\Omega}$ are the reconstructed $\Lambda$ and $\Omega$ momentum vectors. Cuts were optimized for each energy to reduce the combinatorial background.
\label{tab:cuts_Om}}
\begin{tabular}{c|c|c}
\hline
 \begin{minipage}[c][0.5cm][c]{0.33\columnwidth} Cut \end{minipage} & \begin{minipage}[c][0.5cm][c]{0.55\columnwidth} \omm\ \end{minipage} & \begin{minipage}[c][0.5cm][c]{0.55\columnwidth}  \aom\ \end{minipage}\\
\hline
DCA of $\Omega$ to primary vertex & $ < 0.4$ cm; & $< 0.4$ cm; \\
& $< 0.5$ cm (19.6 GeV)  & $< 0.5$ cm (19.6 and 27 GeV); \\
& & $< 0.6$ cm (7.7 GeV)\\
DCA of bachelor $\mbox{K}$ to primary vertex & $ > 1.0$ cm & $>1.0$ cm \\
DCA of $\Lambda$ to primary vertex & $>0.4$ cm; & $>0.4$ cm; \\
& $>0.3$ cm (7.7 GeV) & $>0.3$ cm (7.7 GeV) \\
DCA of $\Lambda$-daughter $p$ to primary vertex & $> 0.6$ cm & $>0.6$ cm \\
DCA of $\Lambda$-daughter $\pi$ to primary vertex & $> 2.0$ cm & $>2.0$ cm \\
DCA between $\Lambda$ and bachelor $\mbox{K}$ & $< 0.7$ cm & $< 0.7$ cm; \\
& & $< 1.0$ cm (7.7 GeV) \\
DCA between $\Lambda$-daughters & $< 0.7$ cm & $< 0.7$ cm; \\
& & $< 1.0$ cm (7.7 GeV) \\
$\Omega$ decay length & $> 3.0$ cm; & $> 3.0$ cm; \\
& $> 2.0$ cm (7.7 GeV) & $> 2.0$ cm (7.7 and 11.5 GeV) \\
$\Lambda$ decay length & $> 5.0$ cm; & $> 5.0$ cm; \\
& $> 4.0$ cm (7.7 GeV) & $> 4.0$ cm (7.7 and 11.5 GeV) \\
$\Lambda$ decay length$ - \Omega$ decay length & $>0$ & $>0$\\
$(\vec{r}_{\Lambda}-\vec{r}_{\Omega})\cdot\vec{p}_{\Lambda}$ & $>0$ & $>0$\\
$(\vec{r}_{\Omega}-\vec{r}_{\rm PV})\cdot\vec{p}_{\Omega}$ & $>0$ & $>0$ \\
$(\vec{r}_{\Omega}-\vec{r}_{\rm PV})\times\vec{p}_{\Omega}/|\vec{r}_{\Omega}-\vec{r}_{\rm PV}|/|\vec{p}_{\Omega}|$ & $<0.12$; & $<0.12$; \\ 
& $<0.15$ (7.7 and 11.5 GeV) & $<0.15$ (7.7 and 11.5 GeV)  \\  
\hline
\end{tabular}
\end{center}
\end{table*}

\begin{figure*} [hbt]
\centering
\begin{minipage}[t]{1.0\textwidth}
\vspace{0pt}
\centering
\includegraphics[width=0.38\textwidth]{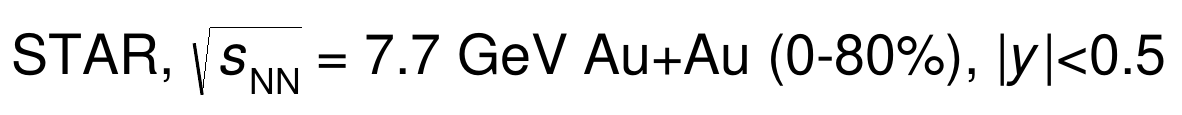}
\end{minipage}
\begin{minipage}[t]{0.246\textwidth}
\vspace{0pt}
\centering
\includegraphics[width=\textwidth]{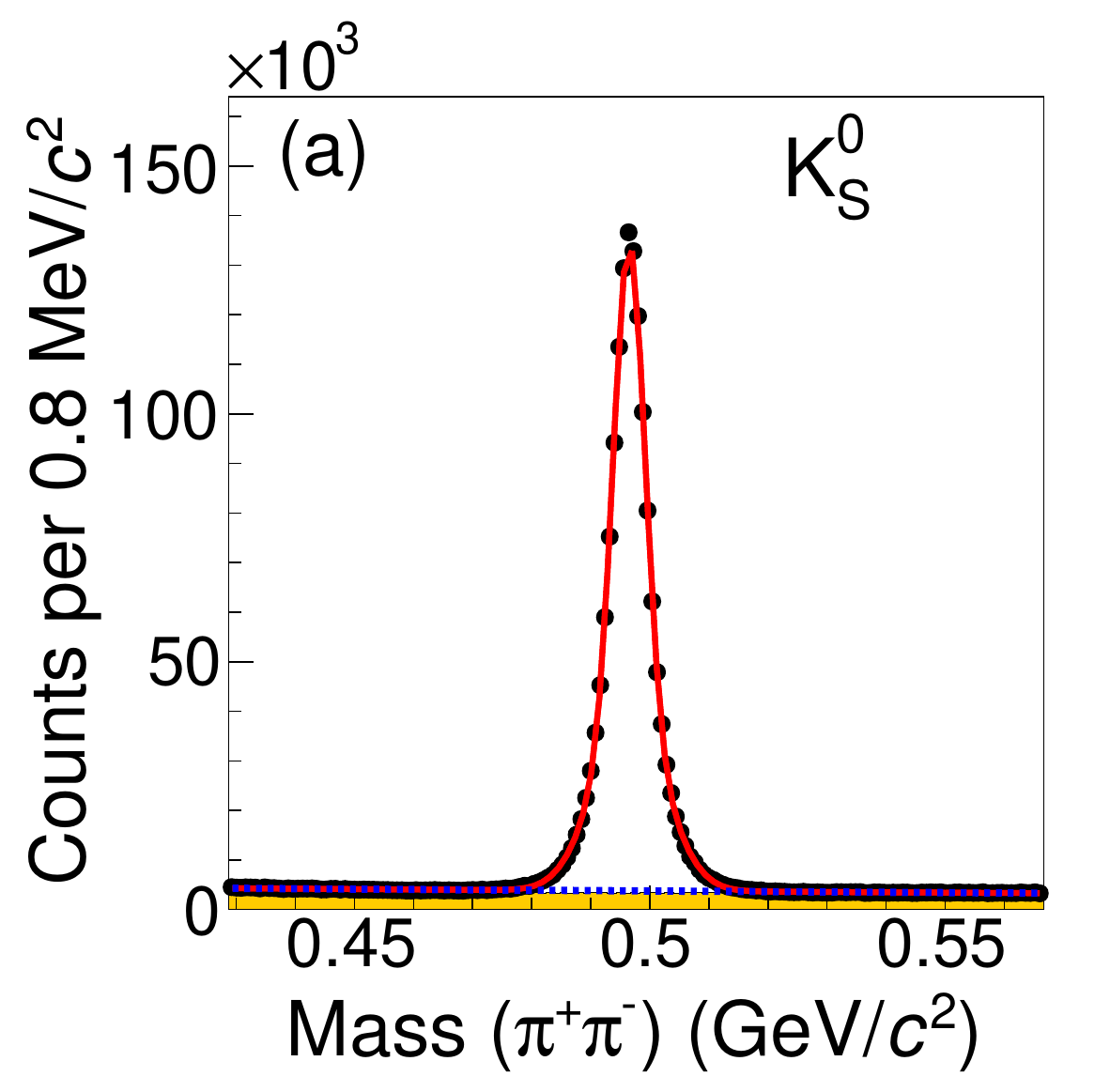}
\end{minipage}
\begin{minipage}[t]{0.246\textwidth}
\vspace{0pt}
\centering
\includegraphics[width=\textwidth]{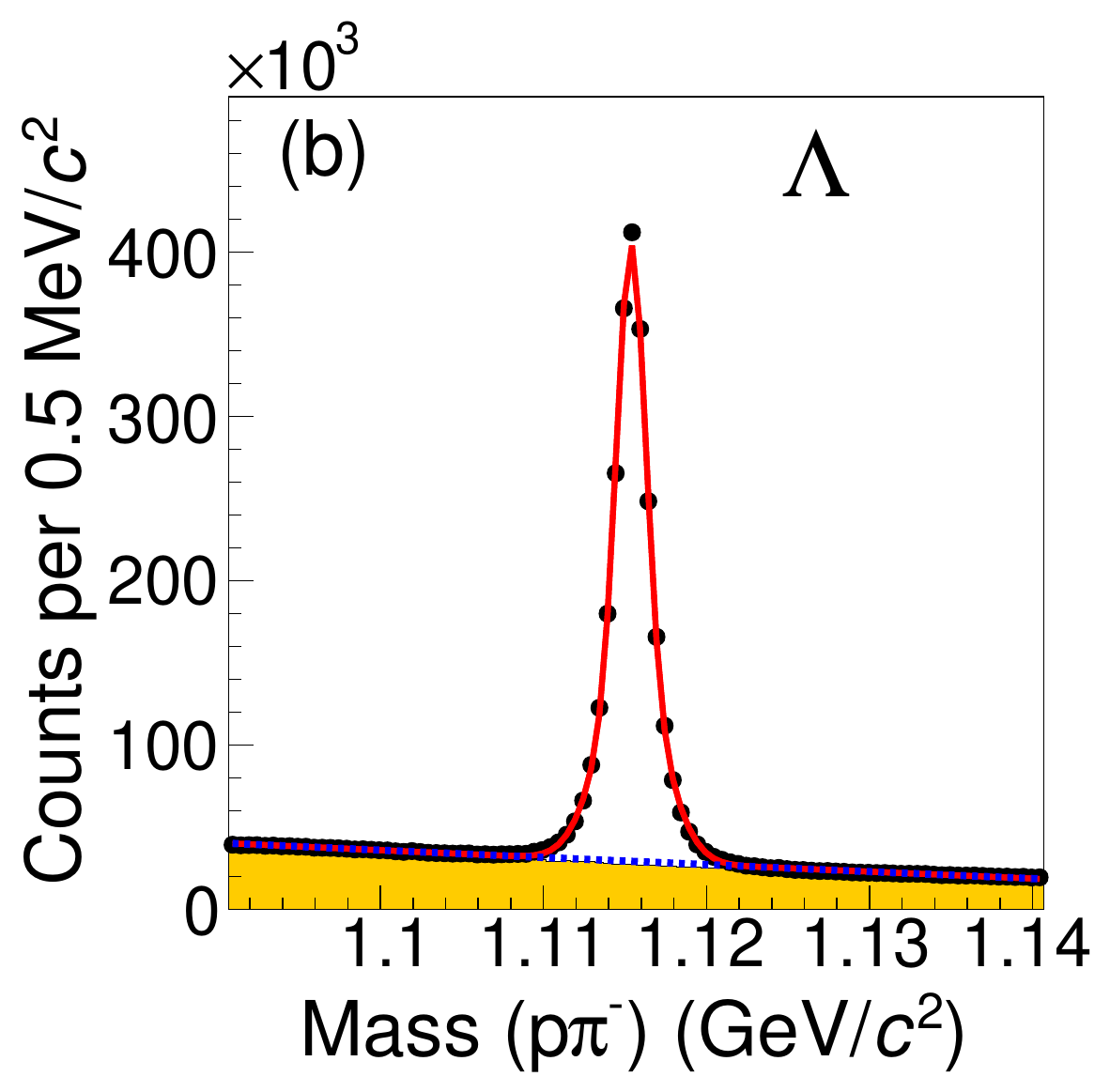}
\end{minipage}
\begin{minipage}[t]{0.246\textwidth}
\vspace{0pt}
\centering
\includegraphics[width=\textwidth]{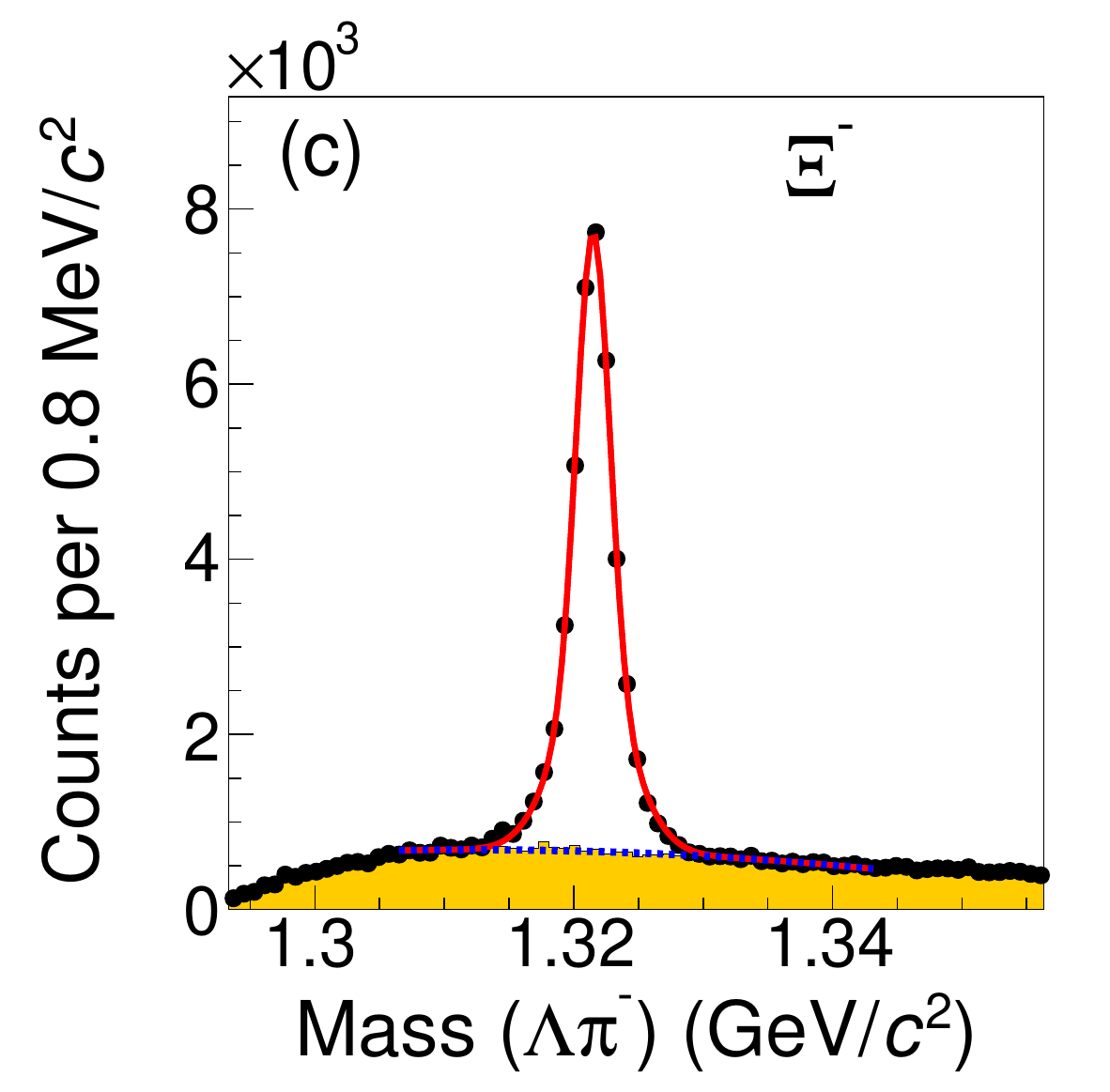}
\end{minipage}
\begin{minipage}[t]{0.246\textwidth}
\vspace{0pt}
\centering
\includegraphics[width=\textwidth]{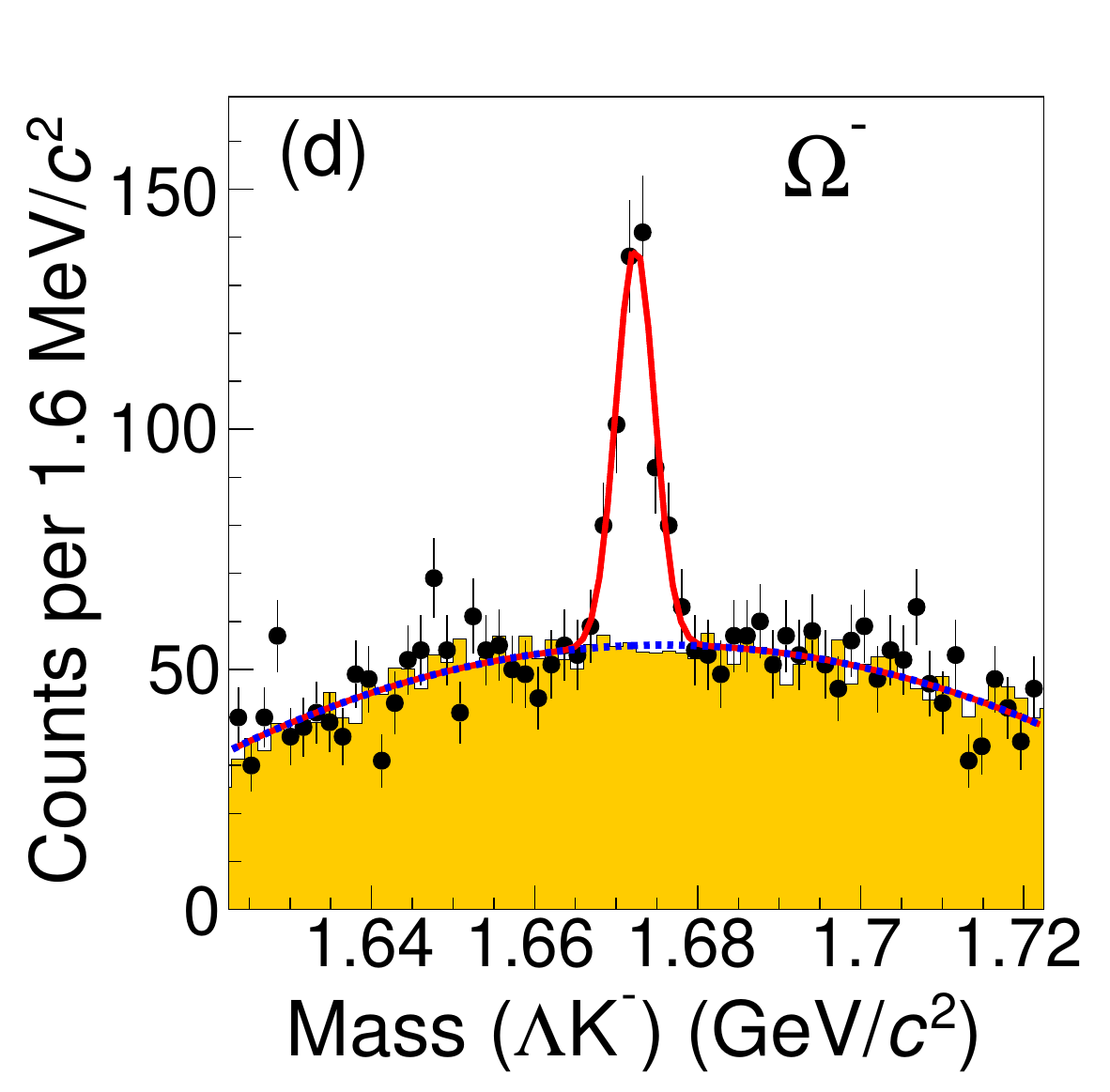}
\end{minipage}
\caption{\ks, \lam, \xim, and \omm\ invariant mass distributions in Au+Au collisions at \sqrtsNN\ = 7.7~GeV. The red solid lines represent the function fit results (double Gaussian plus polynomial for \ks, \lam, and \xim; single Gaussian plus polynomial for \omm), and the blue dashed lines are the fitted background contributions. The orange area shows the corresponding rotational background distribution.} \label{fig_invmass}
\end{figure*}

\begin{figure} [htb]
\centering
\vspace{0pt}
\includegraphics[width=6.5cm]{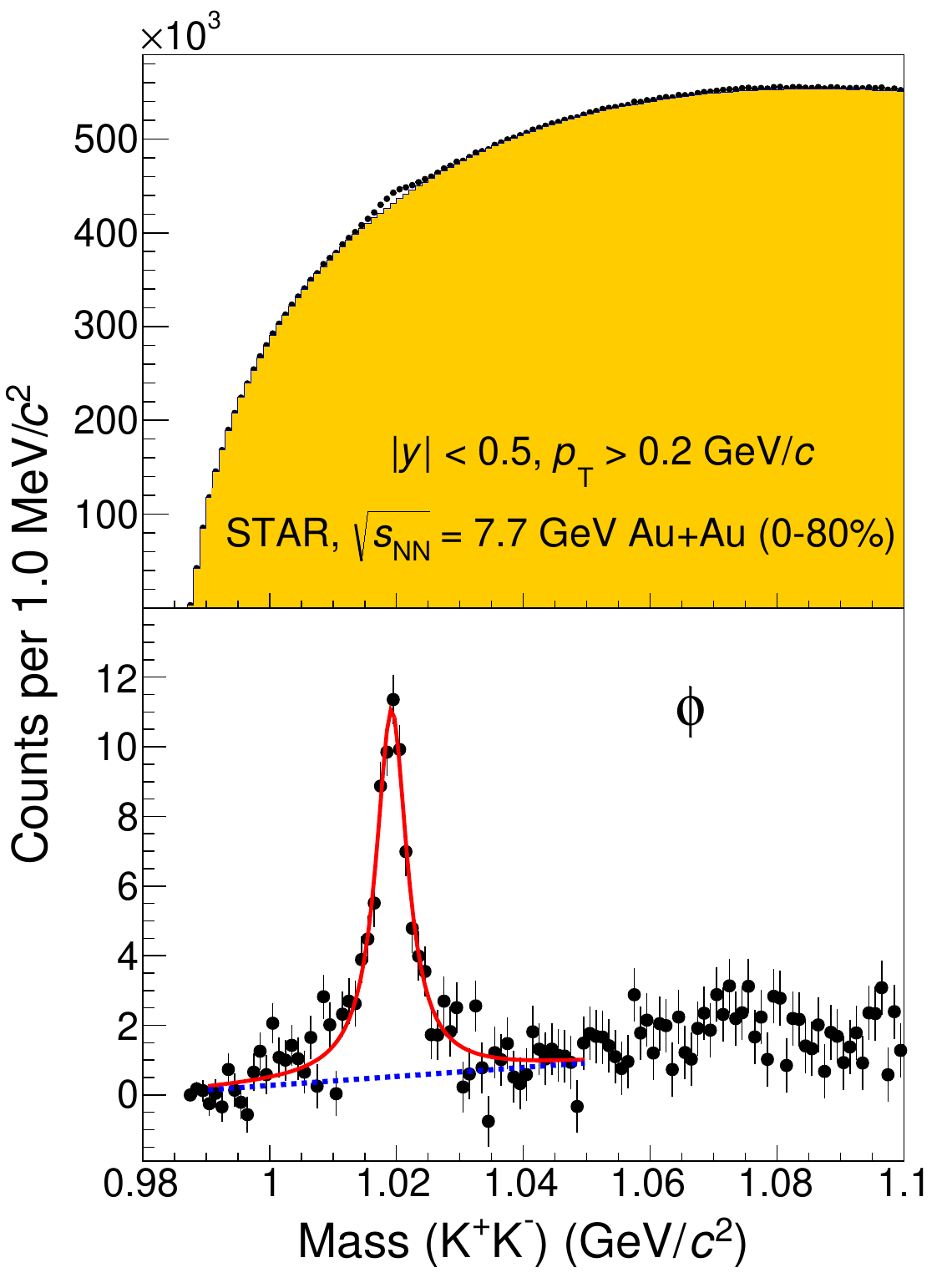}
\caption{$\phi$ invariant mass distributions in Au+Au collisions at \sqrtsNN\ = 7.7~GeV. The upper panel shows the unlike-charge invariant mass distribution (full points) and the mixed-event background (orange area). The lower panel shows the invariant mass distribution after subtracting the background. The red solid line represents the function fit result (Breit-Wigner plus polynomial), and the blue dashed line is the fitted residual background contributions. } \label{fig_invmass_phi}
\end{figure}

After applying the corresponding selection cuts, the resulting invariant mass distributions are shown in Fig.~\ref{fig_invmass} (a) for \ks, (b) for $\Lambda$, (c) for $\Xi$, (d) for $\Omega$, and in Fig.~\ref{fig_invmass_phi} for $\phi$. Even with these cuts, some background remains under the mass peak of each hadron. The random combinatorial background was estimated using a rotation method \cite{huilong} for \ks, $\Lambda$, $\Xi$, and $\Omega$. In the rotation method, one daughter particle track was picked, $\pi^-$, for example, in the case of \ks\ reconstruction. Then both the two-dimensional (2D) position vector of the track's first hit (originating from the primary vertex) and its 2D momentum vector in the transverse plane were rotated by an angle of $\pi$ in this plane. The rotated track was used in the strange hadron reconstruction to break possible correlations between the daughter particles and mimic the random combination contribution. In the $\Omega$ analysis, the bachelor $\mathrm{K}^\pm$ tracks were rotated by five different angles from $\pi/3$ to $5\pi/3$ to increase the background statistics. For $\phi$ meson analysis, in order to minimize the statistical errors, the combinatorial background was estimated with the mixed-event technique. The detailed description can be found in Refs.~\cite{Abelev:2007rw,Abelev:2008aa,Adams:2004ux,Abelev:2008zk}. The invariant mass distribution, the mixed-event background, and the background subtracted distribution are shown in Fig.~\ref{fig_invmass_phi}.

Besides the combinatorial background, there is also a residual background for each particle. This residual background originates from unavoidable particle misidentification. For example, a proton from a $\Lambda$ decay misidentified as a $\pi^+$ may be combined with the $\pi^-$ daughter of $\Lambda$, thereby contributing to the residual background in \ks\ reconstruction. In $\Xi$ reconstruction, a proton from a real $\Lambda$ decay can be combined with another random pion to form a fake $\Lambda$ candidate, which is then combined with the pion daughter of the real $\Lambda$ (as the bachelor pion) to form a fake $\Xi$. The bachelor pion of a $\Xi$ can be misidentified as a kaon, thereby contributing to the residual background in $\Omega$ reconstruction. In order to remove these kinds of residual background, veto cuts were introduced. In \ks\ reconstruction, the $\pi^+$($\pi^-$) daughter of a \ks\ candidate is assumed to be the $p$($\bar{p}$) daughter of a \lam(\alam) to re-calculate the invariant mass. If it falls inside the invariant mass peak of \lam(\alam), then the \ks\ candidate is rejected. For $\Xi$ reconstruction, the proton daughter will be combined with the pion bachelor to calculate the invariant mass. If it falls inside the $\Lambda$ invariant mass peak, the $\Xi$ candidate will be rejected. In $\Omega$ reconstruction, the bachelor kaon is assumed to be a pion to re-calculate the invariant mass. If it falls inside the invariant mass peak of $\Xi$, then the $\Omega$ candidate will be rejected. There also exists minor residual background in the $\Lambda$ invariant mass distribution due to the misidentification of the \ks\ daughters. However, the veto of this residual background would produce a significant drop in $\Lambda$ reconstruction efficiency due to the large width of \ks\ invariant mass peak, hence no veto cuts were applied for $\Lambda$. The veto cuts for \ks, $\Xi$, and $\Omega$ were applied both in signal reconstruction and in construction of the rotational background. The \ks, $\Xi$, and $\Omega$ invariant mass distributions shown in Fig.~\ref{fig_invmass} were obtained after applying the corresponding veto cuts.

The background distributions estimated via rotational or mixed-event methods for \ks, $\Lambda$, $\Xi$, $\Omega$, and $\phi$ are subtracted from the corresponding signal distributions. For \ks, $\Lambda$, and $\Xi$, a double or single Gaussian plus polynomial fitting to the resulting invariant mass distribution around the signal peaks was used to determine the signal peak width as well as the shape of the remaining residual background. The signal peak was defined as $[\mu-4\sigma,\mu+4\sigma]$ and $[\mu-4.5\sigma,\mu+4.5\sigma]$ for $V^0$ and $\Xi$, respectively, where $\mu$ and $\sigma$ are the corresponding Gaussian mean and variance parameters. The number of signal candidates was then obtained by subtracting the total background contribution inside the signal peak from the total number of candidates inside the peak. The background was estimated by one of two methods, either by integrating the polynomial functions from the fitting, or by using a side-band method at higher \ppt\ bins where a reasonable fitting cannot be achieved due to low statistics. The two side-bands on either side of the signal peak are selected symmetrically. Both have the half-peak width and are $5\sigma$($6\sigma$) away from the mean for $V^0$($\Xi$). For $\Omega$, the total number of signal counts was obtained by the side-band method, with the invariant mass peak position and width determined from the embedding simulation data. For the $\phi$ meson, the invariant mass distribution was fitted with the non-relativistic Breit-Wigner function for signal plus a polynomial function (up to second order) for the residual background. 

\subsection{Acceptance and reconstruction efficiency}\label{section-analysis-eff}

\begin{figure*}[htb]
\centering \vspace{0cm}
\includegraphics[width=14.5cm]{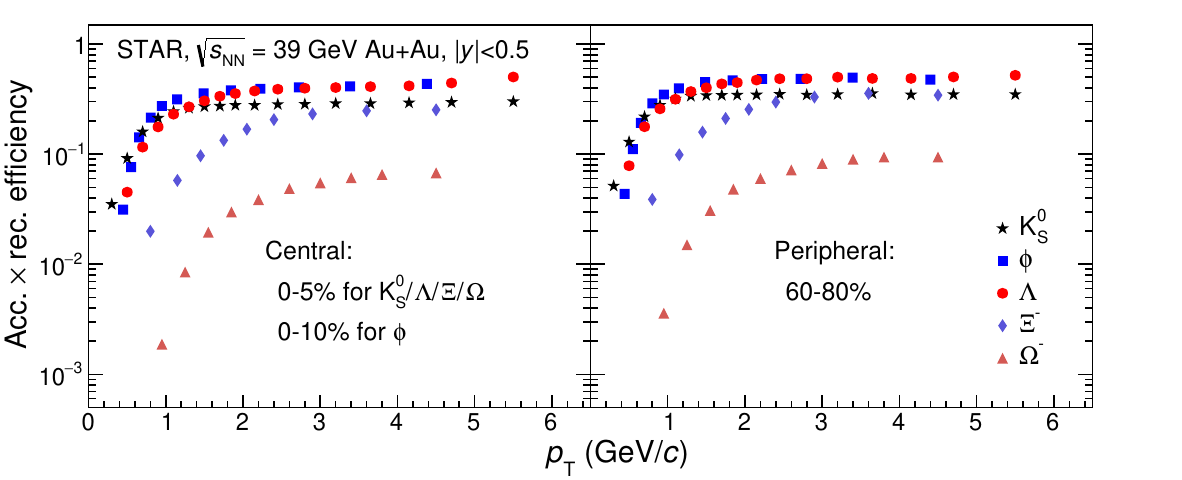}
\vspace{0cm} \caption{Geometrical acceptance and reconstruction efficiency of various strange hadrons at mid-rapidity ($|y|<0.5$) in central (left) and peripheral (right) Au+Au collisions at \sqrtsNN\ = 39~GeV. The branching ratios of measured decay channels are not taken into account here.}\label{fig_eff}
\end{figure*}

Whether a strange hadron can be observed by STAR is determined by the geometrical TPC acceptance for its decay daughter particles, their tracking efficiencies, and the efficiency of subsequent strange hadron reconstruction with the daughter particle tracks. The tracking efficiency and the strange hadron reconstruction efficiency depend on the final state particle multiplicity, which ranges from a few tracks in peripheral collisions to about a thousand tracks in central collisions. Therefore, in STAR, the geometrical acceptance and reconstruction efficiencies for each analyzed particle species were calculated using an embedding technique, in which the simulated Monte Carlo (MC) particles sampled in a given kinematic range were embedded into real events where their efficiency was studied. The number of embedded particles per event was about 5\% of the measured charged particle multiplicity for a given event. Embedded particles were all taken to originate from the real primary vertex in an event. The subsequent strange hadron propagation through STAR, strange hadron decay, and daughter particle propagation were simulated with the GEANT package \cite{geant}. The TPC detector response to the charged daughter particles was simulated with the STAR TPC response simulator (TpcRS). The simulated electronics signals were mixed with those from the real event and processed with the STAR tracking, event reconstruction, and strange hadron reconstruction algorithms. The acceptance and reconstruction efficiency were obtained by dividing the number of reconstructed MC strange hadrons by that of input MC in a certain kinematic range. As an example, the calculated efficiencies for different strange particles in central and peripheral Au+Au collisions at 39 GeV are shown as a function of \ppt\ in Fig.~\ref{fig_eff}. Generally, the efficiencies increase toward peripheral collisions and lower collision energies due to decreasing track multiplicities. As shown in Fig.~\ref{fig_eff}, at 39 GeV, from 0--5\% to 60--80\% collisions,  the \ks\ efficiency increases by $\approx$47\% at \ppt\ $\approx 0.3$ \GeVc\ and by $\approx$16\% at \ppt\ $\approx 5.5$ \GeVc, while the \xim\ efficiency increases by $\approx$95\% at \ppt\ $\approx 0.8$ \GeVc\ and by $\approx$36\% at \ppt\ $\approx 4.5$ \GeVc. With the same analysis cuts, the efficiencies for antibaryons (\alam, \axi\ and \aom, not shown in Fig.~\ref{fig_eff} for clarity) are very similar to those of the respective baryons. The efficiency for $\Omega$ is much smaller than for $\Xi$ is due to both the relatively tighter analysis cuts in $\Omega$ reconstruction and the decay-in-flight of the bachelor kaon.

\subsection{Weak decay feed-down correction for $\Lambda$}\label{section-analysis-fd}

The reconstructed $\Lambda$ hyperons with the cuts listed in Table~\ref{tab:cuts_V0} contain both the prompt components originating from the primary vertex and the secondary components from the weak decays of $\Xi$, $\Xi^0$, and $\Omega$ hyperons. The $\Lambda$ hyperons from the electromagnetic decay of $\Sigma^0$ hyperons are also considered to be prompt since they are not experimentally distinguishable from those directly originating from the Au+Au collisions. Naturally, the $\Lambda$ hyperons from secondary weak-decay vertices have different distributions in the topological cut variables\,---\,for example, the DCA of $V^0$ to the primary vertex. Hence it is mandatory to subtract their contributions to the reconstructed $\Lambda$ yields before applying the acceptance and reconstruction efficiency corrections described in Sec.~\ref{section-analysis-eff}. As shown in Table~\ref{tab:cuts_V0}, a tight cut on the DCA of $\Lambda$ candidates to the primary vertex was used to reduce the secondary contributions. However, some fraction of the secondary $\Lambda$ hyperons still passed this criterion, especially in the high \ppt\ regions. Their contribution to the prompt $\Lambda$ sample was further evaluated with the help of the $\Xi$ and $\Xi^0$ MC embedding data. With these MC data, the prompt $\Lambda$ reconstruction cuts (in Table~\ref{tab:cuts_V0}) were applied to the reconstruction of the secondary MC $\Lambda$ particles from the MC $\Xi$ or $\Xi^0$ decays. Then the total number of reconstructed secondary MC $\Lambda$ particles was scaled according to the corrected yields of the measured $\Xi$ and $\Xi^0$ particles (assuming that $\Xi^0$ has the same yield as $\Xi$ since it cannot be measured by STAR). Those scaled values represent the feed-down contribution and are subtracted from the raw  $\Lambda$ yields. The relative contribution of the secondary $\Lambda$ from $\Xi$ and $\Xi^0$ decays was calculated for each \ppt\ interval in each collision centrality at each collision energy. For Au+Au collisions at 39 GeV, in 0--5\% central collisions, the relative feed-down contribution of $\Xi$ and $\Xi^0$ to \lam\ (\alam) ranges from $\approx$23\% (30\%) at \ppt\ = 0.5 \GeVc\ to $\approx$6\% (15\%) at \ppt\ = 5 \GeVc. For \lam, the feed-down contribution from \omm\ decay was not considered since it is expected to be negligible ($< 1$\%) due to the low yield of \omm\ relative to that of \lam\ in the BES energy range. For \alam, the feed-down contribution from \aom\ cannot be neglected due to the significantly larger ratio of \aom\ to \alam\ yield in more central collisions, at lower \ppt, and especially at lower BES energies. The \aom\ feed-down contribution to \alam\ was evaluated with the \aom\ embedding data and the corrected yields of the measured \aom\ particle at five BES energies, and subtracted from the corresponding raw \alam\ yields. In 0--5\% central collisions and at \ppt\ = 0.5 \GeVc, the relative feed-down contribution of \aom\ to \alam\ increases from $\lesssim3$\% to $\approx$6\% with the collision energy decreasing from 39 to 7.7 GeV.

\subsection{\ppt\ spectra extrapolation at low \ppt}\label{section-analysis-extra}

The \ppt\ spectra of each strange hadron were obtained by dividing the raw yield in a certain \ppt\ interval by the corresponding acceptance and reconstruction efficiencies presented in Sec.~\ref{section-analysis-eff}.
Due to limited detector acceptance at low \ppt, and finite statistics at high \ppt, the spectra were not measured in these regions and hence needed to be extrapolated to these two regions in order to obtain the \ppt\ integrated yield ($dN/dy$) as well as the averaged transverse mass ($\langle m_{\rm T} \rangle-m_0$), where $m_{\rm T}=\sqrt{p_{\rm T}^2+m_0^2}$ is the transverse mass and $m_0$ is the rest mass. The extrapolation to low \ppt\ is particularly important, since it contributes significantly to both observables, while the extrapolation to high \ppt\ usually provides a much smaller contribution. The blast-wave model \cite{Schnedermann:1993ws} can be used for fitting individually the low \ppt\ spectra and extrapolating them to the unmeasured lower \ppt\ region. This model assumes the particles are emitted from a radially expanding thermal source. A common kinetic freeze-out temperature $T$ and a transverse radial flow velocity profile $\beta=\beta_{S}(r/R)^n$ are used to characterize the source, where $\beta_S$ is the surface velocity, $r/R$ is the relative radial position in the source, and $n$ is the exponent of flow velocity profile. The \ppt\ disctribution of the particles is given by
\begin{eqnarray} 
\frac{d^2 N}{2 \pi p_{\rm T} dp_{\rm T} dy} \propto \int_0^R r\,dr\,m_{\rm T}\,I_0\left(\frac{p_{\rm T}\sinh\rho(r)}{T}\right)\nonumber\\
\times K_1\left(\frac{m_{\rm T}\cosh\rho(r)}{T}\right),
\end{eqnarray}
where $\rho(r)=\tanh^{-1}\beta$, $I_0$ and $K_1$ are the modified Bessel functions. The velocity profile parameter $n$ is set to 1 for all the blast-wave model fitting in this analysis.

The \ks\ low \ppt\ spectra can be well fitted and hence extrapolated to unmeasured lower \ppt\ regions ($<$0.2~\GeVc) with the blast-wave model and other two functions,
the exponential function
\begin{eqnarray} 
\frac{d^2 N}{2 \pi p_{\rm T} dp_{\rm T} dy} \propto e^{-\frac{m_{\rm T}}{T}},
\end{eqnarray}
and the Levy function
\begin{eqnarray}
\frac{d^2 N}{2 \pi p_{\rm T} dp_{\rm T} dy} \propto (1+\frac{m_{\rm T}-m_0}{nT})^{-n}.
\end{eqnarray}
In this analysis, all three functions were used to fit the low \ppt\ \ks\ spectra at all energies. For the blast-wave model, the fit range was [0.2, 1.4]~\GeVc. For the exponential function, the fit range was [0.2, 1.8]~\GeVc\ at 7.7 and 11.5 GeV, and [0.2, 1.2]~\GeVc\ at 19.6, 27, and 39 GeV. For the Levy function, the fit range was [0.2, 1.4]~\GeVc\ at 7.7 and 11.5 GeV, and [0.2, 2.0]~\GeVc\ at 19.6, 27, and 39 GeV. The difference between the results from these three functions were considered in the systematic errors for $dN/dy$ and $\langle m_{\rm T} \rangle-m_0$ due to low \ppt\ extrapolation. The Levy function produced a slightly better fit (lower $\chi^2$) for centralities within 30--80\%, while the blast-wave model was better for centralities within 0--30\% at all five collision energies. Therefore, the default \ks\ $dN/dy$ and $\langle m_{\rm T} \rangle-m_0$ values were calculated with those functions correspondingly. Although the contribution is almost negligible ($<10^{-5}$), the \ks\ spectra at intermediate-and-above \ppt\ were fitted separately and extrapolated to unmeasured higher \ppt\ regions (up to 10~\GeVc) with the Levy function for energies above 19.6 GeV or exponential function for energies below 11.5 GeV.

For the $\Lambda$ and $\Xi$ hyperons, the low \ppt\ spectra can be well fit and hence extrapolated to unmeasured lower \ppt\ regions ($<$0.4~\GeVc\ for $\Lambda$ and $<$0.6~\GeVc\ for $\Xi$) with the blast-wave model. A Boltzmann function
\begin{eqnarray}
\frac{d^2 N}{2 \pi p_{\rm T} dp_{\rm T} dy} \propto m_{\rm T} e^{-\frac{m_{\rm T}}{T}},
\end{eqnarray}
and the exponential function were used as alternatives for estimation of the systematic error due to low \ppt\ extrapolation. For $\Lambda$ and \alam, the blast-wave model fit range was [0.4, 3.0]~\GeVc\ for centralities within 0--60\% and [0.4, 2.0]~\GeVc\ for the centrality of 60--80\%. The fit ranges for Boltzmann and exponential functions were both [0.4, 1.4]~\GeVc\ at all energies and slightly narrowed to [0.4, 1.2]~\GeVc\ for $\Lambda$ in central collisions (0--10\%) at 7.7 GeV. For \xim\ and \axi, the blast-wave model fit range was [0.6, 3.2]~\GeVc\ for most centralities, and narrowed to [0.6, 2.2]~\GeVc\ in 60--80\% for all energies and in 40--60\% for 7.7 and 11.5 GeV. The fit ranges for Boltzmann and exponential functions were [0.6, 2.2]~\GeVc\ at 11.5 GeV, and [0.6, 2.6]~\GeVc\ at 19.6, 27, and 39 GeV. At 7.7 GeV,  the fit range was [0.6, 3.2]~\GeVc\ for \xim, and [0.6, 2.6]~\GeVc\ for \axi, when data are available. The high \ppt\ extrapolation was done using the Levy function for 19.6 GeV and higher energies, and the exponential function for 7.7 and 11.5 GeV. 

For the $\Omega$ hyperon, the exponential function was used to fit the spectra over the full measured \ppt\ range ($>$0.7~\GeVc\ for 7.7 GeV and $>$0.8~\GeVc\ for $\ge$11.5 GeV), while the Boltzmann function and the blast-wave model were used alternatively for estimation of the systematic error due to low \ppt\ extrapolation. 
For the $\phi$ meson, the Boltzmann function was found to fit better over the full measured \ppt\ range ([0.4, 1.7]~\GeVc) at 7.7 GeV, and in the low \ppt\ range ([0.4, 2.0]~\GeVc) at 11.5 GeV and higher energies, and was therefore used to extrapolate the spectra into unmeasured low \ppt\ regions ($<$0.4~\GeVc). The Levy function was used as the alternative to fit the full measured \ppt\ spectra for estimation of the systematic error due to low \ppt\ extrapolation.

The integrated $dN/dy$ was obtained by integrating the \ppt\ spectra data in the measured \ppt\ range and the fitted functions in the low \ppt\ and high \ppt\ extrapolated ranges. 
The contributions of low \ppt\ extrapolation in the integrated $dN/dy$ are $\approx$10--14\% for \ks, $\approx$13--28\% for \lam\ and \alam, $\approx$23--46\% for \xim\ and \axi, $\approx$39--46\% for \omm\ and \aom, and $\approx$17--28\% for $\phi$. These contributions are larger for lower energy and more peripheral collisions due to steeper \ppt\ spectra there.

The $\langle m_{\rm T} \rangle-m_0 $ is obtained by integrating the whole range of the $p_{T}$ spectra as follows
\begin{eqnarray}
\label{mtm0}\langle m_{\rm T} \rangle-m_0 = \frac{\int (m_{\rm T}-m_0) \frac{dN}{dp_{\rm T}} dp_{\rm T}}{\int \frac{dN}{dp_{\rm T}}dp_{\rm T}}.
\end{eqnarray} 
The same extrapolation functions used for the integrated $dN/dy$ were used again to calculate the numerator integral of $\langle m_{\rm T} \rangle-m_0 $.

\subsection{Systematic uncertainties}\label{section-analysis-sys}

\begin{table*}[hbt]
\begin{center}
\caption{Summary of systematic uncertainties for \ppt\ spectra. The range indicates the variation between \ppt\ bins, centralities and energies.
\label{tab:systematic}}
\begin{tabular}{c|c|c|c}
\hline
\begin{minipage}[c][0.5cm][c]{0.33\columnwidth} Components \end{minipage} & \begin{minipage}[c][0.5cm][c]{0.33\columnwidth} \ks\ \end{minipage} & \begin{minipage}[c][0.5cm][c]{0.33\columnwidth} $\Lambda$ and \alam\ \end{minipage} & \begin{minipage}[c][0.5cm][c]{0.33\columnwidth} \xim\ and \axi\ \end{minipage}\\
\hline
Signal extraction & 0.5--8\% & 0.5--9\% & 0.5--3\%\\
Particle identification & $< 0.5\%$ & $< 0.7\%$ & $< 1.1\%$ \\
Tracking & 1.5--7\% & 1.5--7\% & 3--15\%\\
Topological reconstruction & 1--4\% & 2--10\% & 3--8\% \\
Detector uniformity & 1--4\% & 1--7\% & 1--8\%\\
Weak decay feed-down correction & n.a. & \lam: 0.2--4\%; \alam: 0.5--10\%  & n.a. \\
Total uncertainty &2--9\% & \lam: 3--12\%; \alam: 3--13\% & 4--17\% \\ 
\hline
\end{tabular}
\end{center}
\end{table*}

Many possible sources which can contribute to the systematic uncertainties of the \ppt\ spectra were evaluated bin-by-bin in this analysis. 

In the signal extraction for \ks\ and $\Xi$, the side-band method and the fitting method with double-Gaussian-plus-polynomial functions were used in the estimation of background in the signal peak region. The difference was factored into the systematic uncertainty. The width of the signal peak was first determined with the double-Gaussian-plus-polynomial fitting and then varied to estimate its contribution to the systematic error. The shape of the signal peak in high \ppt\ bins deviated from the symmetric Gaussian shape due to cuts on decay length, which cause a systematic deviation in our signal counting method. The embedding data were used to estimate this deviation, which reaches $\approx$7\% (3\%) at $\approx5$ \GeVc\ for \ks\ ($\Xi$). For $\Lambda$, the side-band method cannot be used in estimating the systematic error due to the non-linear residual background shape. Hence the fitting ranges have been changed to account for the possible uncertainty in the background shape. Furthermore, the width of the signal peak was varied in the estimation of the systematic uncertainty. The deviation of the signal peak shape from Gaussian was estimated with MC simulation and its contribution to the systematic uncertainty found to be $\approx 8$\% for \ppt\ at $\approx5$ \GeVc. Different sources were assumed to be uncorrelated and hence summed quadratically to obtain the total systematic uncertainty in signal extraction, summarized in Table~\ref{tab:systematic}.

The daughter particle identification cuts, $|n\sigma|$, were varied from their default value of 4.0 to 3.6. The contribution of this cut to the systematic uncertainty in the \ppt\ spectra is small, as listed in Table~\ref{tab:systematic}. The systematic uncertainties due to tracking were estimated by varying the cuts on the minimum number of hit points from the default 16 to 26, the minimum number of hit points used for $dE/dx$ calculation from 1 to 10, and the minimum ratio of the number of hit points to the number of possible hit points from 0.45 to 0.55. The cut values were changed one at a time, and the raw yields and the efficiencies were both re-calculated accordingly to obtain the corrected \ppt\ spectra. The maximum deviations from the default spectra due to these three tracking cuts were accounted for in the systematic uncertainties. The minimum number of daughter hit points contributes the majority of the systematic error in tracking, while the other two cuts contribute minimally. The tracking uncertainties are larger at lower \ppt\ in more central collisions. For \ks\ and \alam\ in the most central Au+Au collisions at 39 GeV, the uncertainties are $\approx$ 6\% and $\approx$ 7\%, respectively, at \ppt\ = 0.5 \GeVc, while for \axi, the uncertainty is $\approx$ 15\% at \ppt\ = 0.8 \GeVc, as listed in Table~\ref{tab:systematic}.

\begin{figure*} [htb]
\centering
\begin{minipage}[t]{1.0\textwidth}
\vspace{0pt}
\centering
\includegraphics[width=0.59\textwidth]{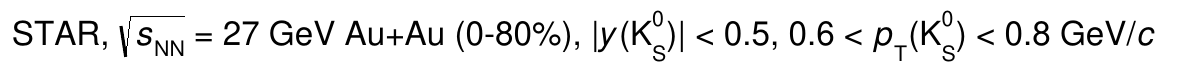}
\end{minipage}
\begin{minipage}[t]{0.245\textwidth}
\vspace{0pt}
\centering
\includegraphics[width=\textwidth]{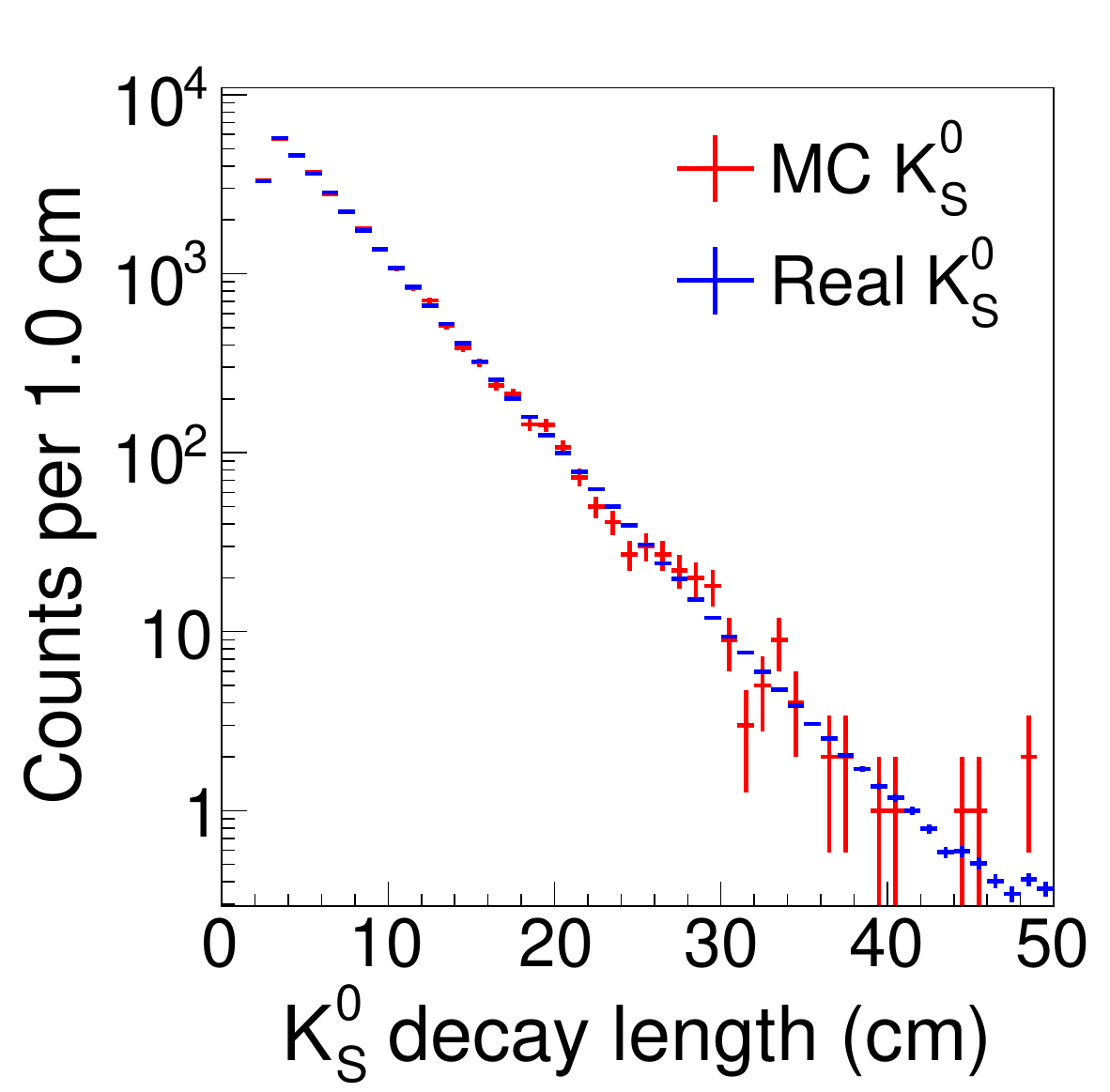}
\end{minipage}
\begin{minipage}[t]{0.245\textwidth}
\vspace{0pt}
\centering
\includegraphics[width=\textwidth]{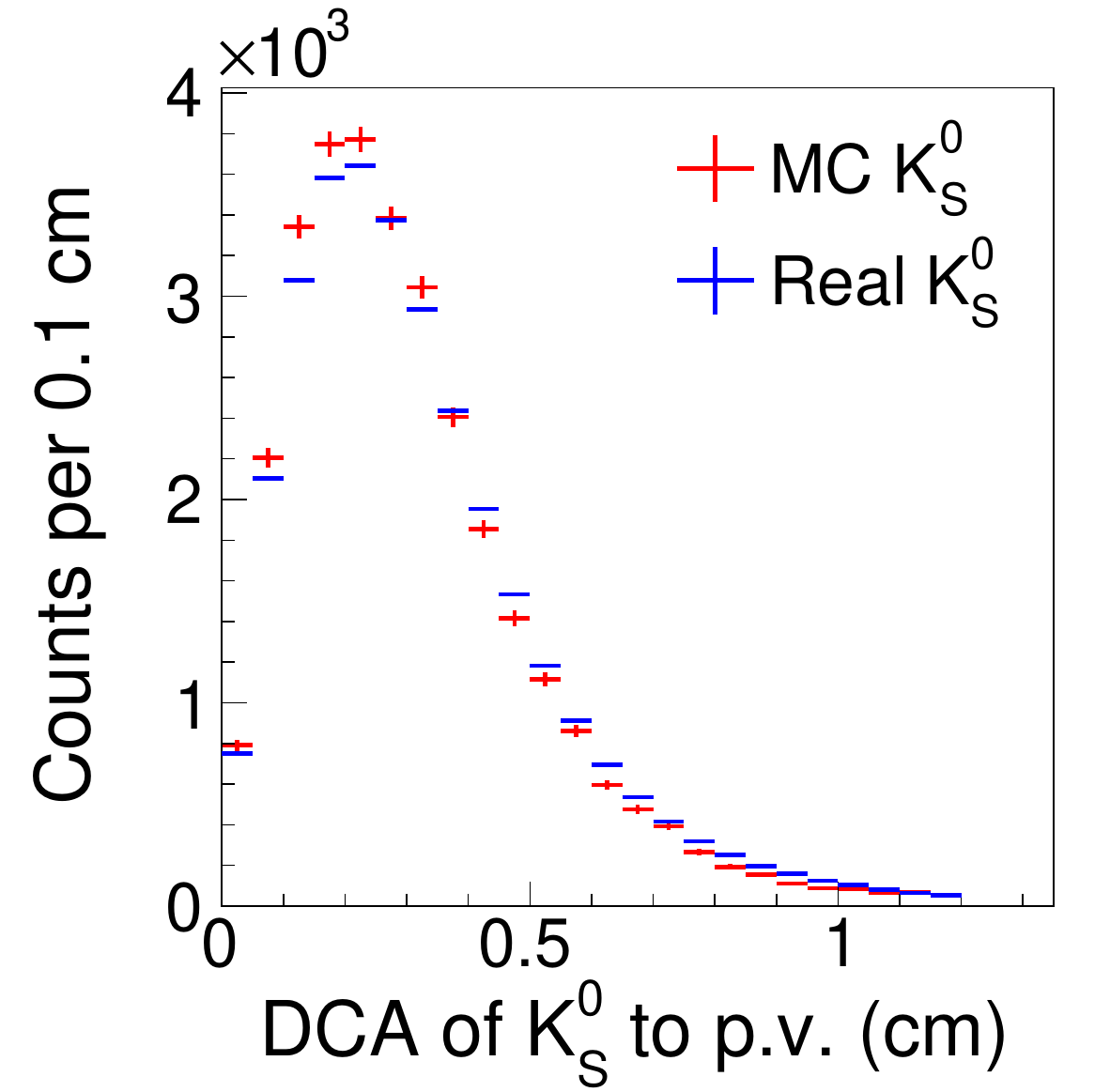}
\end{minipage}
\begin{minipage}[t]{0.245\textwidth}
\vspace{0pt}
\centering
\includegraphics[width=\textwidth]{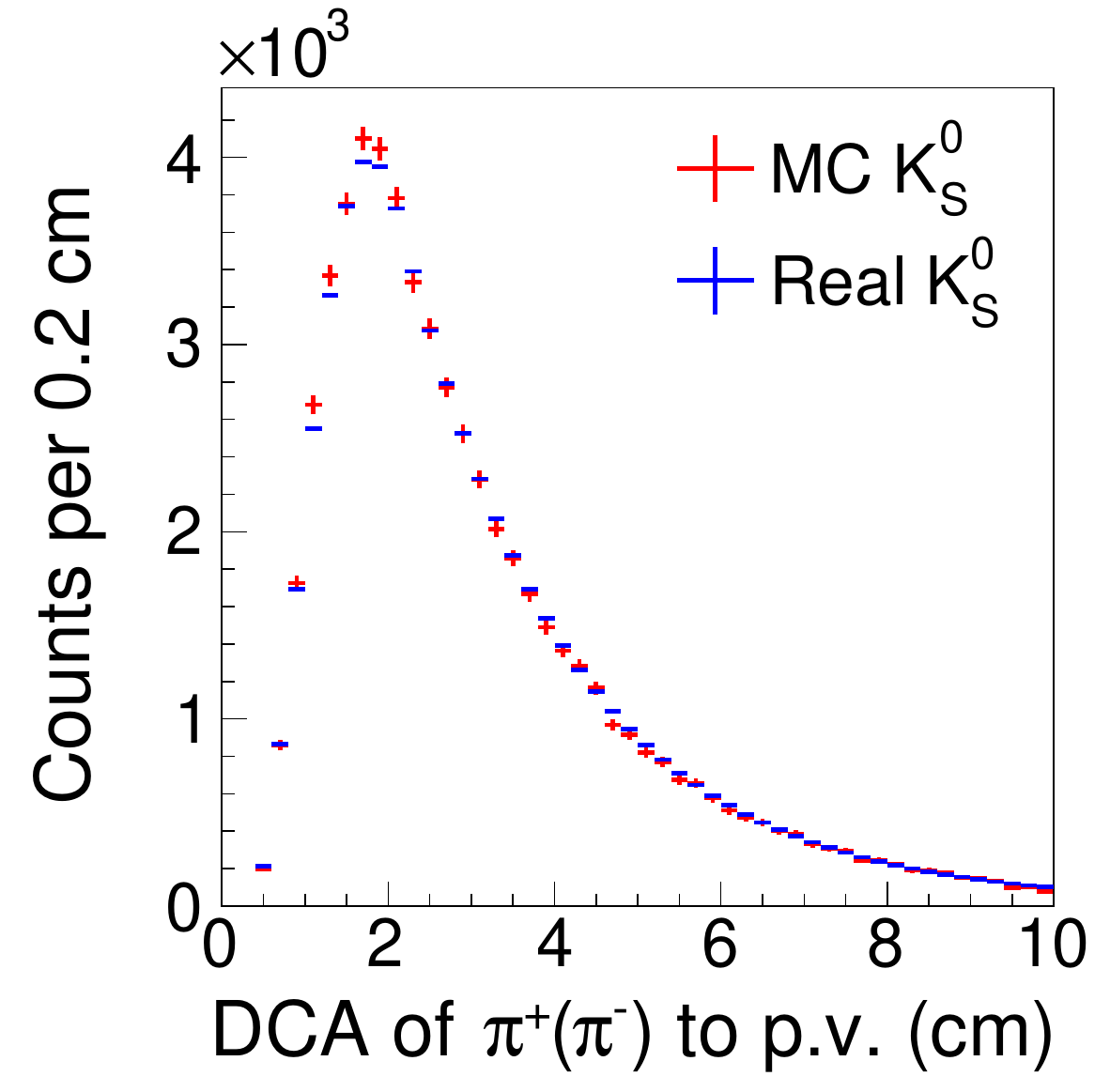}
\end{minipage}
\begin{minipage}[t]{0.245\textwidth}
\vspace{0pt}
\centering
\includegraphics[width=\textwidth]{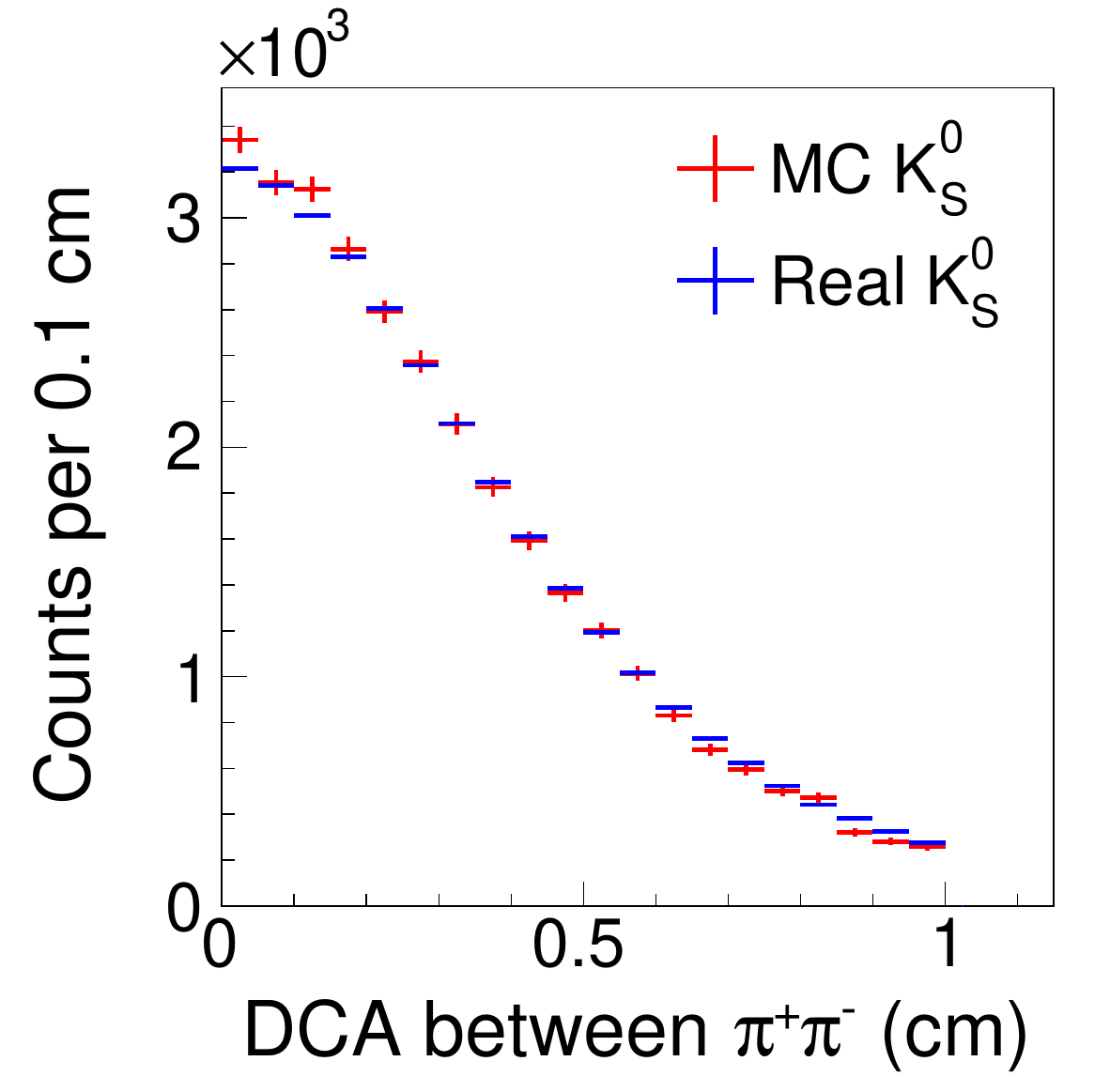}
\end{minipage}
\caption{MC and data comparison on \ks\ topological variable distributions in Au+Au collisions at \sqrtsNN\ = 27~GeV. A same set of loose selection cuts have been applied to both MC and data. The distributions of combinatorial backgrounds in data were estimated with the rotation method and subtracted, and then the resulting data distributions were scaled down to match with the MC statistics. } \label{fig_datamc}
\end{figure*}

The topological cuts were also varied one after another to study the systematic deviations of the \ppt\ spectra. For example, for \ks, the radial decay length cut value was varied in the range of $[2.5, 3.3]$ cm; the DCA of daughters in $[0.55, 0.8]$ cm; the DCA of \ks\ in $[0.6, 1.2]$ cm; and the DCA between daughters in $[0.5, 1.0]$ cm. As listed in Table~\ref{tab:systematic}, the systematic uncertainties from these geometric cuts were generally small compared to the tracking uncertainties, except in the most peripheral collisions, where the primary vertex resolution is worse due to low track multiplicity. The same method was applied to the systematic error studies for $\Xi$ and $\Lambda$ spectra despite there being more topological cuts involved. For these two hyperons, the systematic errors due to these cuts were also small compared to the tracking uncertainties except in the most peripheral collisions. Furthermore, a detailed comparison of the distributions of the topological cut variables from the embedded strange particles and those from strange particles reconstructed in real data was performed. A good agreement between the MC simulation and the data was achieved. As an example, Fig.~\ref{fig_datamc} shows the comparison of \ks\ MC topological variable distributions and those from the real data at 27 GeV. The systematic uncertainties due to the veto cuts in the \ks\ and $\Xi$ analyses were also studied and found to be negligible for both particles.

Non-uniformity of the detector acceptance for collisions at different primary vertex positions along the beam direction and detector asymmetry between forward and backward halves may contribute to the systematic uncertainty due to imperfect detector response simulation and limited real data sample size ($\approx10^5$ MB events) in embedding simulation. The $|z_{\rm vertex}|$ cut was varied to study the acceptance uniformity. At 39 GeV, the cut was changed within $[25, 40]$ cm. The resulting change in the \ppt\ spectra and hence the systematic error from this source is negligible. The default rapidity range $[-0.5, 0.5]$ was also divided into a forward half, $[0, 0.5]$, and a backward half, $[-0.5, 0]$. The maximum deviations of the resulting \ppt\ spectra from the default were accounted for in the systematic errors, which can be as large as 8\% for $\Xi$ in some \ppt\ bins.

For $\Lambda$, the systematic errors due to feed-down corrections were evaluated by propagating the $\Xi$ systematic errors to the $\Lambda$ raw yield, according to the fractions of those feed-down contributions. For simplicity, the $\Xi$ systematic errors were assumed to be uncorrelated with those for $\Lambda$. For \alam, the systematic error due to feed-down contribution from \aom\ was evaluated similarly and directly summed with that from $\overline{\Xi}$ in each \ppt\ bin.

The final bin-by-bin systematic error for a \ppt\ spectrum was a quadratic sum of all the above sources, assuming that they are fully uncorrelated, and is summarized in Table~\ref{tab:systematic}. For the details on the systematic uncertainties for the $\phi$ and $\Omega$ \ppt\ spectra, please refer to Ref.~\cite{Adamczyk:2015lvo}.

For the integrated yield, $dN/dy$, the systematic error in the measured \ppt\ range is simply the sum of the bin-by-bin systematic errors assuming that they are fully correlated. In the extrapolated low \ppt\ region, the systematic errors were estimated considering several potential sources. First, if a change in the analysis cuts produces a change in the \ppt\ spectra, then the extrapolation will also change. Second, different fit functions will have different extrapolations even under the same analysis cuts. Both sources were studied in this analysis by changing the analysis cuts and fitting functions in the extrapolation. The final extrapolated systematic errors are a quadratic sum of both contributions. The systematic error of $\langle m_{\rm T} \rangle-m_0$ was estimated in a similar manner by varying analysis cuts, extrapolation functions, and signal extraction methods. The systematic uncertainties of antibaryon-to-baryon ratios and nuclear modification factors were estimated independently by considering all the sources mentioned above, and hence the possible correlations of the systematic errors in the numerator and the denominator in these observables were consistently treated.

\section{Results and discussions}\label{section-results}

\subsection{Transverse momentum spectra}

\begin{figure*}[htbp]
\centering \vspace{0cm}
\includegraphics[width=12.5cm]{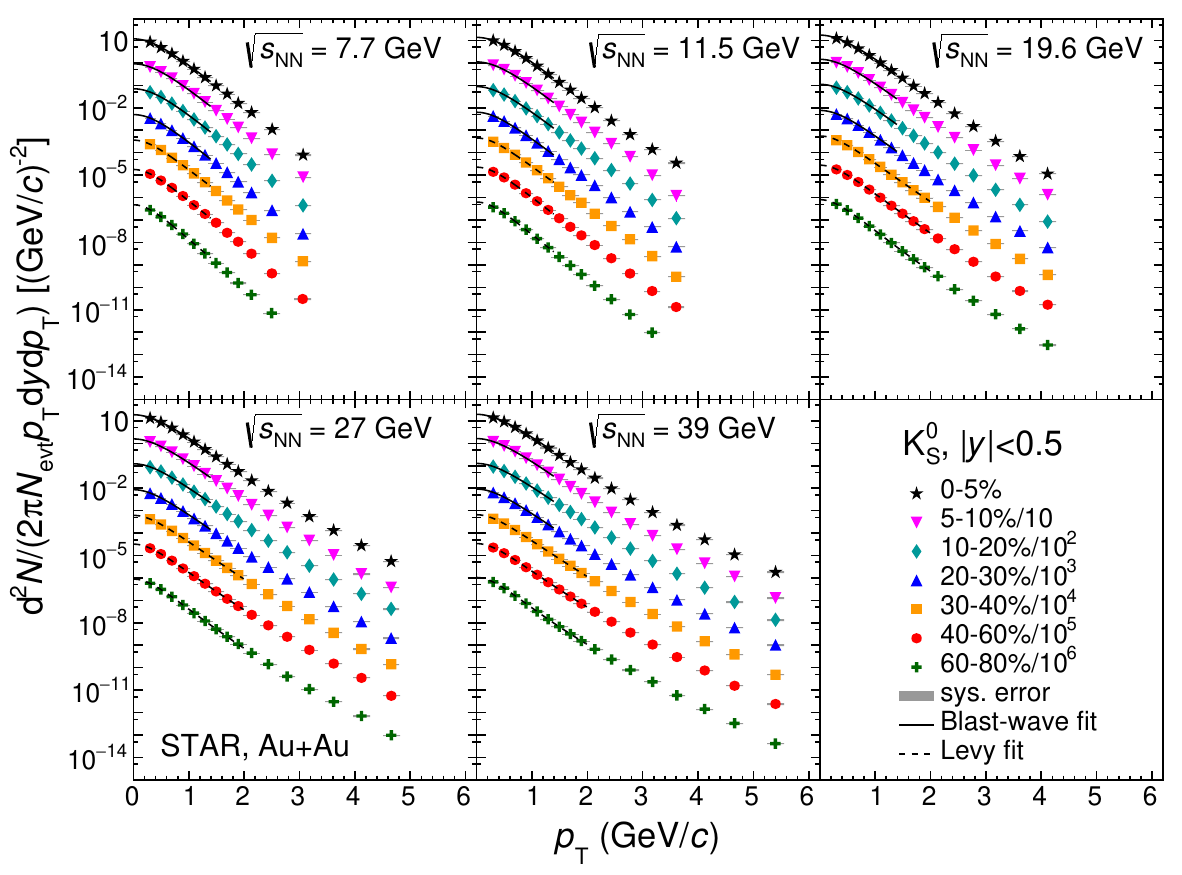}
\vspace{0cm} \caption{The transverse momentum spectra of \ks\ at mid-rapidity ($|y|<$ 0.5) from Au+Au collisions at different centralities and energies (\sqrtsNN = 7.7--39 GeV). The data points are scaled by factors of 10 from central to peripheral collisions for clarity. The vertical gray bands represent the systematic errors, which are small hence the bands look like horizontal bars. The blast-wave model (or Levy function) fit results are shown in the fit range and the low \ppt\ extrapolation range as solid (dashed) lines for the centrality bins within 0--30\% (30--80\%).}\label{fig:spectra_k0s} 
\end{figure*}

\begin{figure*}[htbp]
\centering \vspace{0cm}
\includegraphics[width=12.5cm]{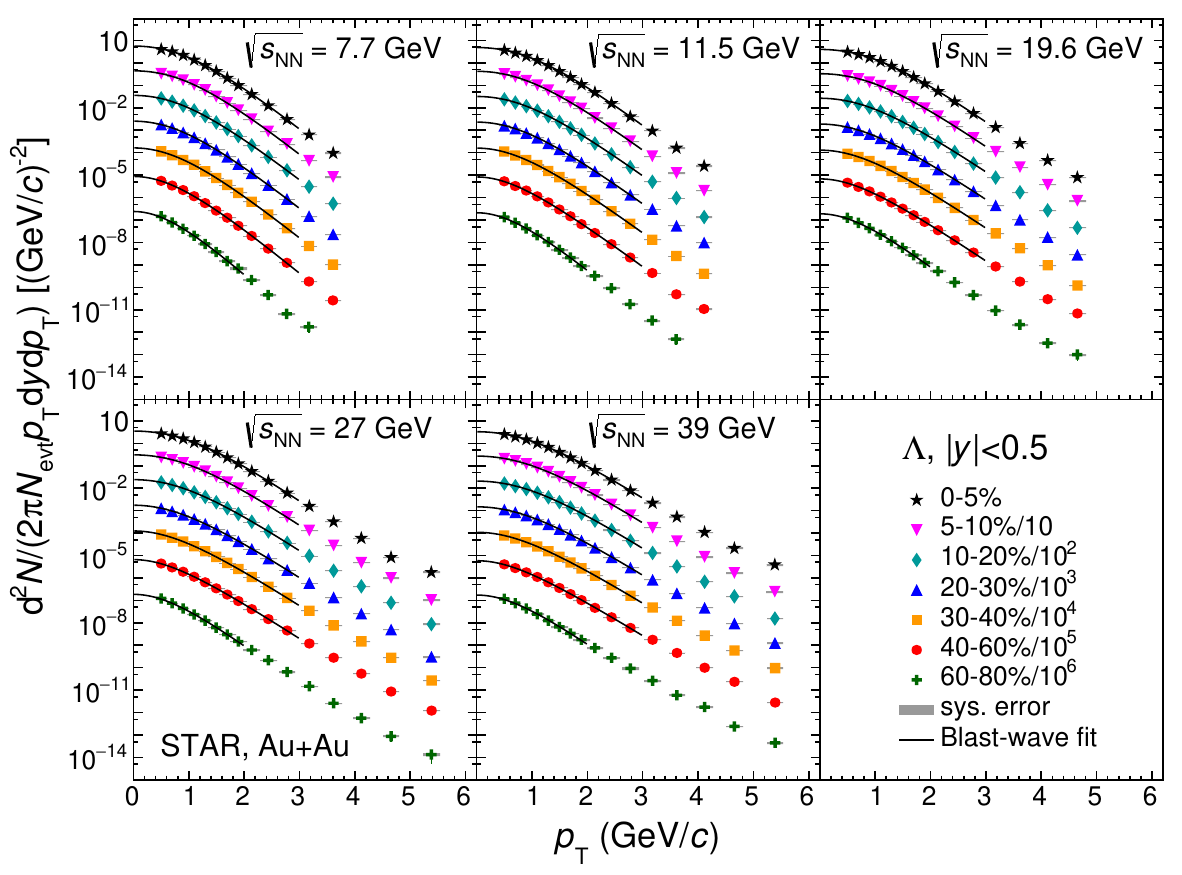}
\vspace{0cm} \caption{The transverse momentum spectra of \lam\ at mid-rapidity ($|y|<$ 0.5) from Au+Au collisions at different centralities and energies (\sqrtsNN = 7.7--39 GeV). The spectra are corrected for the feed-down of \xim\ and $\Xi^0$ decays. The data points are scaled by factors of 10 from central to peripheral collisions for clarity. The vertical gray bands represent the systematic errors, which are small hence the bands look like horizontal bars. The blast-wave model fit results are shown in the fit range and the low \ppt\ extrapolation range as solid lines for all centrality bins.}\label{fig:spectra_la} 
\end{figure*}

\begin{figure*}[htbp]
\centering \vspace{0cm}
\includegraphics[width=12.5cm]{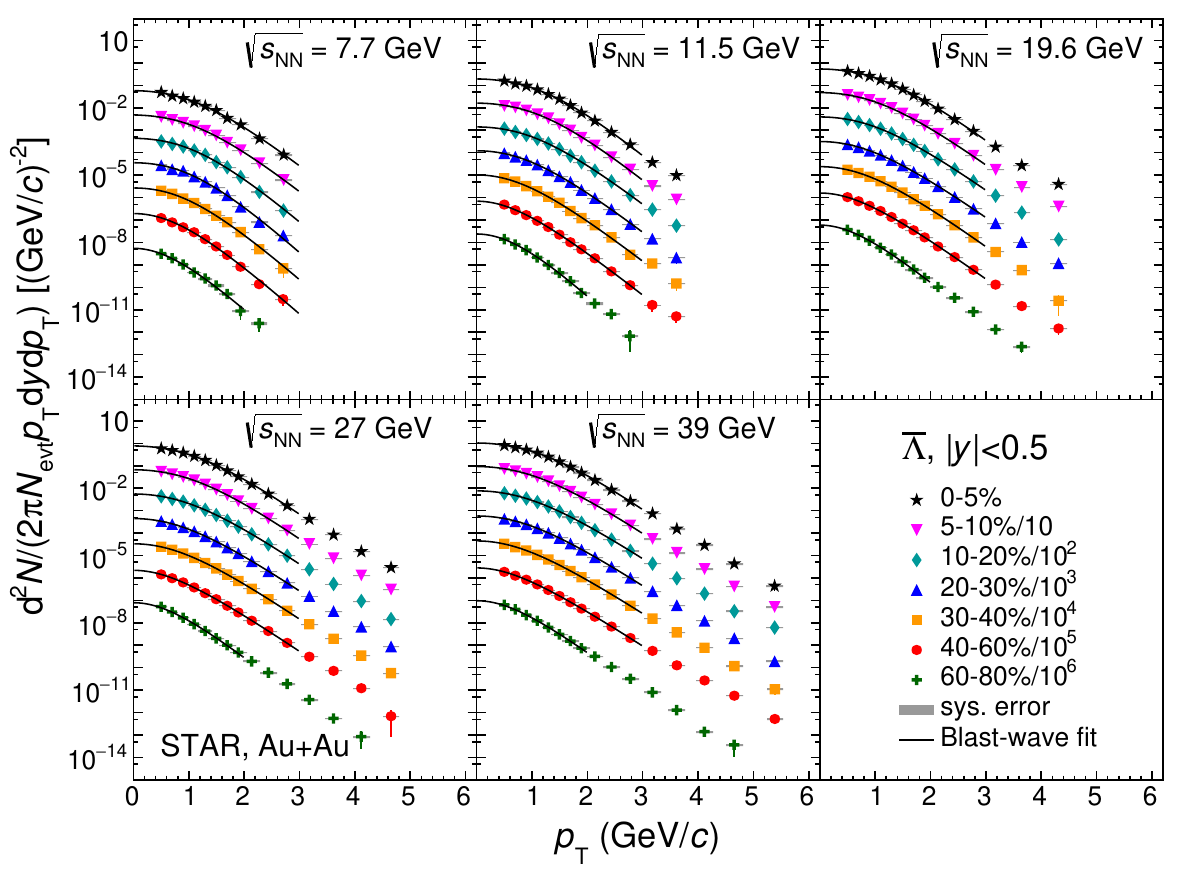}
\vspace{0cm} \caption{The transverse momentum spectra of \alam\ at mid-rapidity ($|y|<$ 0.5) from Au+Au collisions at different centralities and energies (\sqrtsNN = 7.7--39 GeV). The spectra are corrected for the feed-down of \axi, $\overline{\Xi}^0$, and \aom\ decays. The data points are scaled by factors of 10 from central to peripheral collisions for clarity. The vertical gray bands represent the systematic errors, which are small hence the bands look like horizontal bars. The blast-wave model fit results are shown in the fit range and the low \ppt\ extrapolation range as solid lines for all centrality bins.}\label{fig:spectra_ala} 
\end{figure*}

\begin{figure*}[htbp]
\centering \vspace{0cm}
\includegraphics[width=12.5cm]{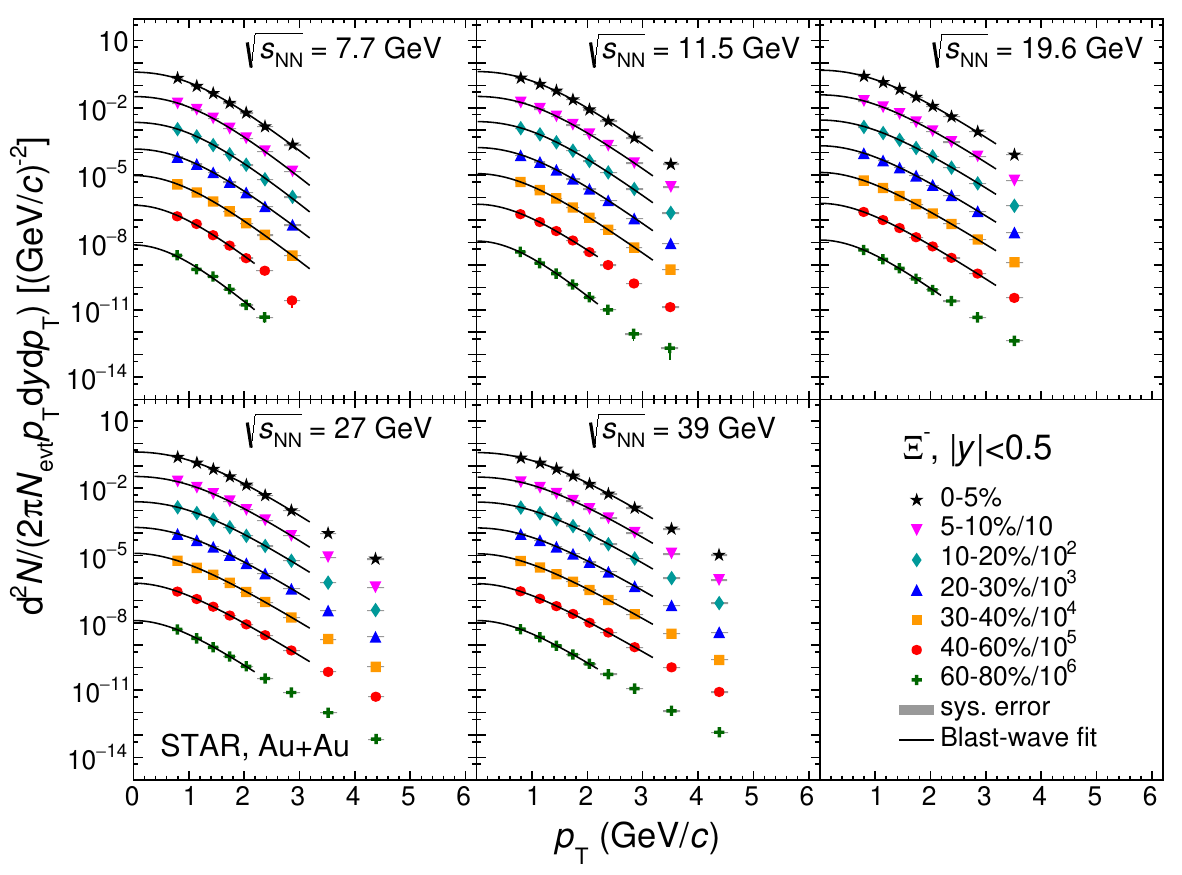}
\vspace{0cm} \caption{The transverse momentum spectra of \xim\ at mid-rapidity ($|y|<$ 0.5) from Au+Au collisions at different centralities and energies (\sqrtsNN = 7.7--39 GeV). The data points are scaled by factors of 10 from central to peripheral collisions for clarity. The vertical gray bands represent the systematic errors, which are small hence the bands look like horizontal bars. The blast-wave model fit results are shown in the fit range and the low \ppt\ extrapolation range as solid lines for all centrality bins.}\label{fig:spectra_xi} 
\end{figure*}

\begin{figure*}[htbp]
\centering \vspace{0cm}
\includegraphics[width=12.5cm]{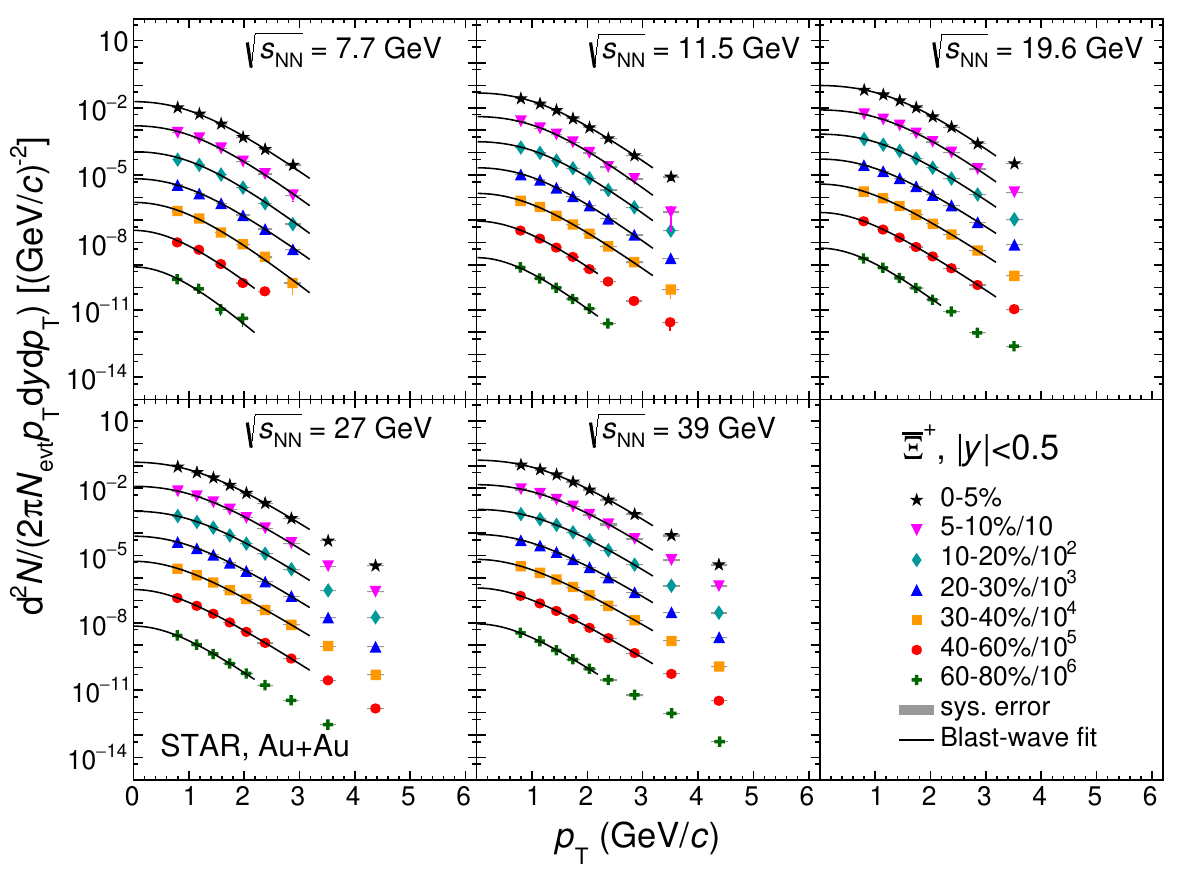}
\vspace{0cm} \caption{The transverse momentum spectra of \axi\ at mid-rapidity ($|y|<$ 0.5) from Au+Au collisions at different centralities and energies (\sqrtsNN = 7.7--39 GeV). The data points are scaled by factors of 10 from central to peripheral collisions for clarity. The vertical gray bands represent the systematic errors, which are small hence the bands look like horizontal bars. The blast-wave model fit results are shown in the fit range and the low \ppt\ extrapolation range as solid lines for all centrality bins.}\label{fig:spectra_axi} 
\end{figure*}

\begin{figure*}[htbp]
\centering \vspace{0cm}
\includegraphics[width=12.5cm]{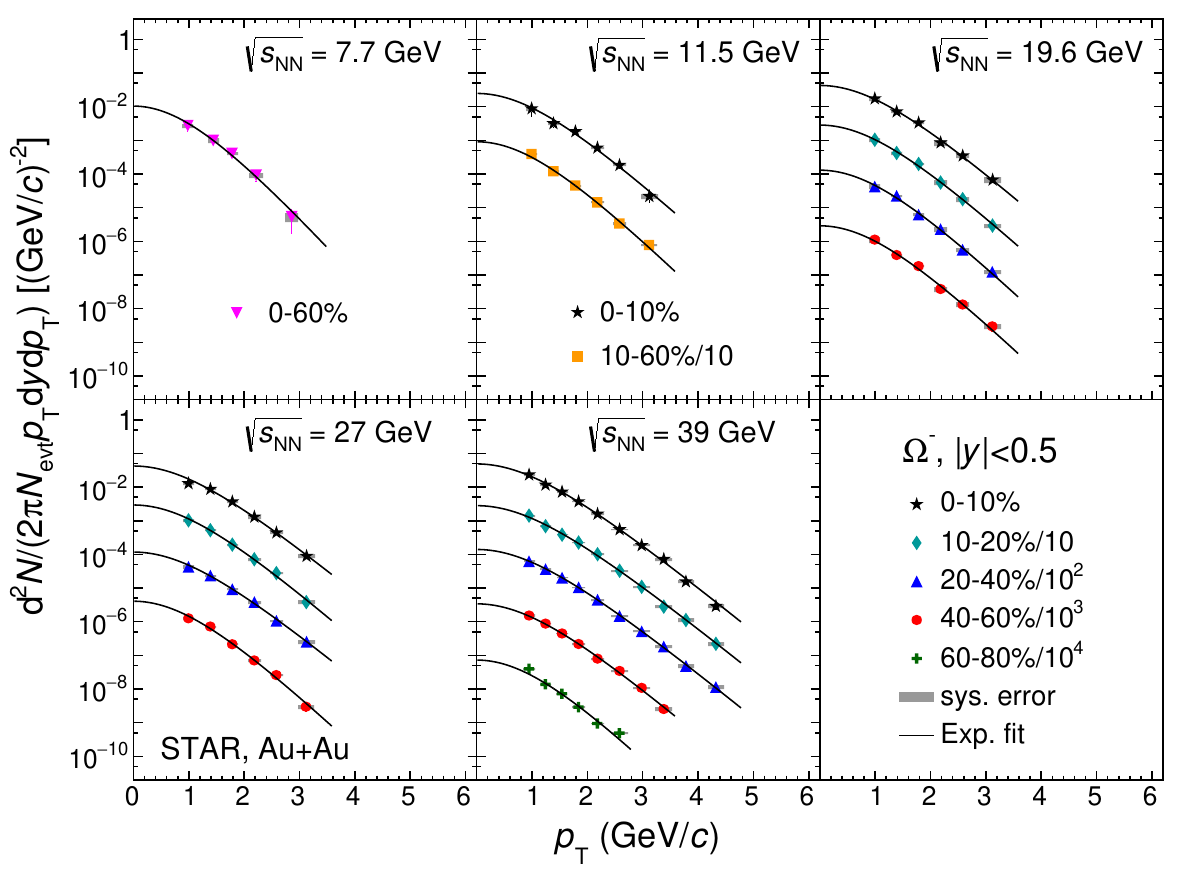}
\vspace{0cm} \caption{The transverse momentum spectra of \omm\ at mid-rapidity ($|y|<$ 0.5) from Au+Au collisions at different centralities and energies (\sqrtsNN = 7.7--39 GeV). The data points are scaled by factors of 10 from central to peripheral collisions for clarity. The vertical gray bands represent the systematic errors, which are small hence the bands look like horizontal bars. The exponential function fit results are shown in the fit range and the low \ppt\ extrapolation range as solid lines for all centrality bins.}\label{fig:spectra_om} 
\end{figure*}

\begin{figure*}[htbp]
\centering \vspace{0cm}
\includegraphics[width=12.5cm]{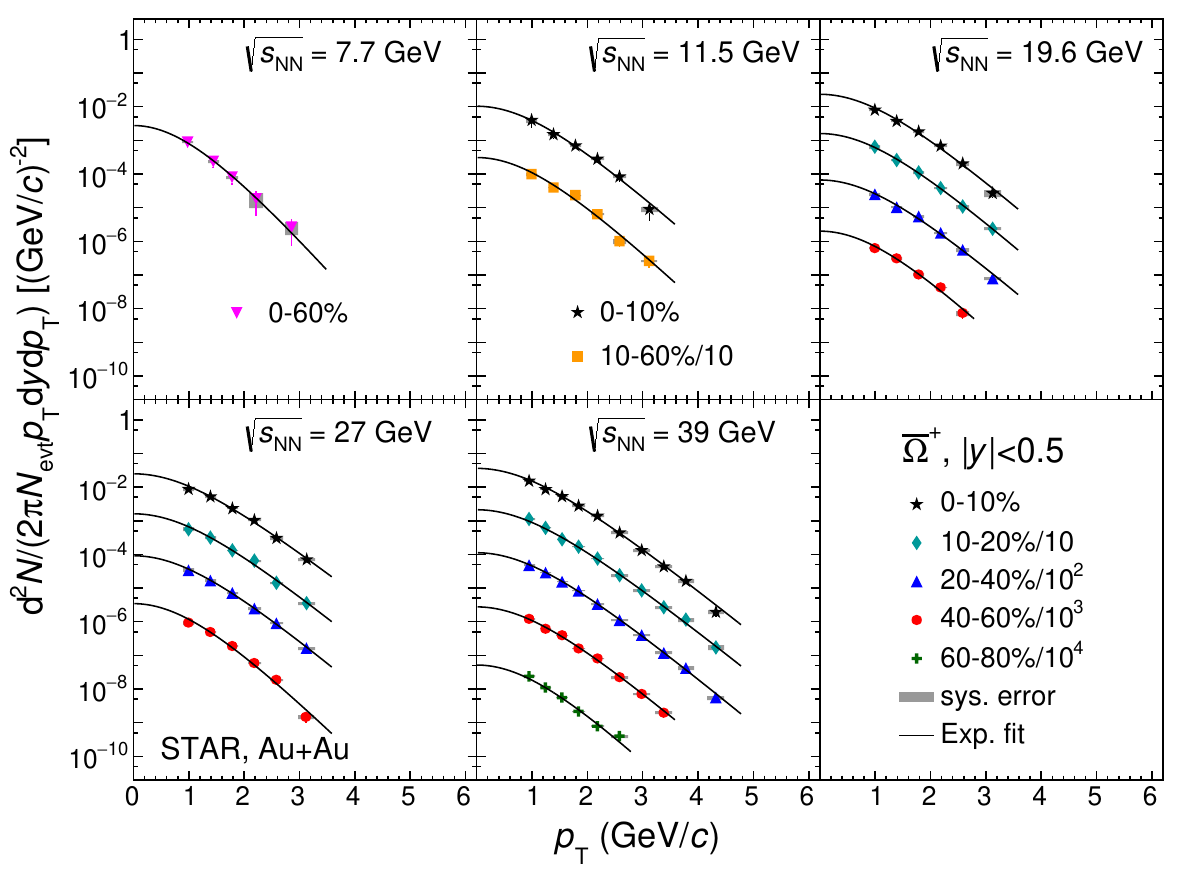}
\vspace{0cm} \caption{The transverse momentum spectra of \aom\ at mid-rapidity ($|y|<$ 0.5) from Au+Au collisions at different centralities and energies (\sqrtsNN = 7.7--39 GeV). The data points are scaled by factors of 10 from central to peripheral collisions for clarity. The vertical gray bands represent the systematic errors, which are small hence the bands look like horizontal bars. The exponential function fit results are shown in the fit range and the low \ppt\ extrapolation range as solid lines for all centrality bins.}\label{fig:spectra_aom} 
\end{figure*}

\begin{figure*}[htbp]
\centering \vspace{0cm}
\includegraphics[width=12.5cm]{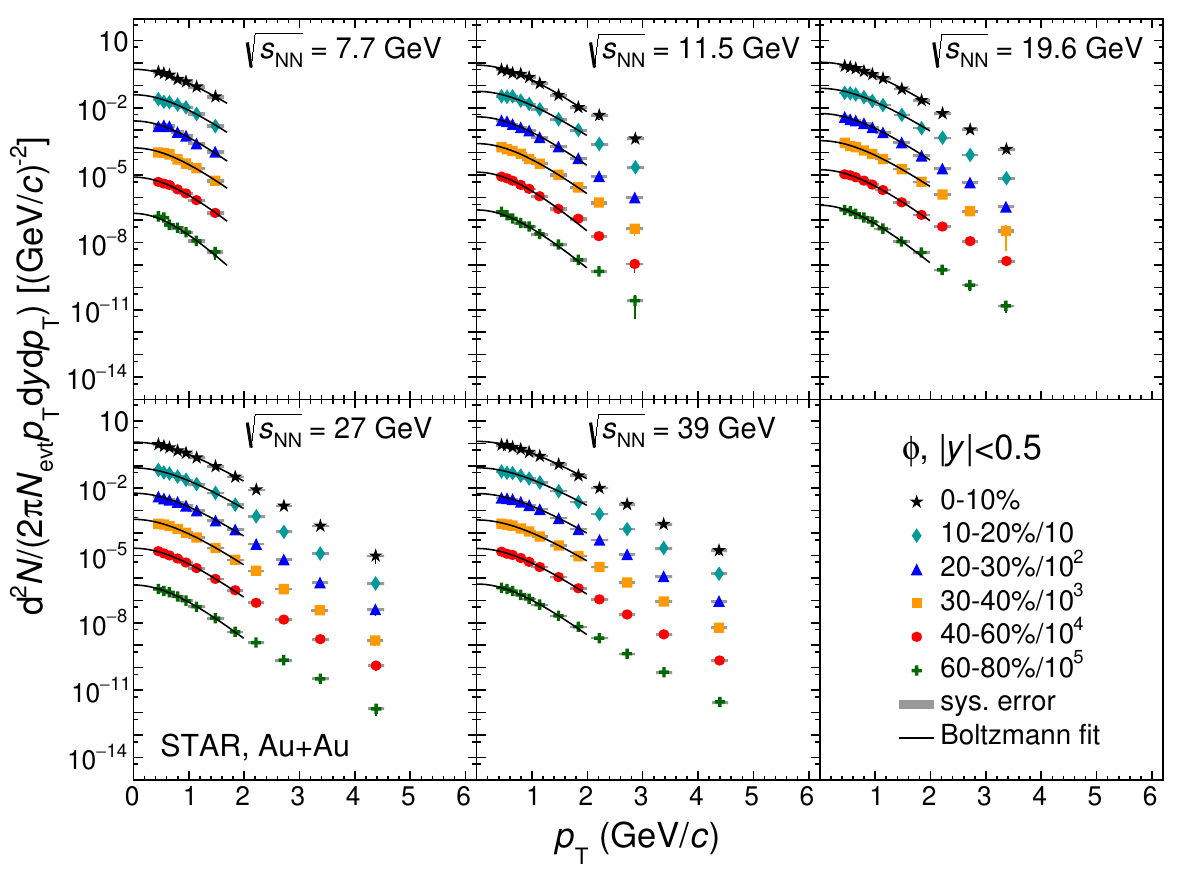}
\vspace{0cm} \caption{The transverse momentum spectra of $\phi$ at mid-rapidity ($|y|<$ 0.5) from Au+Au collisions at different centralities and energies (\sqrtsNN = 7.7--39 GeV). The data points are scaled by factors of 10 from central to peripheral collisions for clarity. The vertical gray bands represent the systematic errors, which are small hence the bands look like horizontal bars. The Boltzmann function fit results are shown in the fit range and the low \ppt\ extrapolation range as solid lines for all centrality bins.}\label{fig:spectra_phi} 
\end{figure*}

Figures~\ref{fig:spectra_k0s},~\ref{fig:spectra_la},~\ref{fig:spectra_ala},~\ref{fig:spectra_xi},~\ref{fig:spectra_axi},~\ref{fig:spectra_om},~\ref{fig:spectra_aom},~and~\ref{fig:spectra_phi} show the transverse momentum spectra of \ks, \lam, \alam, \xim, \axi, \omm, \aom, and $\phi$ at mid-rapidity ($|y|<$ 0.5) in different collision centralities from Au+Au collisions at \sqrtsNN\ = 7.7, 11.5, 19.6, 27, and 39 GeV. The \ppt\ spectra of \omm, \aom\ and $\phi$ are the same as those shown in Ref.~\cite{Adamczyk:2015lvo}. All the \ppt\ spectra shown here have been corrected for geometrical acceptance and reconstruction efficiency, as discussed in Sec.~\ref{section-analysis-eff}. For better visualization, the spectra are scaled by factors of 10 from central to peripheral collisions. The \lam (\alam) spectra are corrected for the feed-down contribution from weak decays of $\Xi$ and $\Xi^0$ baryons using the measured $\Xi$ spectra shown in Fig.~\ref{fig:spectra_xi} or \ref{fig:spectra_axi}. The \alam\ spectra are further corrected for the feed-down contribution from weak decays of \aom\ baryons using the measured \aom\ spectra shown in Fig.~\ref{fig:spectra_aom}. For more details on feed-down correction, please see Sec.~\ref{section-analysis-fd}. 
The systematic errors of invariant yields, described in details in Sec.~\ref{section-analysis-sys}, are shown as vertical gray bands in these figures for each \ppt\ bin. They become generally larger toward more central collisions due to larger track multiplicities. The systematic uncertainties at 19.6 and 27 GeV are less than those at the other three energies due to the better data taking conditions in 2011 than in 2010.
The default function fit results at low \ppt, described in details in Sec.~\ref{section-analysis-extra}, are plotted on top of each \ppt\ spectrum in the corresponding fit range and the low \ppt\ extrapolation range.

\subsection{Averaged transverse mass}

\begin{figure}[htbp]
\centering \vspace{0cm}
\includegraphics[width=8.5cm]{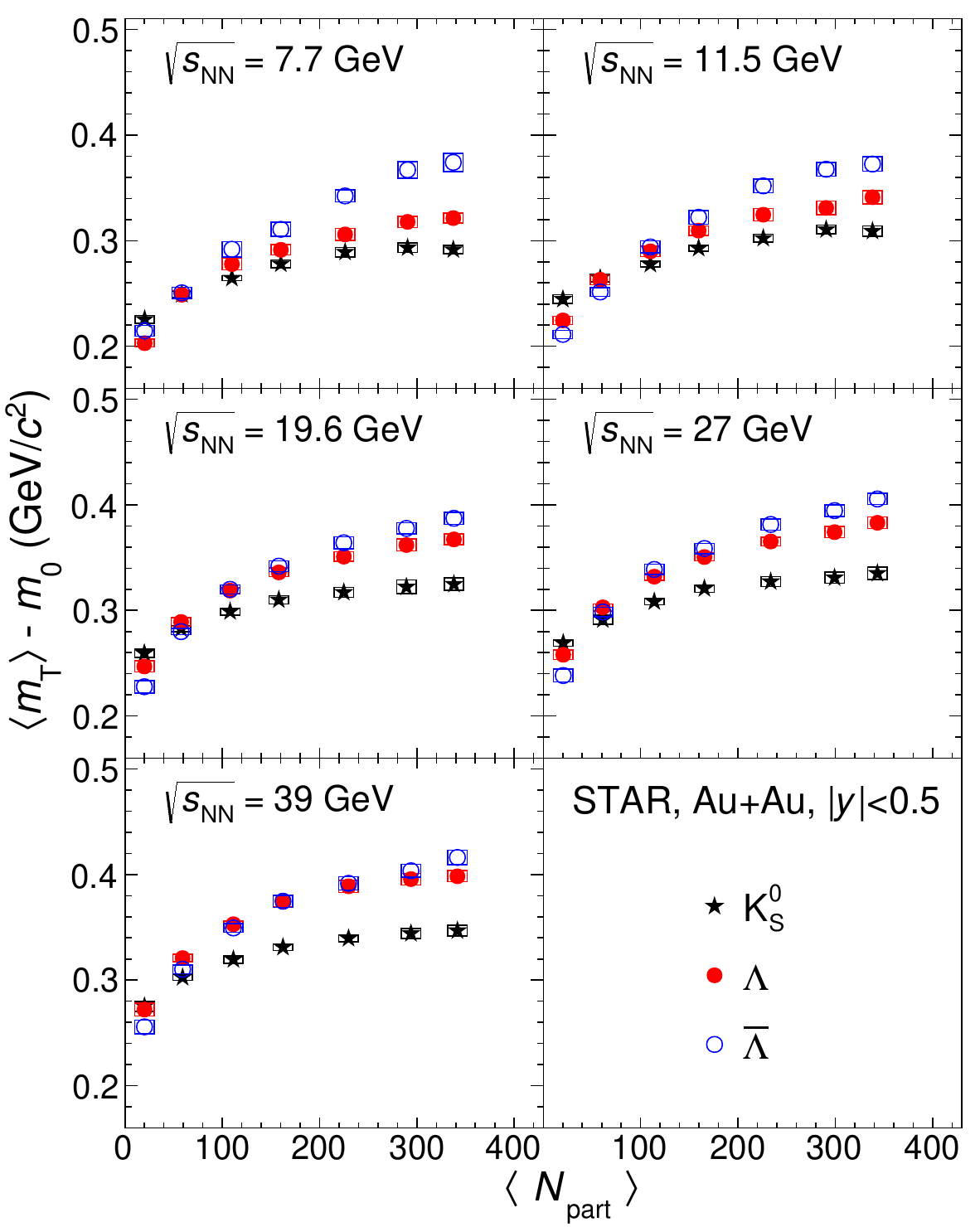}
\vspace{0cm} \caption{The averaged transverse mass, $\left<m_{\rm T}\right>-m_0$, at mid-rapidity ($|y|<$ 0.5) for \ks, \lam, and \alam\ as a function of $\langle N_{\rm part} \rangle$ for Au+Au collisions at \sqrtsNN\ = 7.7--39 GeV. The box on each data point denotes the systematic error.}\label{fig:meanpt_ksla_npart} 
\end{figure}

\begin{figure}[htbp]
\centering \vspace{0cm}
\includegraphics[width=8.5cm]{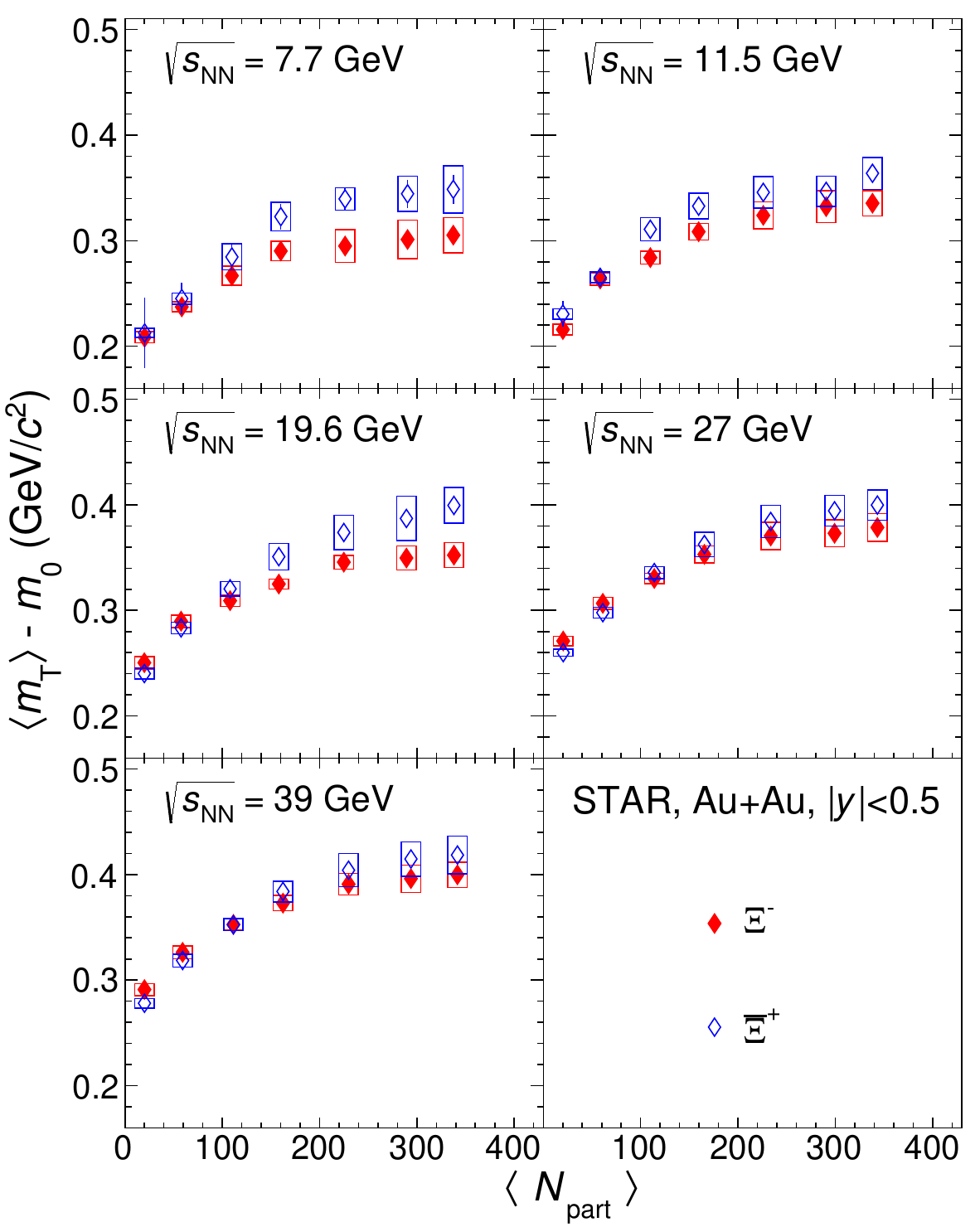}
\vspace{0cm} \caption{The averaged transverse mass, $\left<m_{\rm T}\right>-m_0$, at mid-rapidity ($|y|<$ 0.5) for \xim\ and \axi\ as a function of $\langle N_{\rm part} \rangle$ for Au+Au collisions at \sqrtsNN\ = 7.7--39 GeV. The box on each data point denotes the systematic error.}\label{fig:meanpt_xi_npart}
\end{figure}

\begin{figure}[htbp]
\centering \vspace{0cm}
\includegraphics[width=8.5cm]{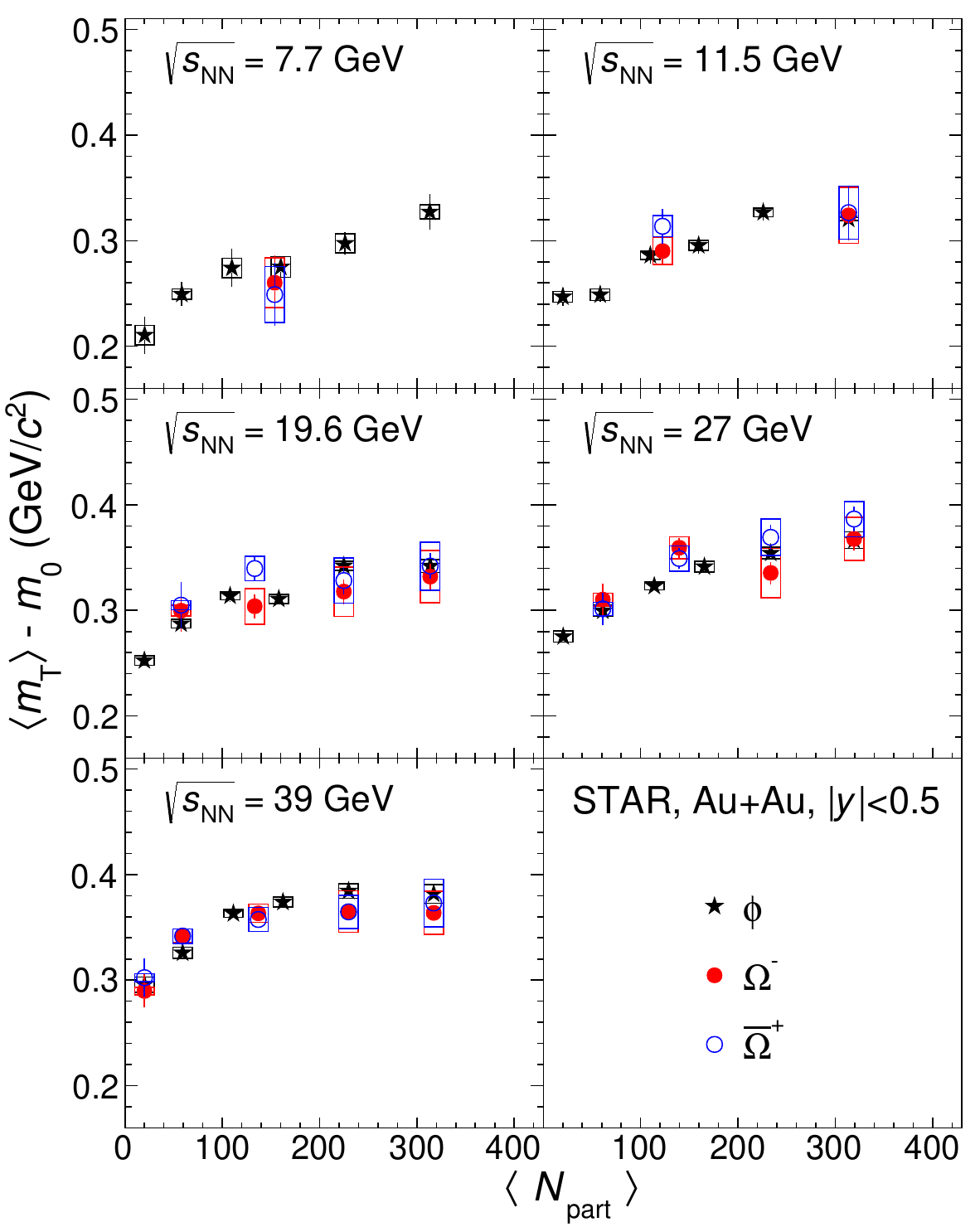}
\vspace{0cm} \caption{The averaged transverse mass, $\left<m_{\rm T}\right>-m_0$, at mid-rapidity ($|y|<$ 0.5) for $\phi$, $\Omega$, and $\overline{\Omega}$ as a function of $\langle N_{\rm part} \rangle$ for Au+Au collisions at \sqrtsNN\ = 7.7--39 GeV. The box on each data point denotes the systematic error.}\label{fig:meanpt_phi_npart} 
\end{figure}

The averaged transverse mass,  $\left< m_{\rm T}\right>-m_0$, can be calculated by Eq.~\ref{mtm0} with the measured and extrapolated \ppt\ spectra of a certain particle species. Its energy and centrality dependence reflects the change of the \ppt\ spectra shapes with collision conditions, and hence provides information regarding the reaction dynamics among the constituents of the colliding systems. Figures~\ref{fig:meanpt_ksla_npart}, \ref{fig:meanpt_xi_npart}, and \ref{fig:meanpt_phi_npart} show the $\left< m_{\rm T}\right>-m_0$ at mid-rapidity ($|y|<$ 0.5) for \ks, \lam, \alam, \xim, \axi, \omm, \aom, and $\phi$ as a function of $\langle N_{\rm part} \rangle$ for Au+Au collisions at \sqrtsNN\ = 7.7, 11.5, 19.6, 27, and 39 GeV. The $\left< m_{\rm T}\right>-m_0$ value of each particle species increases with the increasing $\langle N_{\rm part} \rangle$ at all energies, indicating the gradual development of collective motion with the increasing medium volume. The $\left< m_{\rm T}\right>-m_0$ increases faster toward central collisions for hyperons in general, and for $\Lambda$ in particular, than that of \ks. In contrast, the $\left< m_{\rm T}\right>-m_0$ values of multistrange hyperons, $\Xi$ and $\Omega$, seem to be consistent with that of $\phi$ meson within the uncertainties. Additionally, at lower collision energies and toward central collisions, the $\left< m_{\rm T}\right>-m_0$ of anti-hyperons becomes larger than that of hyperons. The difference is most sizable for \alam\ and \lam\ in the most central Au+Au collisions at \sqrtsNN\ = 7.7 GeV. This phenomenon could be explained by the larger possibility for a lower \ppt\ \alam\ to be annihilated in a \lam-hyperon-rich medium created in such collisions. Generally, in a thermodynamic system with bulk expansion, the radial flow of a particle is only dependent on its mass. In contrast, the split of $\left<m_{\rm T}\right>-m_0$ between antibaryon and baryon indicates that the spectra may not be driven only by the bulk expansion with a common velocity. Hadronic processes, such as baryon-antibaryon annihilations, might also have significant impacts in the final hadron productions at this lower energy region.

\begin{figure}[htbp]
\centering \vspace{0cm}
\includegraphics[width=6.9cm]{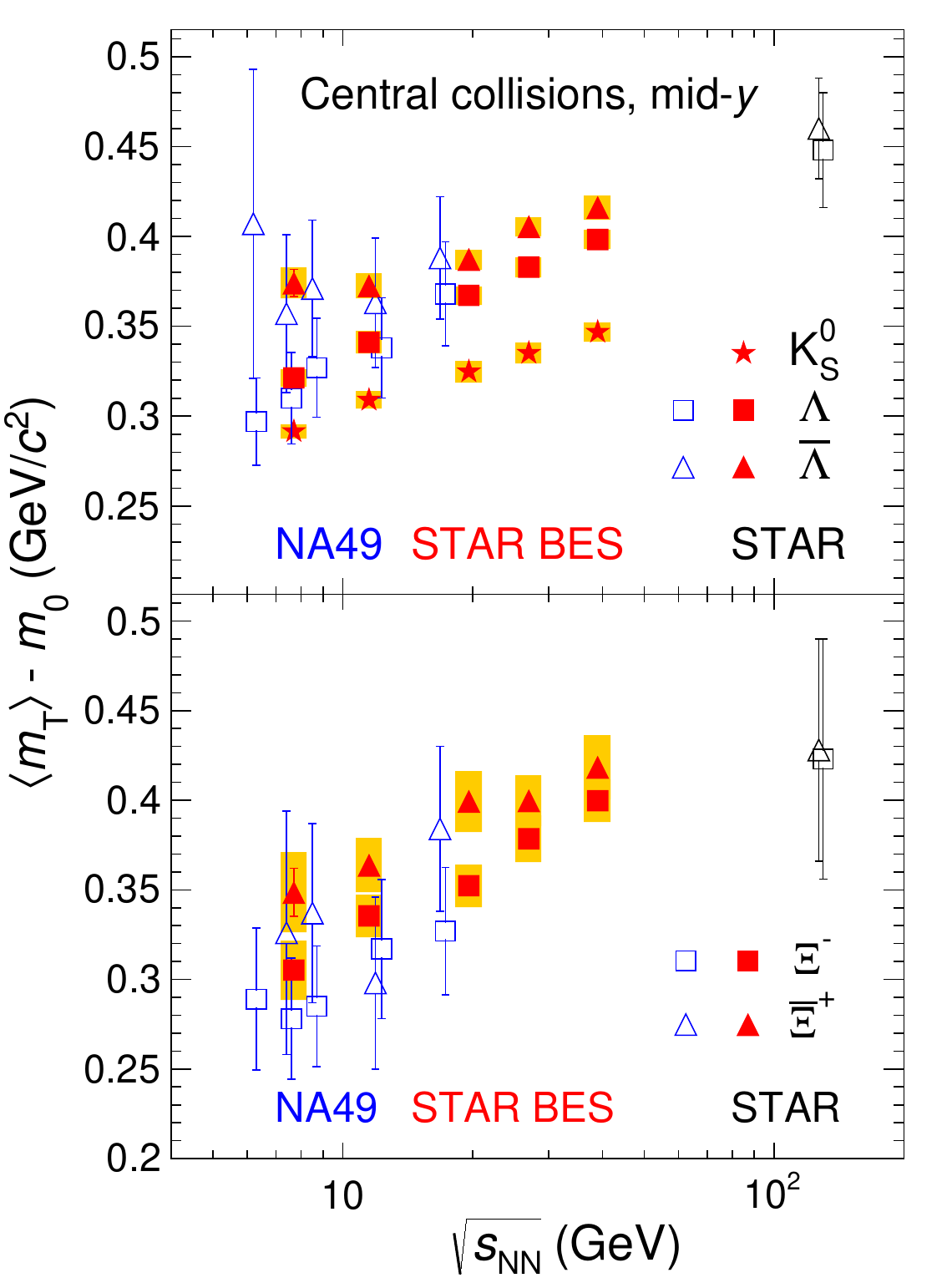}
\vspace{0cm} \caption{The averaged transverse mass, $\left<m_{\rm T}\right>-m_0$, at mid-rapidity ($|y|<$ 0.5) for \ks, \lam, \alam, \xim, and \axi\ as a function of energy from 0--5\% central Au+Au collisions at \sqrtsNN\ = 7.7--39 GeV. For comparison, previous results from central Pb+Pb collisions at \sqrtsNN\,=\,6.3--17.3~GeV at SPS~\cite{na49prc} and from central Au+Au collisions at \sqrtsNN\,=\,130~GeV at RHIC are shown as open markers~\cite{Adler:2002uv,Adams:2003fy}. The NA49 and STAR 130 GeV \alam\ and \axi\ data points are slightly shifted to the left for clarity. The orange shaded bands on the STAR BES data points represent the systematic errors.}\label{fig:meanpt_bes}
\end{figure}

Figure~\ref{fig:meanpt_bes} shows the $\left<m_{\rm T}\right>-m_0$ at mid-rapidity ($|y|<$ 0.5) for \ks, \lam, \alam, \xim, and \axi\ from 0--5\% central Au+Au collisions at \sqrtsNN\ = 7.7--39 GeV. The previously measured $\left<m_{\rm T}\right>-m_0$ for \lam, \alam, \xim, and \axi\ in central Pb+Pb collisions at \sqrtsNN\ = 6.3--17.3~GeV at SPS ~\cite{na49prc} and in central Au+Au collisions at \sqrtsNN\,=\,130~GeV at RHIC~\cite{Adler:2002uv,Adams:2003fy} are also shown for comparison. In general, the STAR BES $\left<m_{\rm T}\right>-m_0$ for \lam, \alam, \xim, and \axi\ shows a trend similar to previous measurements but with much smaller uncertainties. The $\left<m_{\rm T}\right>-m_0$ values for \ks, \lam, \xim, and \axi\ show an increasing trend with the increasing collision energy. However, the $\left<m_{\rm T}\right>-m_0$ value for \alam\ at \sqrtsNN\,=\,7.7 GeV seems to be as large as the value at \sqrtsNN\,=\,11.5 GeV, and apparently breaks the monotonous increasing trend. This observation again may indicate the significant impact of annihilation processes on antibaryon production in a baryon-rich QCD matter.

\subsection{Particle yields}

\begin{figure}[htbp]
\centering \vspace{0cm}
\includegraphics[width=8.5cm]{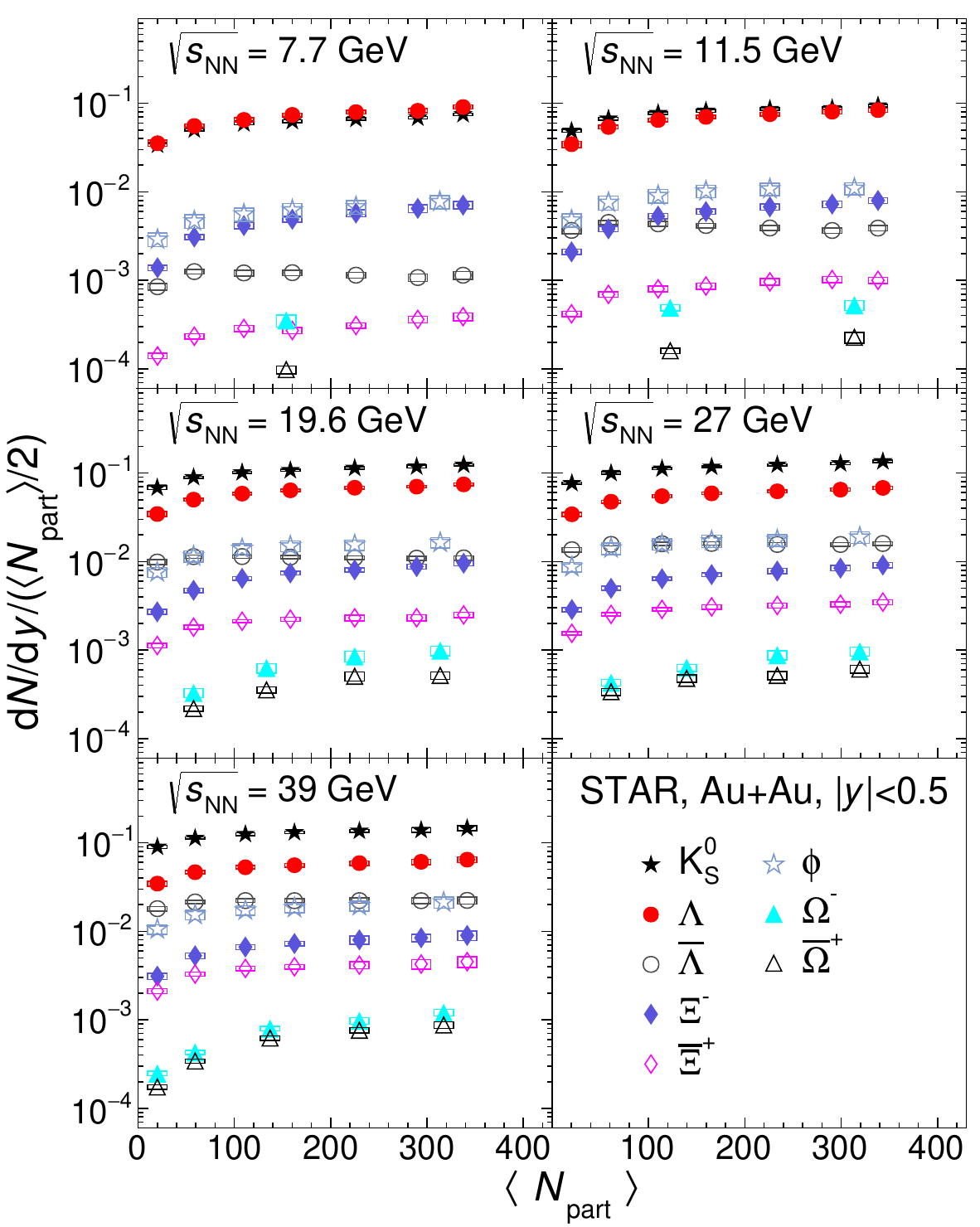}
\vspace{0cm} \caption{The integrated yield, $dN/dy$, per average number of participating nucleon pairs ($\left<N_{\rm part}\right>/2$), of various strange hadrons (\ks, $\phi$, $\Lambda$, \alam, \xim, \axi, \omm, \aom) at mid-rapidity ($|y|<0.5$) as a function of number of participating nucleons, $\langle N_{\rm part} \rangle$, from Au+Au collisions at \sqrtsNN\ = 7.7--39 GeV. The box on each data point denotes the systematic error. For clarity, uncertainties in $\langle N_{\rm part} \rangle$ are not included.}\label{fig1.fig}
\end{figure}

Figure~\ref{fig1.fig} shows the collision centrality dependence of the integrated yield, $dN/dy$, per average number of participating nucleon pairs ($\left<N_{\rm part}\right>/2$), of various strange hadrons (\ks, $\Lambda$, \alam, \xim, \axi, \omm, \aom, and $\phi$) at mid-rapidity ($|y|<0.5$) from Au+Au collisions at \sqrtsNN\ = 7.7, 11.5, 19.6, 27, and 39 GeV. These normalized yields increase from peripheral to central collisions for all particle species except \alam\ at all collision energies. The \alam\ normalized yield has weak centrality dependence, and it even slightly decreases toward central collisions at lower collision energies. This is similar to the case of $\bar{p}$~\cite{bes_pid}, indicating the larger impact of annihilation processes on antibaryon production in more central collisions.

\begin{figure}[htbp]
\centering \vspace{0cm}
\includegraphics[width=6.0cm]{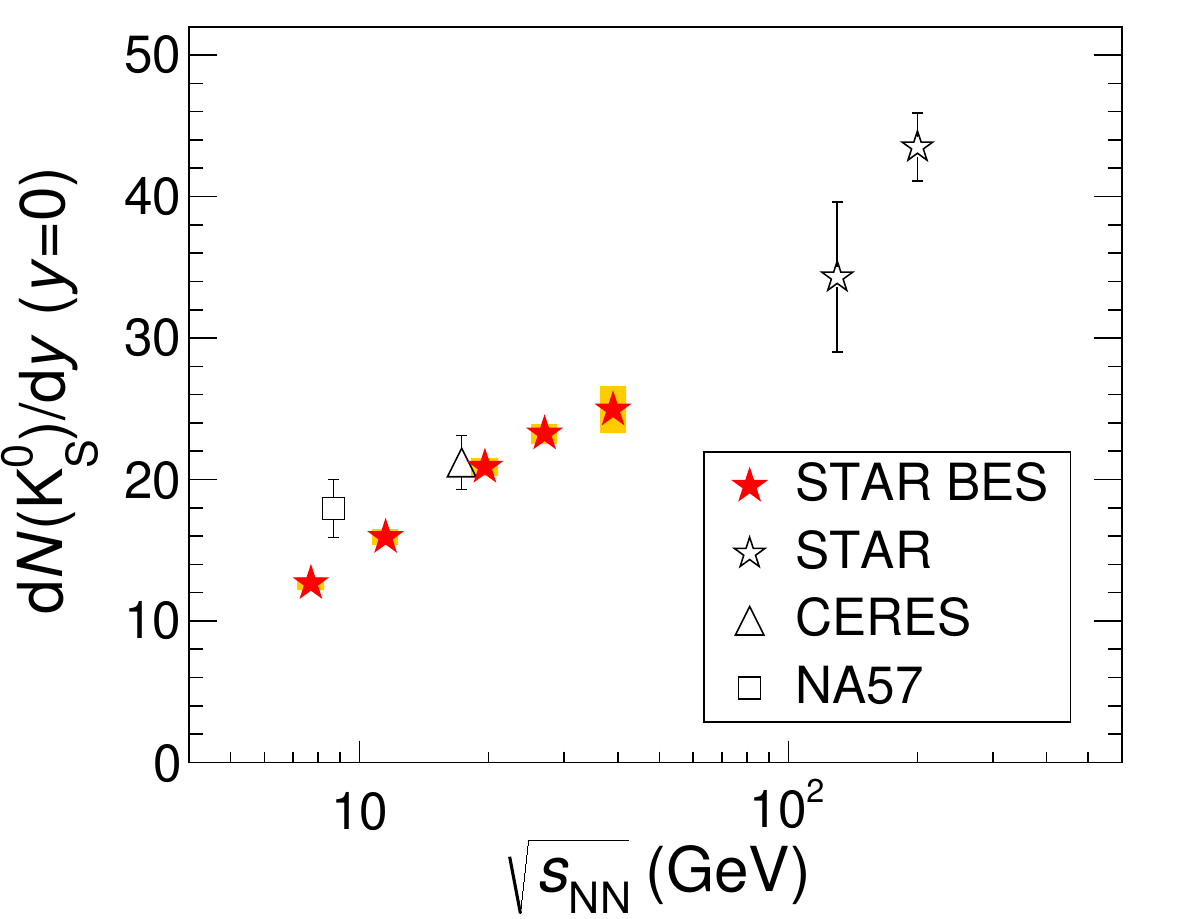}
\vspace{0cm} \caption{The \ks\ integrated yield, $dN/dy$, at mid-rapidity ($|y|<0.5$) as a function of collision energy from 0--5\% central Au+Au collisions at \sqrtsNN\ = 7.7--39 GeV. The orange shaded bands on the STAR BES data points represent the systematic errors. Also shown are the previous mid-rapidity results from 0--5\% central Au+Au collisions at \sqrtsNN\,=\,130 and 200 GeV ($|y|<0.5$) from STAR \cite{starpid_130,Agakishiev:2011ar}, from 0--5\% central Pb+Pb collisions at \sqrtsNN\,=\,8.7 GeV ($|y|<0.5$) from NA57 \cite{Antinori:2004ee,Antinori:2006ij}, and from 0--7\% central Pb+Au collisions at \sqrtsNN\,=\,17.3 GeV from CERES \cite{ceres}. CERES mid-rapidity data are the extrapolated values based on the measurements at backward rapidity.}\label{fig_yield_ks}
\end{figure}

\begin{figure}[htbp]
\centering \vspace{0cm}
\includegraphics[width=8.7cm]{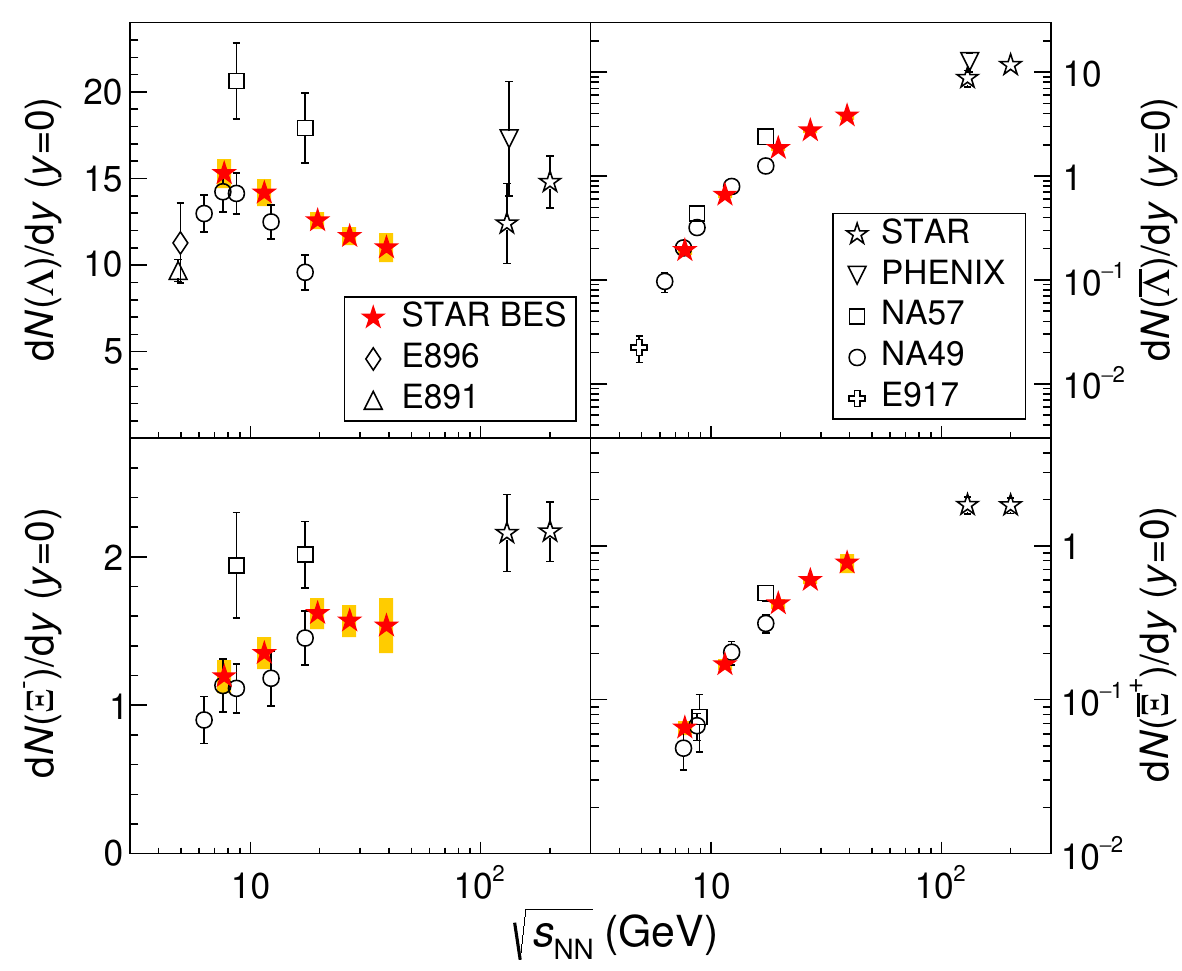}
\vspace{0cm} \caption{Collision energy dependence of the \lam, \alam, \xim, and \axi\ integrated yields, $dN/dy$, at mid-rapidity ($|y|<0.5$) in 0--5\% central Au+Au collisions at \sqrtsNN\ = 7.7--39 GeV. $\Lambda(\overline{\Lambda})$ yields are corrected for weak decay feed-down. The orange shaded bands on the STAR BES data points represent the systematic errors. Also shown are the results from central Au+Au collisions at AGS \cite{e896,e891,e891_2,e917}, PHENIX \cite{phenix_lambda} and STAR \cite{Adler:2002uv,Adams:2003fy,Adams:2006ke,Agakishiev:2011ar} and central Pb+Pb collisions at NA57 \cite{Antinori:2004ee,Antinori:2006ij} and NA49 \cite{na49prc}. The rapidity ranges are $|y|<0.5$ for NA57, PHENIX, NA49 \xim(\axi), and STAR $\Lambda$ at 130 and 200 GeV, $|y|<0.75$ for STAR $\Xi$ at 130 and 200 GeV, $|y|<0.4$ for AGS and NA49 \lam(\alam). The \lam\ and \alam\ results from AGS and PHENIX are inclusive, and those from NA49 and from STAR at higher energies are corrected for weak decay feed-down, while those from NA57 are not significantly affected by weak decay feed-down ($<5\%$ for \lam\ and $<10\%$ for \alam). The E896, PHENIX, and NA57 8.7 GeV \axi\ data points are slightly shifted to the right for clarity. }\label{fig_yield_laxi}
\end{figure}

Figures \ref{fig_yield_ks} and \ref{fig_yield_laxi} show the collision energy dependence of the particle yield ($dN/dy$) at mid-rapidity ($|y|<0.5$) for
\ks, \lam, \alam, \xim, and \axi\ from 0--5\% central Au+Au collisions at \sqrtsNN\ = 7.7, 11.5, 19.6, 27, and 39 GeV, compared to the corresponding data from AGS E896/E891/E917, CERES, NA49, and NA57 in the similar energy range, as well as to the STAR and PHENIX data at higher collision energies. 
The NA49 and NA57 data are from central Pb+Pb collisions, and have been re-scaled according to the estimated numbers of wounded nucleons, $\left<N_{\rm W}\right>$. The scale factor is $\langle N_{\rm part} \rangle/\left<N_{\rm W}\right>$, where $\langle N_{\rm part} \rangle$ is the average number of participants in 0--5\% central Au+Au collisions in STAR (see Table~\ref{tab:npart}). The E917 data and the STAR \ks\ and $\Xi$ data at \sqrtsNN\ = 130 GeV have been re-scaled in a similar manner to account for the centrality difference between these measurements and the STAR BES.
Figures \ref{fig_yield_ks} and \ref{fig_yield_laxi} show that the STAR BES data lie on a trend established by the corresponding data from AGS, NA49, NA57, CERES and previous STAR data, though there seems to be an obvious non-monotonic energy dependence in the \lam\ $dN/dy$ when connecting the STAR BES data with the previous STAR measurements at higher energies. In the BES energy range, while the STAR BES data and the NA49 data are consistent within uncertainties in general except the slight difference in \lam\ yield at \sqrtsNN\,=\,17.3 GeV, the NA57 data are significantly higher for all particle species at both energies except the \axi\ yield with large uncertainty at \sqrtsNN\,=\,8.7\,\,GeV. 

As shown in Fig.~\ref{fig_yield_laxi}, the yields of anti-hyperons increase rapidly with increasing collision energy. However, there seems to be a non-trivial energy dependence in the \lam\ and \xim\ $dN/dy$. The \xim\ $dN/dy$ first slightly increases with energy from 7.7 to 19.6 GeV, then remains almost constant for energies between 19.6 and 39 GeV, finally rising again toward higher energies. The \lam\ $dN/dy$ decreases first when energy increases from 7.7 to 39 GeV, then rises up toward higher energies. It should be noted that the proton $dN/dy$ shows a similar minimum at 39 GeV~\cite{bes_pid}. The proton $dN/dy$ is almost doubled when the collision energy decreases from 39 to 7.7 GeV, reflecting a significant increase in baryon density due to baryon stopping at lower collision energy. Therefore, in terms of hadronic rescatterings, the observed energy dependency of \lam\ $dN/dy$ in the RHIC BES and higher energies (\sqrtsNN\ $>7.7$ GeV) might originate from the interplay between \lam-\alam\ pair production, which strongly increases with the increasing collision energy, and the associated production of \lam\ along with kaons in nucleon-nucleon scatterings \cite{Cleymans:2004bf}, which strongly increases with increasing net baryon density and/or decreasing beam energy.

\subsection{Antibaryon-to-baryon ratios}

\begin{figure}[htbp]
\centering \vspace{0cm}
\includegraphics[width=8.5cm]{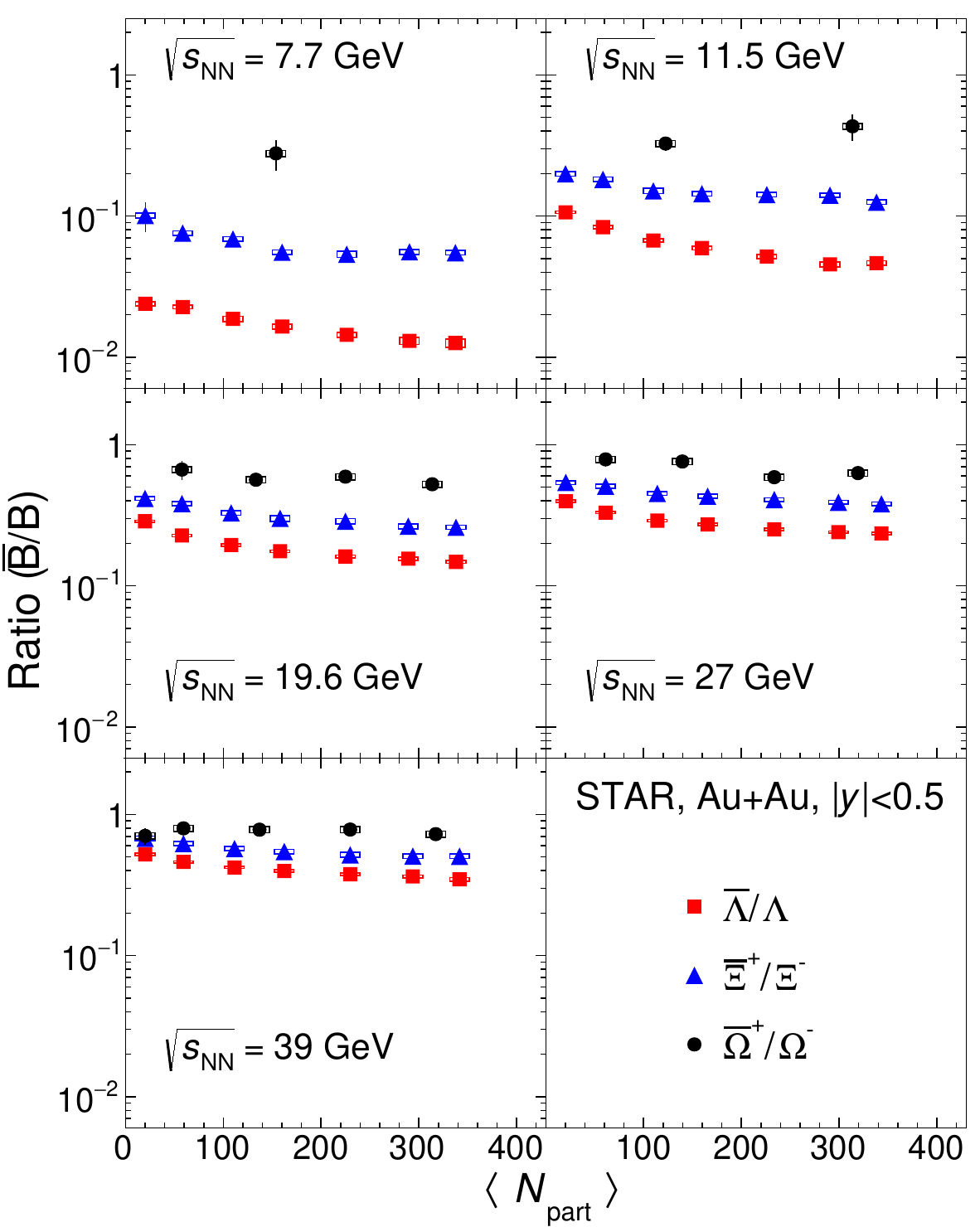}
\vspace{0cm} \caption{The antibaryon-to-baryon ratios, \alam/\lam, \axi/\xim, and \aom/\omm, as functions of $\langle N_{\rm part} \rangle$ from Au+Au collisions at \sqrtsNN\ = 7.7--39 GeV. The box on each data point denotes the systematic error. }\label{fig_bbarb_bes} 
\end{figure}

The difference in $\left<m_{\rm T}\right>-m_0$ between antibaryons and baryons shown in Fig.~\ref{fig:meanpt_bes} might be explained by the absorption of antibaryons due to annihilation at low momentum in a baryon-rich environment. This may result in a decrease of antibaryon yields relative to baryon yields from peripheral to central collisions. Figure~\ref{fig_bbarb_bes} shows the antibaryon-to-baryon ratios, \alam/\lam, \axi/\xim, and \aom/\omm, as functions of $\langle N_{\rm part} \rangle$ from Au+Au collisions at \sqrtsNN\ = 7.7--39 GeV. Indeed, the ratios of \alam/\lam\ and \axi/\xim\ show significant decreases from peripheral to central collisions, especially at lower collision energies. A similar centrality and energy dependence was also observed in the $\bar{p}/p$ ratio \cite{bes_pid}. On the other hand, at lower collision energies, strange baryons (not antibaryons) can also be produced in association with kaons through the nucleon-nucleon interactions, which become more important in central collisions due to the increase of binary nucleon-nucleon collisions per participating nucleon pair. This will result in significant baryon stopping at mid-rapidity, but without creating more antibaryons, hence resulting in a decrease of the antibaryon-to-baryon ratio at mid-rapidity with increasing centrality.

\begin{figure}[htbp]
\centering \vspace{0cm}
\includegraphics[width=8.3cm]{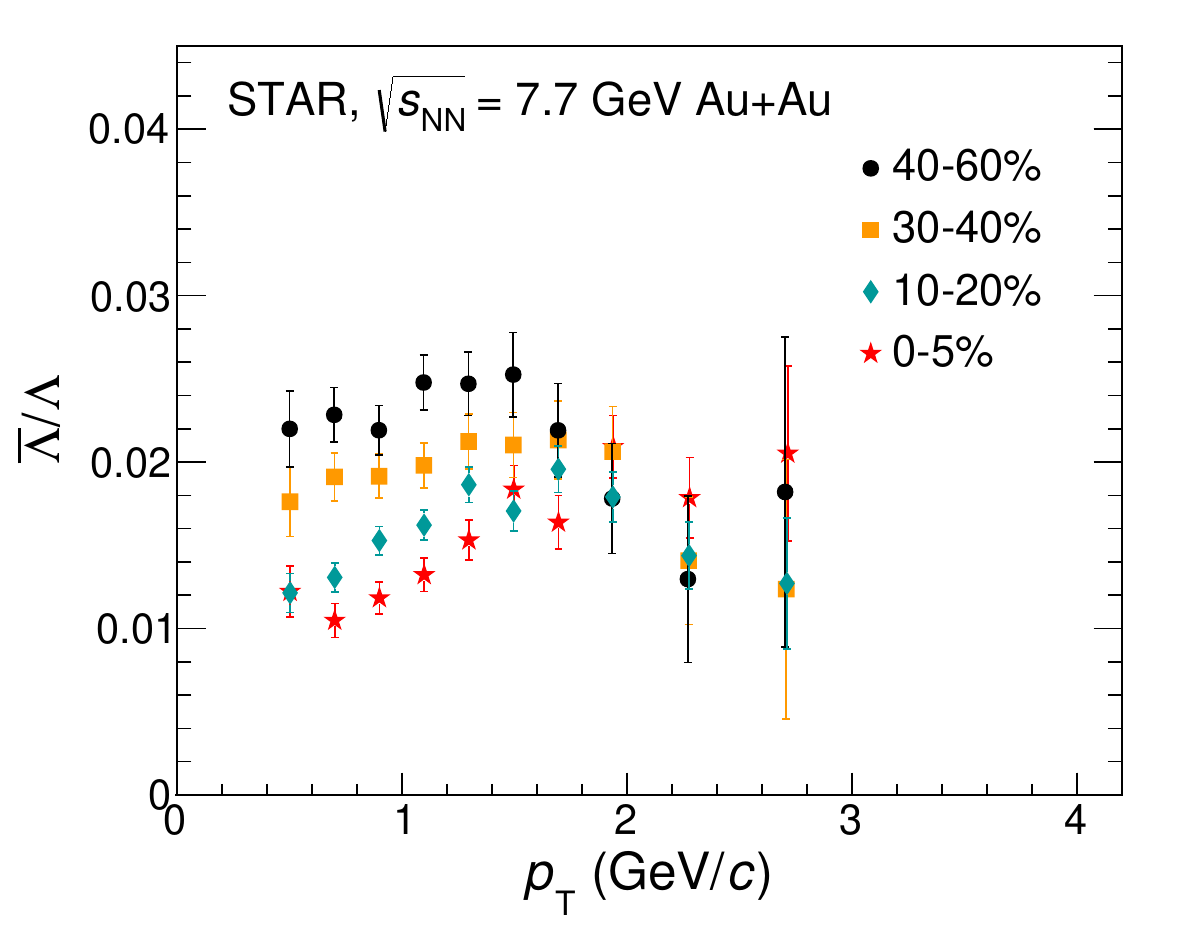}
\vspace{0cm} \caption{\alam/\lam\ ratio as a function of \ppt\ from different centralities of Au+Au collisions at \sqrtsNN\,=\,7.7 GeV. The errors are statistical only.}\label{fig_bbarb_pt} 
\end{figure}

\begin{figure}[htbp]
\centering \vspace{0cm}
\includegraphics[width=8.3cm]{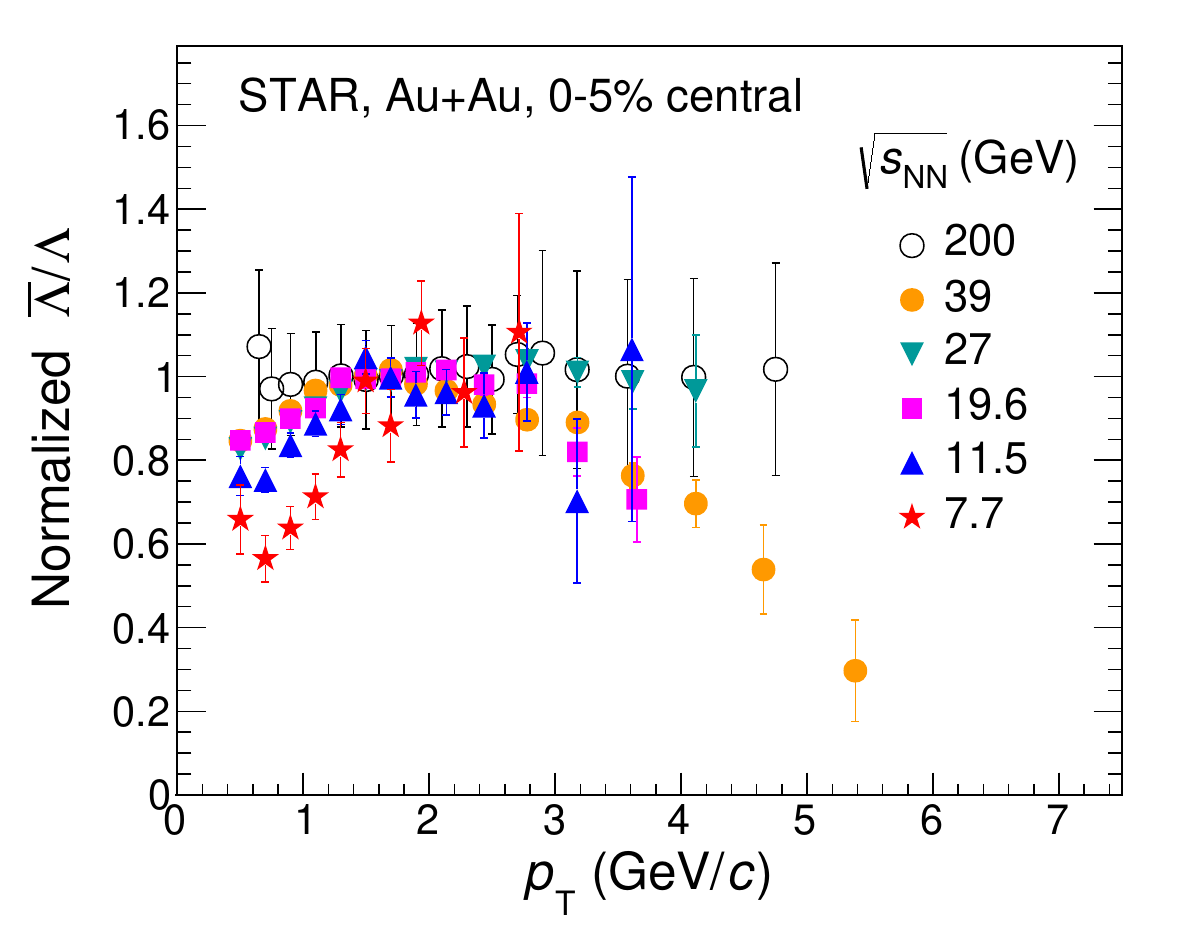}
\vspace{0cm} \caption{Normalized \alam/\lam\ ratio as a function of \ppt\ from 0--5\% central Au+Au collisions at different energies. The STAR results at \sqrtsNN\,=\,200 GeV \cite{Adams:2006ke} are shown as open circles for comparison. The errors are statistical only. All the ratios are normalized according to the average values inside the \ppt\ range of $[1.4, 2.0]$ \GeVc.}\label{fig_bbarb_cent} 
\end{figure}

To examine more closely how the antibaryon and baryon spectra are different, we plot the \alam/\lam\ ratio as a function of \ppt\ in Fig.~\ref{fig_bbarb_pt} for different centrality bins at \sqrtsNN\ =~7.7 GeV and the normalized \alam/\lam\ ratio vs \ppt\ in Fig.~\ref{fig_bbarb_cent} for different energies in central collisions. For \ppt $\gtrsim 2$ \GeVc\ at \sqrtsNN\ =~39 GeV, the ratio decreases with increasing \ppt\ likely due to the semi-hard scattering process dominated by the valence quarks. 
It is evident that the \alam/\lam\ ratio at low \ppt ($\lesssim 2$ \GeVc) increases with increasing \ppt\ and energy and decreasing $\langle N_{\rm part} \rangle$. Hadronic transport model studies could further identify which contributions are dominant: the antibaryon absorption and/or the nucleon-nucleon strangeness association production.

\begin{figure}[htbp]
\centering \vspace{0cm}
\includegraphics[width=8.3cm]{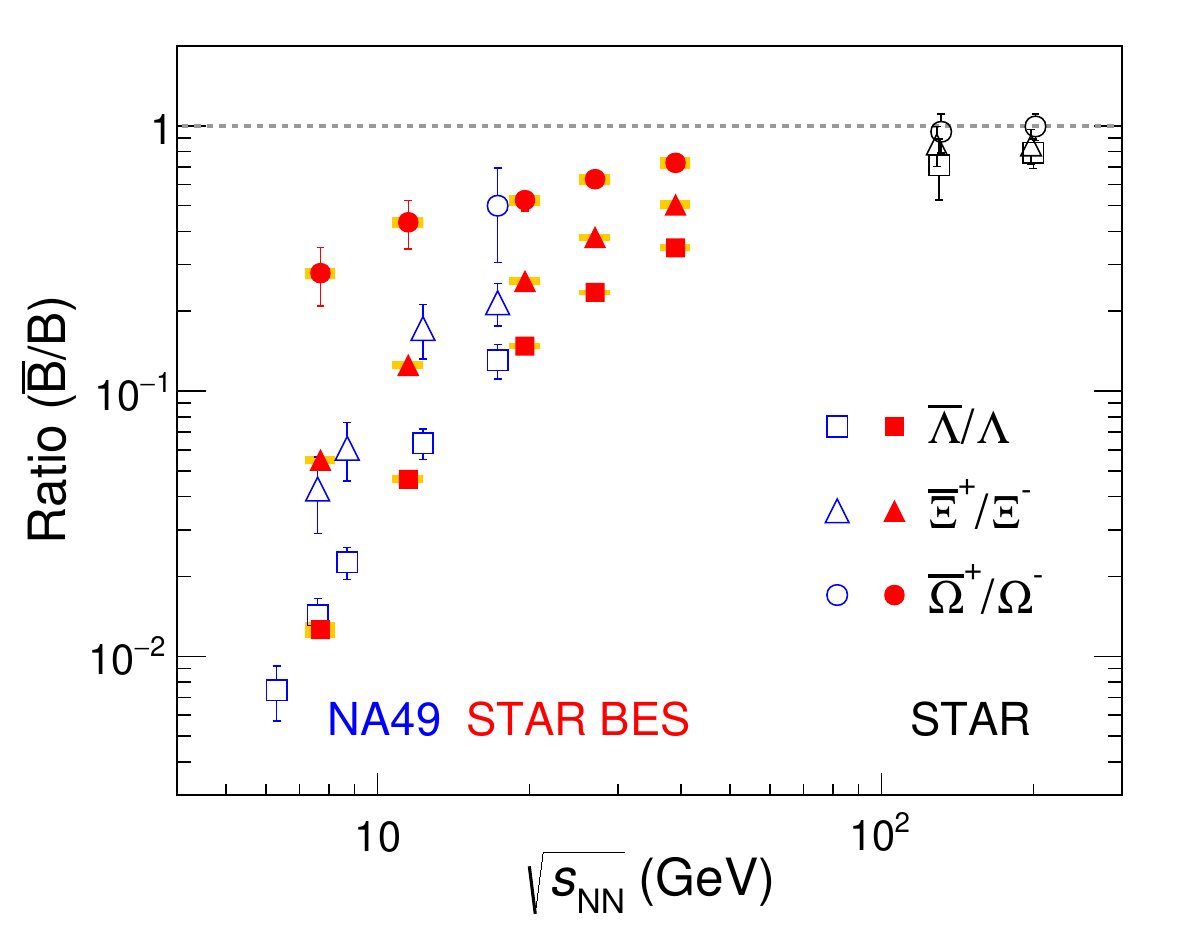}
\vspace{0cm} \caption{The collision energy dependence of $\overline{\Lambda}/\Lambda$, $\overline{\Xi}^+/\Xi^-$, and $\overline{\Omega}^+/\Omega^-$ ratios at mid-rapidity ($|y|<0.5$) in central (0--5\% for $\overline{\Lambda}/\Lambda$ and $\overline{\Xi}^+/\Xi^-$; 0--10\% for $\overline{\Omega}^+/\Omega^-$ for \sqrtsNN$\ \ge 11.5$ GeV; and 0--60\% for $\overline{\Omega}^+/\Omega^-$ for \sqrtsNN$\ = 7.7$ GeV) Au+Au collisions from the STAR Beam Energy Scan (solid symbols). The orange shaded bands represent the systematic errors. The ratios in central Pb+Pb collisions from NA49 \cite{na49prl,na49prc} and in central Au+Au collisions at higher energies ($\ge\,$130\,GeV) from STAR \cite{Adler:2002uv,Adams:2003fy,Adams:2006ke,Agakishiev:2011ar} are also shown as open symbols for comparison. The previous STAR $\overline{\Xi}^+/\Xi^-$ and $\overline{\Omega}^+/\Omega^-$ data points are slightly shifted to the left and to the right respectively for clarity. }\label{fig_bbarb} 
\end{figure}

\begin{figure}[htbp]
\centering \vspace{0cm}
\includegraphics[width=8.5cm]{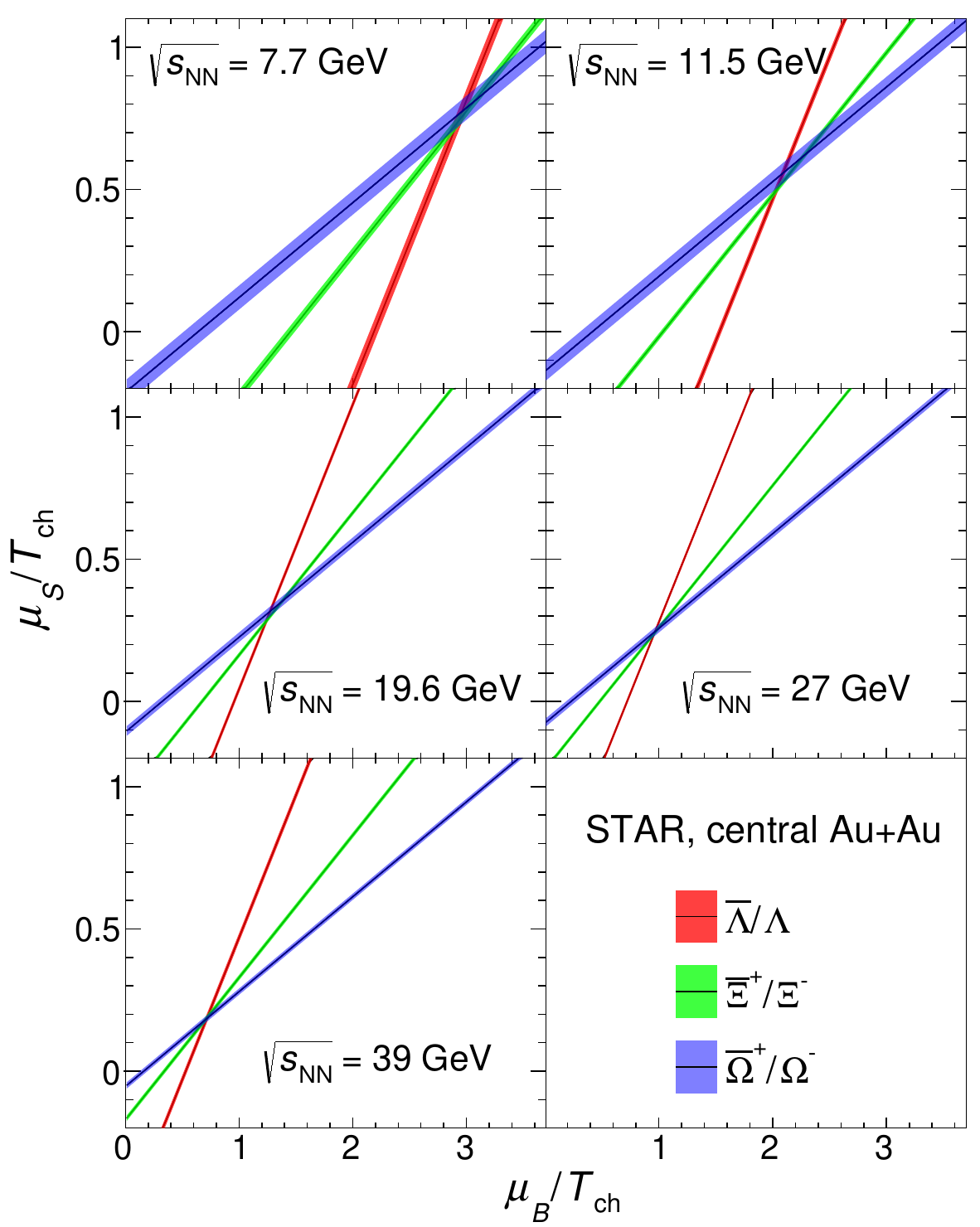}
\vspace{0cm} 
\caption{Testing result of the thermal model in $\mu_B/T_{\rm ch}$ and $\mu_S/T_{\rm ch}$ parameter space with three strange antibaryon-to-baryon ratios in central Au+Au collisions at \sqrtsNN\,=\,7.7--39 GeV. Errors are propagated from the corresponding $\overline{\rm B}/{\rm B}$ ratios, whose errors are the quadratic sum of statistical and systematic errors. } \label{fig_bbarb_stat_check}
\end{figure}

\begin{figure}[htbp]
\centering \vspace{0cm}
\includegraphics[width=8.0cm]{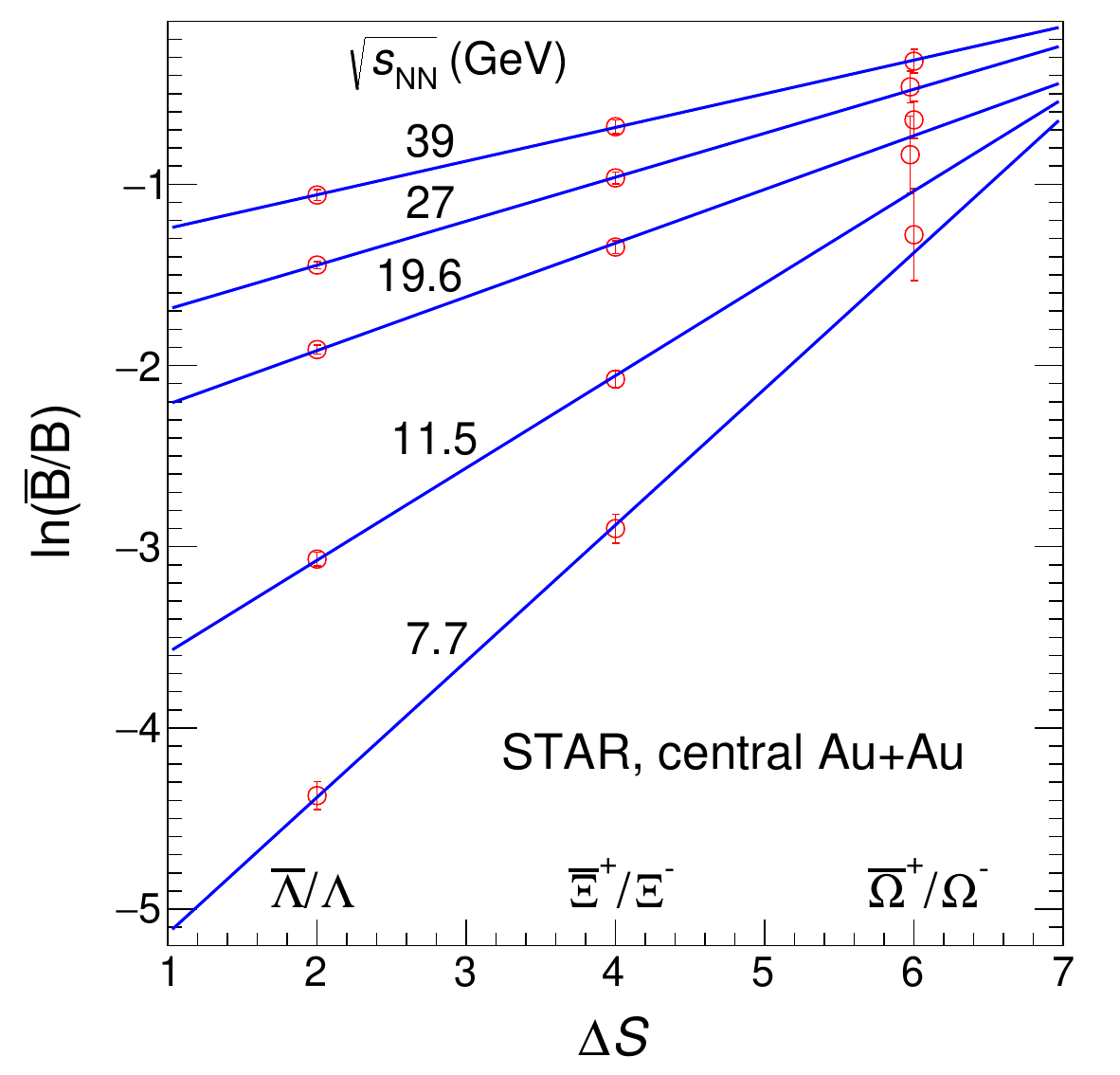}
\vspace{0cm} 
\caption{Thermal model fitting to  $\ln(\overline{\rm B}/{\rm B})$ vs $\Delta S$ with a linear function to determine $\mu_B/T_{\rm ch}$ and $\mu_S/T_{\rm ch}$ for central Au+Au collisions at \sqrtsNN\,=\,7.7--39 GeV. The $\overline{\Omega}^+/\Omega^-$ data points at 11.5 and 27 GeV are slightly shifted to the left for clarity. } \label{fig_bbarb_stat_fit}
\end{figure}

\begin{figure}[htbp]
\centering \vspace{0cm}
\includegraphics[width=8.3cm]{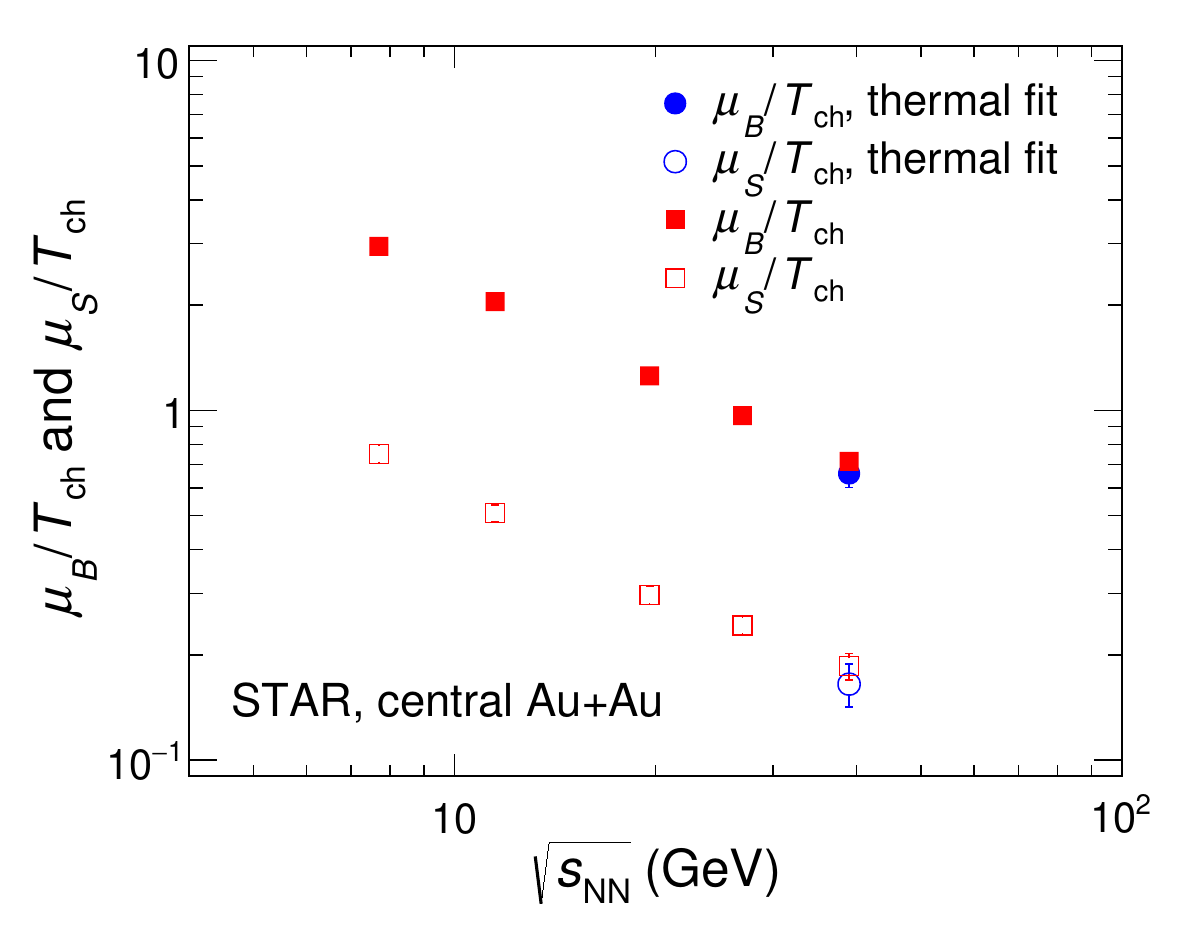}
\vspace{0cm} \caption{The $\mu_B/T_{\rm ch}$ and $\mu_S/T_{\rm ch}$ parameters (red symbols) in central Au+Au collisions at \sqrtsNN\,=\,7.7-39 GeV obtained from the fits shown in Fig.~\ref{fig_bbarb_stat_fit}. The blue symbols are the corresponding results from the thermal model fit to the yields of $\pi$, K, $p$, $\Lambda$, $\Xi$, \ks, and $\Omega$ at \sqrtsNN\,=\,39 GeV~\cite{bes_pid}. } \label{fig_bbarb_mut}
\end{figure}

Figure~\ref{fig_bbarb} shows the \ppt-integrated antibaryon-to-baryon ratios ($\overline{\rm B}/{\rm B}$) in central collisions from the STAR Beam Energy Scan in comparison to those from STAR higher energies and NA49. It seems that the STAR BES data are consistent with the NA49 data and fall within the published energy dependence trend. For all energies, the ratios show a hierarchy of $\overline{\Omega}^+/\Omega^- > \overline{\Xi}^+/\Xi^- > \overline{\Lambda}/\Lambda$, which is consistent with the predictions from statistical thermal models \cite{becattini_prc,pbm_shm,Redlich:2001kb,becattiniprc73}.

In heavy-ion collisions, the baryon and antibaryon multiplicities can be described by thermal models \cite{cleymans} with the parameters of particle mass, degeneracy factor, baryon chemical potential ($\mu_B$), strangeness chemical potential ($\mu_S$), charge chemical potential ($\mu_Q$), strangeness saturation factor ($\gamma_s$), and chemical freeze-out temperature $T_{\rm ch}$. By taking the ratio of antibaryon to baryon yield, one obtains 
\begin{eqnarray}
\label{eq:mubmus}
\ln(\overline{\rm B}/{\rm B}) = - 2\mu_B/T_{\rm ch} + \mu_S/T_{\rm ch}\cdot \Delta S, 
\end{eqnarray}
where $\Delta S$ is the difference of strangeness number between antibaryon and baryon. It shows that most parameters can be canceled out in the $\overline{\rm B}/{\rm B}$ ratios except for $\mu_S/T_{\rm ch}$ and $\mu_B/T_{\rm ch}$. These two parameters are properties of the collision system at chemical freeze-out and should be independent of the particle type according to the thermal model, which assumes that all hadrons originate from the same thermal source. With the three measured antibaryon-to-baryon ratios, $\overline{\Lambda}/\Lambda$, $\overline{\Xi}^+/\Xi^-$, and $\overline{\Omega}^+/\Omega^-$, one can test this thermal model assumption by considering that different antibaryon-to-baryon ratios have different strangeness number difference, $\Delta S$. For a certain antibaryon-to-baryon ratio, Eq.~\ref{eq:mubmus} is effectively a linear function between $\mu_B/T_{\rm ch}$ and $\mu_S/T_{\rm ch}$. With three antibaryon-to-baryon ratios, three straight lines should cross at the same point on the ($\mu_B/T_{\rm ch}$, $\mu_S/T_{\rm ch}$) plane, which provides a good test for the thermal model assumption. Figure~\ref{fig_bbarb_stat_check} shows the test result for central Au+Au collisions at \sqrtsNN\,=\,7.7--39 GeV, which indicates the validity of this model over the BES energy range. Therefore, the two thermal model parameters, $\mu_B/T_{\rm ch}$ and $\mu_S/T_{\rm ch}$, in this collision system, can also be extracted using a linear fit with Eq.~\ref{eq:mubmus} to the three measured antibaryon/baryon ratios at each energy, as shown in Fig.~\ref{fig_bbarb_stat_fit}. The $\mu_B/T_{\rm ch}$ and $\mu_S/T_{\rm ch}$ parameters in central Au+Au collisions at all five BES energies obtained from the fits are shown in Fig.~\ref{fig_bbarb_mut}. Also shown are the corresponding results from the thermal model (grand-canonical ensemble) fitting to the yields of particles including $\pi$, K, $p$, $\Lambda$, $\Xi$, \ks, and $\Omega$ at 39 GeV~\cite{bes_pid}. We see good agreement between the results from these two methods at this collision energy.
Alternatively, the $\mu_B/T_{\rm ch}$ and $\mu_S/T_{\rm ch}$ parameters can be compared with lattice QCD calculations \cite{seewagatoo} to further constrain the strangeness chemical freeze-out temperature $T_{\rm ch}$ in these collisions.

\subsection{Baryon-to-meson ratios}

\begin{figure}[htbp]
\centering \vspace{0cm}
\includegraphics[width=8.7cm]{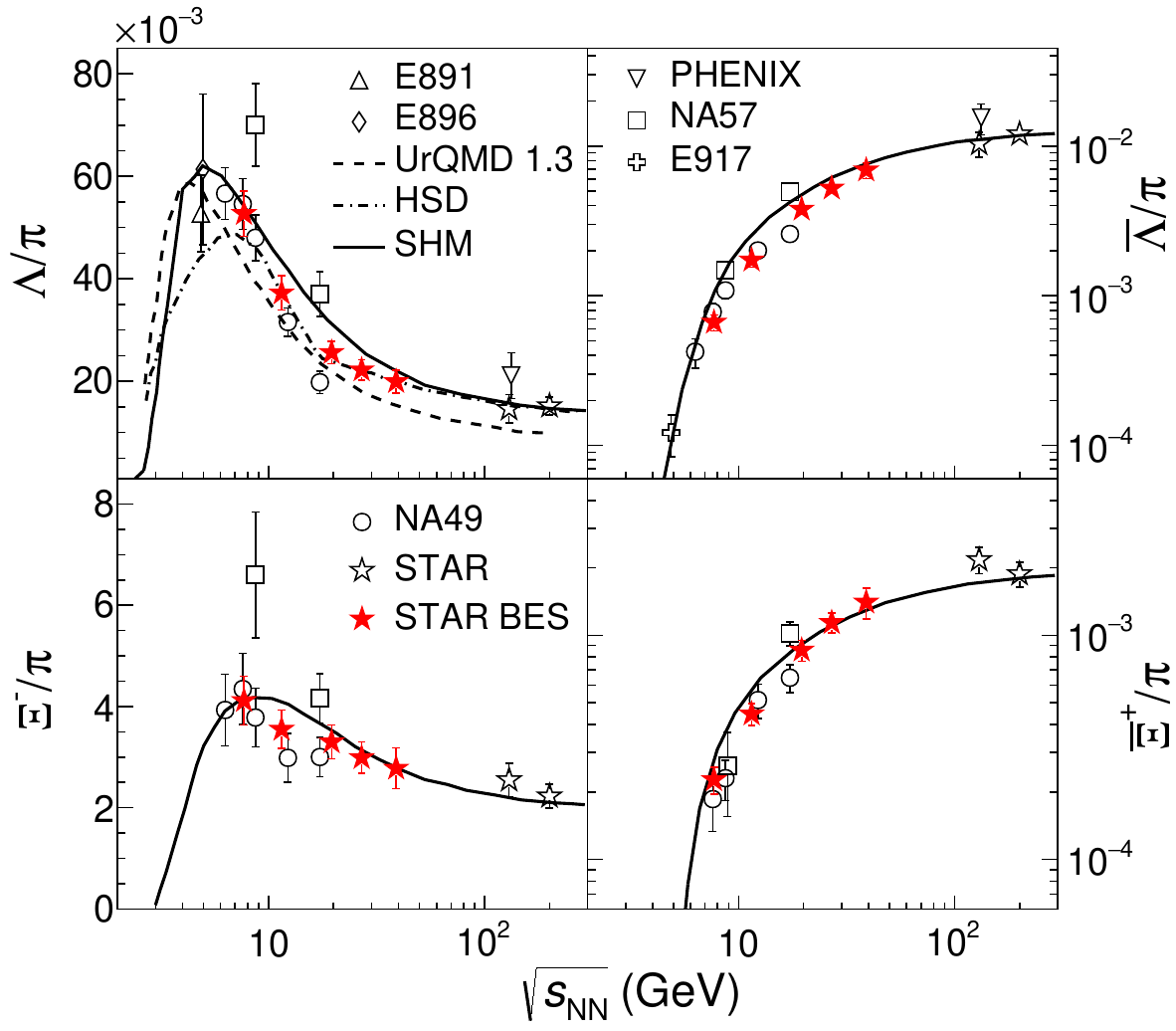}
\vspace{0cm} \caption{Energy dependence of \lam , \alam, \xim, and \axi to pions ($1.5(\pi^++\pi^-)$) ratios at mid-rapidity in central Au+Au collisions from STAR Beam Energy Scan (solid symbols). The STAR BES mid-rapidity pion yields are taken from \cite{bes_pid}. Errors are the quadratic sum of statistical and systematic errors. Also shown are existing AGS \cite{e896,E802,e917,e891,e891_2}, NA49 \cite{na49prc,Afanasiev:2002mx,Alt:2007aa,Alt:2006dk}, PHENIX \cite{phenix_lambda,phenix_pid} and STAR \cite{Adler:2002uv,Adams:2003fy,Adams:2006ke,starpid_200,Abelev:2008ab,Agakishiev:2011ar} data as open symbols, as well as calculations from hadronic transport models (UrQMD 1.3 and HSD) \cite{Bass:1998ca,Bleicher:1999xi,Cassing:1999es,hsdurqmd} and a statistical hadron gas model (SHM) \cite{pbm_shm} as dashed or solid lines. The E896, PHENIX, and NA57 8.7 GeV \axi/$\pi$ data points are slightly shifted to the right for clarity. }\label{fig_ratio} 
\end{figure}

Figure~\ref{fig_ratio} shows the ratios of \lam, \alam, \xim, and \axi\ mid-rapidity yields to that of all pions ($1.5(\pi^++\pi^-)$) in central Au+Au collisions from the STAR Beam Energy Scan. The existing data from various experiments at different energies are also shown for comparison. The data are compared to the calculations from hadronic transport models (UrQMD 1.3 and HSD \cite{Bass:1998ca,Bleicher:1999xi,Cassing:1999es,hsdurqmd}) and a statistical hadron gas model (SHM \cite{pbm_shm}). The STAR Beam Energy Scan data are in good agreement with the trend of the existing experimental data. The hadronic models (UrQMD 1.3 and HSD) seem to reproduce the $\Lambda/\pi$ data, indicating that the hadronic rescatterings might play an important role in hyperon production in heavy-ion collisions in this energy range. On the other hand, the SHM model predictions agree well with data across the whole energy range from AGS to top RHIC energies. SHM is based on a grand canonical ensemble and assumes chemical equilibrium. The energy dependence of the parameters $T_{\rm ch}$ and $\mu_{B}$ in the model were obtained with a smooth parametrization of the original fitting parameters to the mid-rapidity particle ratios from heavy ion experiments at SPS and RHIC.
The $\mathrm{K}^+$/$\pi^+$ \cite{bes_pid}, $\Lambda/\pi$, and $\Xi^-/\pi$ ratios all show a maximum at \sqrtsNN\ $\approx$ 8 GeV, which seems to be consistent with the picture of maximum net-baryon density at freeze-out at this collision energy \cite{Randrup:2006uz}.

\subsection{Nuclear modification factor}

\begin{table*}[hbt]
\begin{center}
\caption{The average number of binary nucleon-nucleon collisions ($\langle N_{\rm coll} \rangle $) for various collision centralities in Au+Au collisions at 7.7--39 GeV, determined using the charged particle multiplicity distributions and the Glauber Monte Carlo simulation \cite{Adamczyk:2013gw}. The errors represent systematic uncertainties. The inelastic $p+p$ cross-sections used in the simulations are 30.8, 31.2, 32, 33, and 34 mb for $\sqrt{\mathrm{s}}$\,=\,7.7, 11.5, 19.6, 27, and 39 GeV, respectively \cite{Miller:2007ri}.}\label{tab:nbin}
\begin{tabular}{c|ccccccccccccc}
\hline
 \sqrtsNN\ (GeV)  & 0--5\% & \hspace{0.25cm} & 5--10\% & \hspace{0.25cm} & 10--20\% & \hspace{0.25cm} & 20--30\% & \hspace{0.25cm} & 30--40\% & \hspace{0.25cm} & 40--60\% & \hspace{0.25cm} & 60--80\% \\
\hline
7.7 & 774 $\pm$ 28 & & 629 $\pm$ 20 & & 450 $\pm$ 22 & & 283 $\pm$ 24 & & 171 $\pm$ 23 & & 74  $\pm$ 16 & & 19.2  $\pm$ 6.3  \\
11.5 & 784 $\pm$ 27 & & 635 $\pm$ 20 & & 453 $\pm$ 23 & & 284 $\pm$ 23 & & 172 $\pm$ 22 & & 75 $\pm$ 16 & & 19.1 $\pm$ 7.8 \\
19.6 & 800 $\pm$ 27 & & 643 $\pm$ 20 & & 458 $\pm$ 24 & & 285 $\pm$ 26 & & 170 $\pm$ 23 & & 74 $\pm$ 15 & & 18.9 $\pm$ 6.9 \\
27 & 841 $\pm$ 28 & & 694 $\pm$ 22 & & 497 $\pm$ 26 & & 312 $\pm$ 28 & & 188 $\pm$ 25 & & 82 $\pm$ 18 & & 20.0 $\pm$ 8.6 \\
39 & 853 $\pm$ 27 & & 687 $\pm$ 21 & & 492 $\pm$ 26 & & 306 $\pm$ 27 & & 183 $\pm$ 24 & & 79 $\pm$ 17 & & 19.4 $\pm$ 7.7 \\
\hline
\end{tabular}
\end{center}
\end{table*}

\begin{figure}[htbp]
\centering \vspace{0cm}
\includegraphics[width=8.5cm]{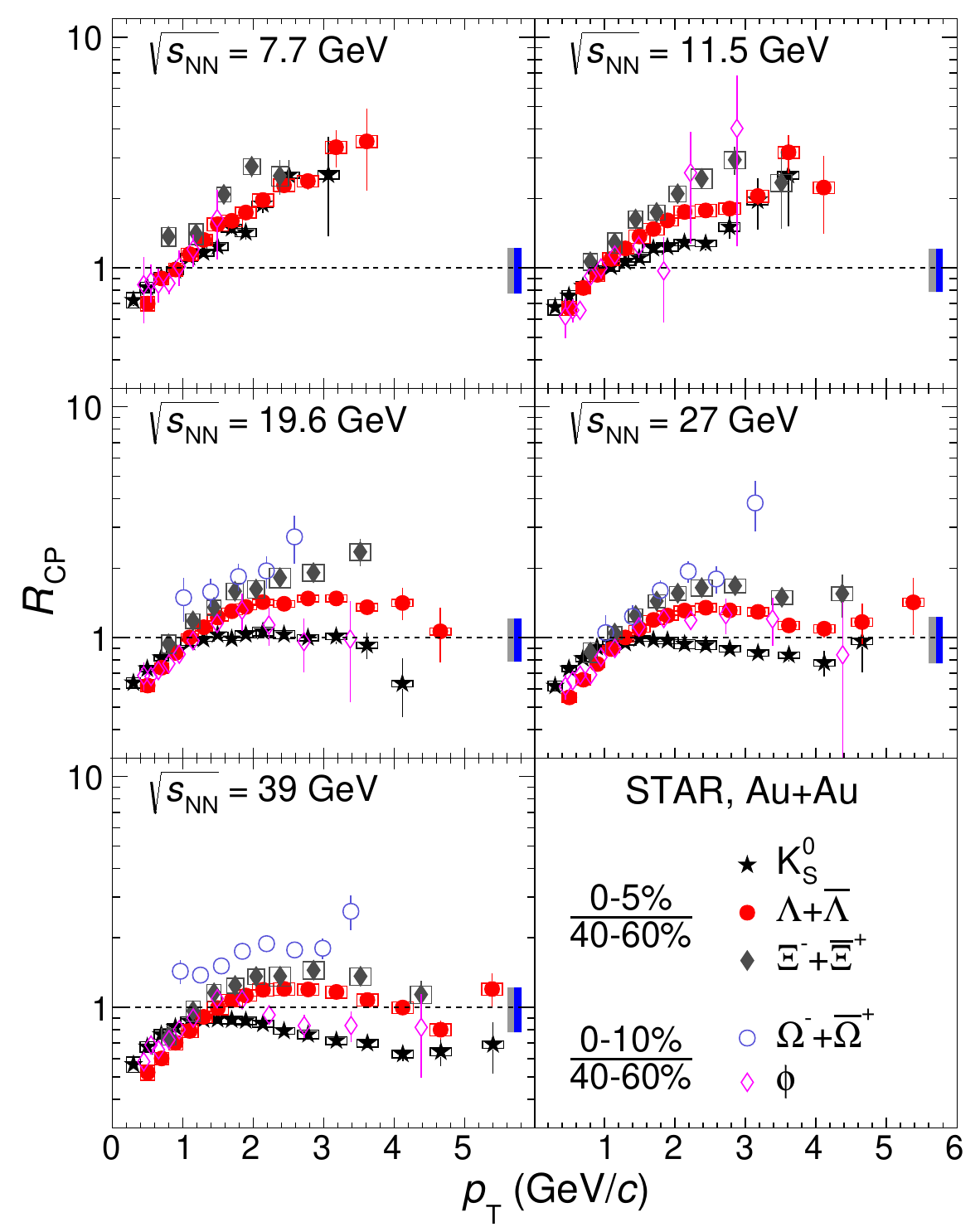}
\vspace{0cm} \caption{\ks, \lam$+$\alam, and \xim$+$\axi\ $R_{\textrm{\tiny{CP}}}$(0--5\%)/(40--60\%), $\phi$ and \omm$+$\aom\ $R_{\textrm{\tiny{CP}}}$(0--10\%)/(40--60\%), at mid-rapidity ($|y|<0.5$) in Au+Au collisions at \sqrtsNN\ = 7.7--39 GeV. The vertical bars denote the statistical errors. The box on each data point of \ks, $\Lambda$, and $\Xi$ denotes the systematic error. There are only statistical errors for $\Omega$ and $\phi$. For \sqrtsNN\ $\leq$ 19.6 GeV, the \lam$+$\alam\ $R_{\textrm{\tiny{CP}}}$ excludes the minor contribution from \alam. The gray and blue bands on the right side of each panel represent the normalization errors from $N_{\textrm{\scriptsize{coll}}}$ for $R_{\textrm{\tiny{CP}}}$(0--5\%)/(40--60\%) and $R_{\textrm{\tiny{CP}}}$(0--10\%)/(40--60\%) respectively.}\label{fig_rcp}
\end{figure}

Figure~\ref{fig_rcp} presents the nuclear modification factor, $R_{\textrm{\tiny{CP}}}$, of
\ks, $\Lambda+\overline{\Lambda}$, $\Xi^{-}+\overline{\Xi}^+$, $\phi$ and \omm$+$\aom\ in Au+Au collisions at \sqrtsNN\,=\, 7.7--39 GeV. $R_{\textrm{\tiny{CP}}}$ is defined as the ratio
of particle yield in central collisions to that in peripheral
ones scaled by the average number of inelastic binary collisions
$N_{\textrm{\scriptsize{coll}}}$, i.e.,
\begin{eqnarray}
R_{\textrm{\tiny{CP}}}=\frac{[(dN/dp_{T})/\langle N_{\textrm{\scriptsize{coll}}} \rangle]_{\textrm{central}}}{[(dN/dp_{T})/\langle N_{\textrm{\scriptsize{coll}}} \rangle]_{\textrm{peripheral}}}.
\end{eqnarray}
Here $N_{\textrm{\scriptsize{coll}}}$ is determined from Glauber Monte Carlo simulations. See Table~\ref{tab:nbin} for the $N_{\textrm{\scriptsize{coll}}}$ values for Au+Au collisions in the STAR Beam Energy Scan.
$R_{\textrm{\tiny{CP}}}$ will be unity if nucleus-nucleus collisions are just simple superpositions
of nucleon-nucleon collisions. Deviation of these ratios from unity
would imply contributions from nuclear or in-medium effects. For \ppt\,$\approx$\,4~\GeVc, one can see from Fig.~\ref{fig_rcp} that the \ks\
$R_{\textrm{\tiny{CP}}}$ is below unity at \sqrtsNN\,=\,39 GeV. This is similar to the observation at top RHIC energy \cite{Adams:2003am} though the lowest $R_{\textrm{\tiny{CP}}}$ value is larger here. Then the \ks\ $R_{\textrm{\tiny{CP}}}$ at \ppt\,$>$\,2~\GeVc\ keeps increasing with decreasing collision energies, indicating that the partonic energy loss effect becomes less important. Eventually, the cold nuclear matter effect (Cronin effect) \cite{Cronin1} starts to take over at \sqrtsNN\,= 11.5 and 7.7
GeV and enhances all the hadron (including \ks) yields at intermediate \ppt\ (up to $\approx$3.5~\GeVc). Similar to the observation for identified charged hadrons \cite{Adamczyk:2017nof}, the energy evolution of strange hadron $R_{\textrm{\scriptsize{CP}}}$
reflects the decreasing partonic effects with decreasing beam
energies. In addition, the particle $R_{\textrm{\tiny{CP}}}$
differences are apparent for \sqrtsNN$\ \geq$ 19.6 GeV. However,
the differences become smaller at \sqrtsNN\,=\,11.5
GeV and eventually vanish at \sqrtsNN\,=\,7.7 GeV, which may also suggest different properties of the system created in Au+Au collisions at \sqrtsNN\ = 11.5 and 7.7 GeV, compared to
those in \sqrtsNN$\ \geq$ 19.6 GeV.

\subsection{Baryon enhancement at intermediate \ppt}

\begin{figure}[htbp]
\centering \vspace{0cm}
\includegraphics[width=8.5cm]{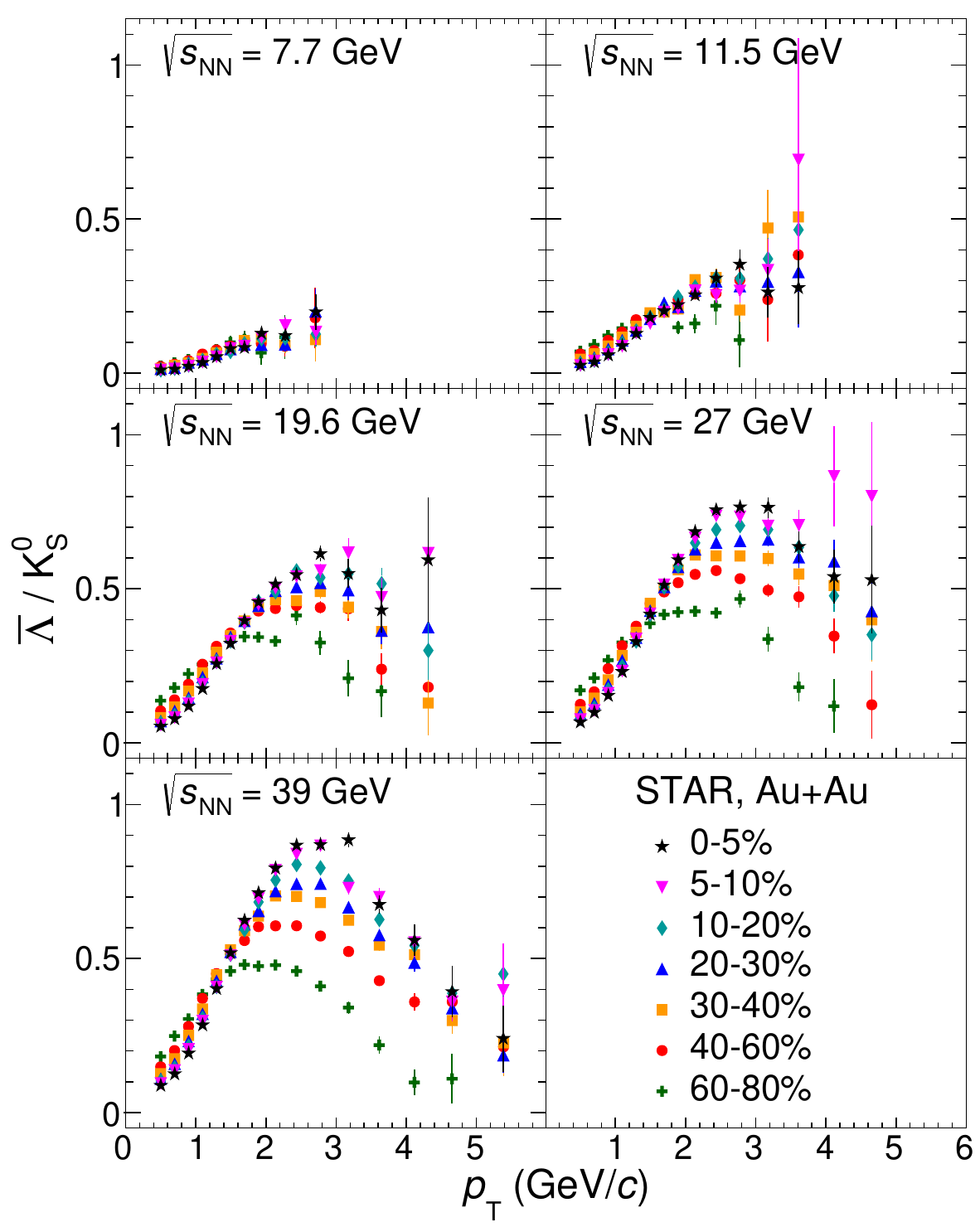}
\vspace{0cm} \caption{\alam/\ks\ ratio as a function of \ppt\ at mid-rapidity ($|y|<0.5$) in different centralities from Au+Au collisions at \sqrtsNN\ = 7.7--39 GeV. Errors are statistical only.}\label{fig_la_k0s}
\end{figure}

The enhancement of baryon-to-meson ratios at intermediate \ppt\ in central {\it A}+{\it A} collisions compared to peripheral {\it A}+{\it A} or {\it p}+{\it p} collisions at the same energy is interpreted as a consequence of hadron formation through parton recombination and parton collectivity in central collisions \cite{coal,coal2,reconbination0,reconbination1,reconbination2,reconbination21,reconbination3,reconbination31,jinhui1}. Therefore, the baryon-to-meson ratios are expected to be sensitive to the parton dynamics of the collision system. The multi-strange baryon-to-meson ratio, $\Omega$/$\phi$, has been described in detail in Ref.~\cite{Adamczyk:2015lvo}. Figure~\ref{fig_la_k0s} shows the \alam/\ks\ ratio as a function of \ppt\ in different centralities from Au+Au collisions at \sqrtsNN\ = 7.7--39 GeV. The \alam\ is chosen instead of \lam, because it is a newly produced baryon in the baryon-rich medium created in lower Beam Energy Scan energies. At \sqrtsNN\ $\geq$ 19.6 GeV, the \alam/\ks\ reaches its maximum value at \ppt\ $\approx$ 2.5 \GeVc\ in central collisions, while in peripheral collisions, the maximum value is significantly lower. This shows that there is baryon enhancement at intermediate \ppt\ for \sqrtsNN\ $\geq$ 19.6 GeV similar to that observed at higher energies. For Au+Au collisions at \sqrtsNN\ $\leq$ 11.5 GeV, the difference between the values of \alam/\ks\ in the measured \ppt\ range in 0--5\% and 40--60\% is much less significant. Unfortunately, the maximum \alam/\ks\ value in each centrality bin cannot be clearly identified due to limited \ppt\ reach and statistics at \sqrtsNN\ $\leq$ 11.5 GeV, hence whether baryon-to-meson enhancement still persists at these energies remains uncertain with the current data.

\section{Conclusions}\label{Conclusions}

In this paper we present STAR measurements of strange hadrons (\ks, $\Lambda$, \alam, \xim, \axi, \omm, \aom, and $\phi$) production at mid-rapidity ($|y|<0.5$) in Au+Au collisions at \sqrtsNN\ = 7.7, 11.5, 19.6, 27, and 39 GeV, from data taken in the first phase of the RHIC Beam Energy Scan Program. 

The \ppt\ spectra, averaged transverse mass ($\left< m_{\rm T}\right>-m_0$), and integrated yield ($dN/dy$) have been extracted with high precision for all strange hadron species and for all centralities and collision energies. Generally, the STAR BES data follow the trend of the previous measurements from AGS, SPS and RHIC. These measurements also exhibit the following features in strange hadron production in this energy range. {\bf (1)} $\left< m_{\rm T}\right>-m_0$ of antibaryons and baryons significantly deviate from each other toward lower collision energies, especially for \alam\ and \lam. {\bf (2)} $dN/dy$ of strange baryons (\lam\ and \xim) show a non-monotonic energy dependence, while the $dN/dy$ of the corresponding antibaryons and \ks\ and $\phi$ mesons increases monotonically toward higher energies.

All the antibaryon-to-baryon ratios decrease toward central collisions due to increased baryon stopping over the BES energy range. The \alam/\lam\ ratio tends to increase significantly with increasing \ppt\ in the low \ppt\ region ($\lesssim 2$ \GeVc) in central collisions, especially for \sqrtsNN\ $\leq$ 11.5 GeV. This can be due to either antibaryon absorption or the increase of stopped baryons without pair production.
The thermal model has been tested with the measured antibaryon-to-baryon ratios, $\overline{\Lambda}/\Lambda$, $\overline{\Xi}^+/\Xi^-$, and $\overline{\Omega}^+/\Omega^-$, and then two chemical freeze-out parameters, $\mu_B/T_{\rm ch}$ and $\mu_S/T_{\rm ch}$, are extracted for central Au+Au collisions at five BES energies. The strange baryon-to-pion ratios are found to be consistent with the calculations of the statistical hadron gas model, and for \lam/$\pi$ ratio, consistent with hadronic transport models as well.

For intermediate-to-high \ppt\ probes, the nuclear modification factors ($R_{\textrm{\tiny{CP}}}$) of various strange hadrons and the \alam/\ks\ ratio have been presented. The \ks\ $R_{\textrm{\tiny{CP}}}$ shows no suppression for \ppt\ up to 3.5 \GeVc\ at energies of 7.7 and 11.5 GeV. The particle-type dependence of $R_{\textrm{\tiny{CP}}}$ also becomes smaller at \sqrtsNN\,$\le$\,11.5 GeV. These observations show that the partonic energy loss effect becomes less significant with decreasing collision energy. The cold nuclear matter effect, such as the Cronin effect, starts to take over at \sqrtsNN\,= 11.5 and 7.7 GeV and enhances the hadron yields at intermediate \ppt. The \alam/\ks\ ratio data show baryon-to-meson enhancement at intermediate \ppt\ ($\approx$2.5 \GeVc) in central collisions at energies above 19.6 GeV. Unfortunately, the precision of current \alam/\ks\ measurements below 11.5 GeV does not allow us to unambiguously conclude regarding possible baryon-to-meson enhancement at these energies. These measurements point to the beam energy region below 19.6 GeV for further investigation of the deconfinement phase transition.

\begin{acknowledgments}
\makeatletter{}We thank the RHIC Operations Group and RCF at BNL, the NERSC Center at LBNL, and the Open Science Grid consortium for providing resources and support.  This work was supported in part by the Office of Nuclear Physics within the U.S. DOE Office of Science, the U.S. National Science Foundation, the Ministry of Education and Science of the Russian Federation, National Natural Science Foundation of China (especially under Contract No.~11890710), Chinese Academy of Science, the Ministry of Science and Technology of China and the Chinese Ministry of Education, the National Research Foundation of Korea, Czech Science Foundation and Ministry of Education, Youth and Sports of the Czech Republic, Hungarian National Research, Development and Innovation Office (FK-123824), New National Excellency Programme of the Hungarian Ministry of Human Capacities (UNKP-18-4), Department of Atomic Energy and Department of Science and Technology of the Government of India, the National Science Centre of Poland, the Ministry  of Science, Education and Sports of the Republic of Croatia, RosAtom of Russia, German Bundesministerium fur Bildung, Wissenschaft, Forschung and Technologie (BMBF), and the Helmholtz Association.
 
\end{acknowledgments}

\makeatletter{}
 

\begin{thebibliography}{56}

%\cite{Stephanov:2004wx}
\bibitem{Stephanov:2004wx} 
  M.~A.~Stephanov,
  %``QCD phase diagram and the critical point,''
  Prog.\ Theor.\ Phys.\ Suppl.\  {\bf 153}, 139 (2004).
%  [Int.\ J.\ Mod.\ Phys.\ A {\bf 20}, 4387 (2005)]
%  doi:10.1142/S0217751X05027965
%  [hep-ph/0402115].
  %%CITATION = doi:10.1142/S0217751X05027965;%%
  %461 citations counted in INSPIRE as of 15 Apr 2020

%\cite{Mohanty:2009vb}
\bibitem{Mohanty:2009vb} 
  B.~Mohanty,
  %``QCD Phase Diagram: Phase Transition, Critical Point and Fluctuations,''
  Nucl.\ Phys.\ A {\bf 830}, 899C (2009).
%  doi:10.1016/j.nuclphysa.2009.10.132
% [arXiv:0907.4476 [nucl-ex]].
  %%CITATION = doi:10.1016/j.nuclphysa.2009.10.132;%%
  %99 citations counted in INSPIRE as of 15 Apr 2020

%\cite{Aggarwal:2010cw}
\bibitem{Aggarwal:2010cw} 
M.~M.~Aggarwal {\it et al.} (STAR Collaboration), arXiv:1007.2613.
  %``An Experimental Exploration of the QCD Phase Diagram: The Search for the Critical Point and the Onset of De-confinement,''  
  %%CITATION = ARXIV:1007.2613;%%
  %76 citations counted in INSPIRE as of 26 Sep 2013

%\cite{Adamczyk:2013dal}
\bibitem{Adamczyk:2013dal} 
  L.~Adamczyk {\it et al.} (STAR Collaboration),
  %``Energy Dependence of Moments of Net-proton Multiplicity Distributions at RHIC,''
  Phys.\ Rev.\ Lett.\  {\bf 112}, 032302 (2014).
%  doi:10.1103/PhysRevLett.112.032302
%  [arXiv:1309.5681 [nucl-ex]].
  %%CITATION = doi:10.1103/PhysRevLett.112.032302;%%
  %408 citations counted in INSPIRE as of 15 Apr 2020

%\cite{Adamczyk:2014ipa}
\bibitem{Adamczyk:2014ipa} 
  L.~Adamczyk {\it et al.} (STAR Collaboration),
  %``Beam-Energy Dependence of the Directed Flow of Protons, Antiprotons, and Pions in Au+Au Collisions,''
  Phys.\ Rev.\ Lett.\  {\bf 112}, 162301 (2014).
%  doi:10.1103/PhysRevLett.112.162301
%  [arXiv:1401.3043 [nucl-ex]].
  %%CITATION = doi:10.1103/PhysRevLett.112.162301;%%
  %171 citations counted in INSPIRE as of 15 Apr 2020

%\cite{Adamczyk:2014fia}
\bibitem{Adamczyk:2014fia} 
  L.~Adamczyk {\it et al.} (STAR Collaboration),
  %``Beam energy dependence of moments of the net-charge multiplicity distributions in Au+Au collisions at RHIC,''
  Phys.\ Rev.\ Lett.\  {\bf 113}, 092301 (2014).
%  doi:10.1103/PhysRevLett.113.092301
%  [arXiv:1402.1558 [nucl-ex]].
  %%CITATION = doi:10.1103/PhysRevLett.113.092301;%%
  %280 citations counted in INSPIRE as of 15 Apr 2020

\bibitem{bes_pid} 
  L.~Adamczyk {\it et al.} (STAR Collaboration),
  %``Bulk Properties of the Medium Produced in Relativistic Heavy-Ion Collisions from the Beam Energy Scan Program,''
  Phys.\ Rev.\ C {\bf 96}, 044904 (2017).
%  doi:10.1103/PhysRevC.96.044904
  %[arXiv:1701.07065 [nucl-ex]].
  %%CITATION = doi:10.1103/PhysRevC.96.044904;%%
  %35 citations counted in INSPIRE as of 05 Jan 2018

%\cite{Adamczyk:2017wsl}
\bibitem{Adamczyk:2017wsl} 
  L.~Adamczyk {\it et al.} (STAR Collaboration),
  %``Collision Energy Dependence of Moments of Net-Kaon Multiplicity Distributions at RHIC,''
  Phys.\ Lett.\ B {\bf 785}, 551 (2018).
%  doi:10.1016/j.physletb.2018.07.066
%  [arXiv:1709.00773 [nucl-ex]].
  %%CITATION = doi:10.1016/j.physletb.2018.07.066;%%
  %74 citations counted in INSPIRE as of 15 Apr 2020

%\cite{Adamczyk:2013gv}
\bibitem{Adamczyk:2013gv}
L.~Adamczyk \textit{et al.} (STAR Collaboration),
%``Observation of an Energy-Dependent Difference in Elliptic Flow between Particles and Antiparticles in Relativistic Heavy Ion Collisions,''
Phys.\ Rev.\ Lett.\ \textbf{110}, 142301 (2013).
%doi:10.1103/PhysRevLett.110.142301
%[arXiv:1301.2347 [nucl-ex]].
%76 citations counted in INSPIRE as of 15 Apr 2020

%\cite{Adamczyk:2013gw}
\bibitem{Adamczyk:2013gw} 
  L.~Adamczyk {\it et al.} (STAR Collaboration),
  %``Elliptic flow of identified hadrons in Au+Au collisions at $\sqrt{s_{NN}}=$ 7.7-62.4 GeV,''
  Phys.\ Rev.\ C {\bf 88}, 014902 (2013).
%  doi:10.1103/PhysRevC.88.014902
% [arXiv:1301.2348 [nucl-ex]].
  %%CITATION = doi:10.1103/PhysRevC.88.014902;%%
  %77 citations counted in INSPIRE as of 05 Nov 2016

%\cite{Adamczyk:2015lvo}
\bibitem{Adamczyk:2015lvo} 
  L.~Adamczyk {\it et al.} (STAR Collaboration),
  %``Probing parton dynamics of QCD matter with $\Omega$ and $\phi$ production,''
  Phys.\ Rev.\ C {\bf 93}, 021903(R) (2016).
%  doi:10.1103/PhysRevC.93.021903
 % [arXiv:1506.07605 [nucl-ex]].
  %%CITATION = doi:10.1103/PhysRevC.93.021903;%%
  %5 citations counted in INSPIRE as of 17 Nov 2016

%\cite{Adamczyk:2017nof}
\bibitem{Adamczyk:2017nof} 
L.~Adamczyk {\it et al.} (STAR Collaboration),
  %``Beam Energy Dependence of Jet-Quenching Effects in Au+Au Collisions at $\sqrt{s_{_{ \mathrm{NN}}}}$ = 7.7, 11.5, 14.5, 19.6, 27, 39, and 62.4 GeV,''
  Phys.\ Rev.\ Lett.\  {\bf 121}, 032301 (2018).
%  doi:10.1103/PhysRevLett.121.032301
%  [arXiv:1707.01988 [nucl-ex]].
  %%CITATION = doi:10.1103/PhysRevLett.121.032301;%%
  %10 citations counted in INSPIRE as of 14 Feb 2019

\bibitem{raf82}
J.~Rafelski, B.~M\"{u}ller, Phys. Rev. Lett. \textbf{48}, 1066 (1982).

%%%%%%% AGS %%%%%%%%%

\bibitem{e891}
S.~Ahmad {\it et al.} (E891 Collaboration),
  %``Lambda production by 11.6-A/GeV/c Au beam on Au target,''
  Phys.\ Lett.\ B {\bf 382}, 35 (1996),
  Erratum: Phys.\ Lett.\ B {\bf 386}, 496 (1996).

\bibitem{e891_2}
S.~Ahmad {\it et al.}, Nucl.\ Phys.\ A {\bf 636}, 507 (1998).
 %%CITATION = PHLTA,B382,35;%%

\bibitem{e802}
L.~Ahle {\it et al.} (E802 Collaboration),
 %``Centrality dependence of kaon yields in Si + A and Au + Au collisions at
 %the AGS,''
Phys.\ Rev.\  C {\bf 60}, 044904 (1999).
 %[arXiv:nucl-ex/9903009].
 %%CITATION = PHRVA,C60,044904;%%

\bibitem{e917}
B.~B.~Back {\it et al.} (E917 Collaboration),
  %``Anti-lambda production in Au+Au collisions at 11.7-AGeV/c,''
  Phys.\ Rev.\ Lett.\  {\bf 87}, 242301 (2001).
 %[arXiv:nucl-ex/0101008].
 %%CITATION = PRLTA,87,242301;%%

\bibitem{e896}
S.~Albergo {\it et al.} (E896 Collaboration),
  %``Lambda spectra in 11.6-A-GeV/c Au Au collisions,''
  Phys.\ Rev.\ Lett.\  {\bf 88}, 062301 (2002). 
%%CITATION = PRLTA,88,062301;%%

\bibitem{e895}
P.~Chung {\it et al.} (E895 Collaboration),
  %``Near threshold production of the multistrange xi- hyperon,''
  Phys.\ Rev.\ Lett.\  {\bf 91}, 202301 (2003).
%%CITATION = PRLTA,91,202301;%%

%%%%% NA49 %%%%%%%%%

%\cite{Afanasiev:2002mx}
\bibitem{Afanasiev:2002mx} 
S.~V.~Afanasiev {\it et al.} (NA49 Collaboration),
  %``Energy dependence of pion and kaon production in central Pb + Pb collisions,''
  Phys.\ Rev.\ C {\bf 66}, 054902 (2002).
  %%CITATION = NUCL-EX/0205002;%%
  %363 citations counted in INSPIRE as of 26 Sep 2013

\bibitem{Anticic:2003ux}
T.~Anticic {\it et al.} (NA49 Collaboration),
  %``Lambda and anti-Lambda production in central Pb - Pb collisions at 40-A-GeV, 80-A-GeV and 158-A-GeV,''
Phys.\ Rev.\ Lett.\ {\bf 93}, 022302 (2004).
  %[arXiv:nucl-ex/0311024 [nucl-ex]].

\bibitem{na49prl}
C.~Alt {\it et al.} (NA49 Collaboration),
 %``Omega- and anti-Omega+ production in central Pb + Pb collisions at 40-AGeV
 %and 158-AGeV,''
Phys.\ Rev.\ Lett.\ {\bf 94}, 192301 (2005).
 %[arXiv:nucl-ex/0409004].
 %%CITATION = PRLTA,94,192301;%%

%\cite{Alt:2007aa}
\bibitem{Alt:2007aa} 
C.~Alt {\it et al.} (NA49 Collaboration),
  %``Pion and kaon production in central Pb + Pb collisions at 20-A and 30-A-GeV: Evidence for the onset of deconfinement,''
  Phys.\ Rev.\ C {\bf 77}, 024903 (2008).
%  [arXiv:0710.0118 [nucl-ex]].
  %%CITATION = ARXIV:0710.0118;%%
  %157 citations counted in INSPIRE as of 26 Sep 2013

\bibitem{na49prc}
C.~Alt {\it et al.} (NA49 Collaboration),
  %``Energy dependence of Lambda and Xi production in central Pb+Pb collisions at A-20, A-30, A-40, A-80, and A-158 GeV measured at the CERN Super Proton Synchrotron,''
Phys.\ Rev.\ C {\bf 78}, 034918 (2008).
 %[arXiv:0804.3770 [nucl-ex]].
 %%CITATION = PHRVA,C78,034918;%%

%\cite{Anticic:2009ie}
\bibitem{Anticic:2009ie}
 T.~Anticic {\it et al.} (NA49 Collaboration),
  %``System-size dependence of Lambda and Xi production in nucleus-nucleus collisions at 40A and 158A-GeV measured at the CERN Super Proton Synchrotron,''
  Phys.\ Rev.\ C {\bf 80}, 034906 (2009).
  %[arXiv:0906.0469 [nucl-ex]].
  %%CITATION = PHRVA,C80,034906;%%

%%%%% NA57 %%%%%%%%%
\bibitem{Antinori:2004ee}
F.~Antinori {\it et al.} (NA57 Collaboration),
  %``Energy dependence of hyperon production in nucleus nucleus collisions at SPS,''
  Phys.\ Lett.\ B {\bf 595}, 68 (2004).
 %%CITATION = PHLTA,B595,68;%%
%\bibitem{na57jpg}
% F.~Antinori {\it et al.}, % [NA57 Collaboration],
 %``Enhancement of hyperon production at central rapidity in 158-A-GeV/c Pb-Pb
 %collisions,''
\bibitem{Antinori:2006ij} 
  F.~Antinori {\it et al.} (NA57 Collaboration),
  %``Enhancement of hyperon production at central rapidity in 158-A-GeV/c Pb-Pb collisions,''
  J.\ Phys.\ G {\bf 32}, 427 (2006);
%%CITATION = JPHGB,G32,427;%%
http://wa97.web.cern.ch/WA97/yields.html.

%%%%%%% CERES %%%%%%%%%%

\bibitem{ceres}
J.~Milosevic (CERES Collaboration),
  %``Strange particle production and elliptic flow from CERES,''
  J.\ Phys.\ G {\bf 32}, S97 (2006).
 %[arXiv:nucl-ex/0606020].
 %%CITATION = JPHGB,G32,S97;%%

%%%%%%% STAR %%%%%%%%%%%

%\cite{Adler:2002uv}
\bibitem{Adler:2002uv}
C.~Adler {\it et al.} (STAR Collaboration),
  %``Midrapidity Lambda and anti-Lambda production in Au + Au collisions at s(NN)**(1/2) = 130-GeV,''
  Phys.\ Rev.\ Lett.\  {\bf 89}, 092301 (2002).
%  [arXiv:nucl-ex/0203016].
  %%CITATION = PRLTA,89,092301;%%

\bibitem{starpid_130}
C.~Adler {\it et al.} (STAR Collaboration),
  %``Kaon production and kaon to pion ratio in Au+Au collisions at s(NN)**1/2 = 130-GeV,''
  Phys.\ Lett.\ B {\bf 595}, 143 (2004).
 %[arXiv:nucl-ex/0206008].
 %%CITATION = PHLTA,B595,143;%%

%\cite{Adams:2003fy}
\bibitem{Adams:2003fy}
J.~Adams {\it et al.} (STAR Collaboration),
  %``Multistrange baryon production in Au-Au collisions at S(NN)**1/2 = 130 GeV,''
  Phys.\ Rev.\ Lett.\  {\bf 92}, 182301 (2004).
%  [arXiv:nucl-ex/0307024].
  %%CITATION = PRLTA,92,182301;%%

%\cite{Adams:2006ke}
\bibitem{Adams:2006ke}
J.~Adams {\it et al.} (STAR Collaboration),
  %``Scaling Properties of Hyperon Production in Au+Au Collisions at s**(1/2) = 200-GeV,''
  Phys.\ Rev.\ Lett.\  {\bf 98}, 062301 (2007).
  %[arXiv:nucl-ex/0606014].
  %%CITATION = PRLTA,98,062301;%%

\bibitem{Abelev:2007xp}
B.~I.~Abelev {\it et al.} (STAR Collaboration),
  %``Enhanced strange baryon production in Au + Au collisions compared to p + p at s(NN)**(1/2) = 200-GeV,''
  Phys.\ Rev.\ C {\bf 77}, 044908 (2008).
  %%CITATION = PHRVA,C77,044908;%%

%\cite{Abelev:2008zk}
\bibitem{Abelev:2008zk} 
B.~I.~Abelev {\it et al.} (STAR Collaboration),
  %``Energy and system size dependence of phi meson production in Cu+Cu and Au+Au collisions,''
  Phys.\ Lett.\ B {\bf 673}, 183 (2009).
%  doi:10.1016/j.physletb.2009.02.037
% [arXiv:0810.4979 [nucl-ex]].
  %%CITATION = doi:10.1016/j.physletb.2009.02.037;%%
  %78 citations counted in INSPIRE as of 05 Nov 2016

%\cite{Abelev:2010rv}
\bibitem{Abelev:2010rv} 
  B.~I.~Abelev {\it et al.} (STAR Collaboration),
  %``Observation of an Antimatter Hypernucleus,''
  Science {\bf 328}, 58 (2010).
%  doi:10.1126/science.1183980
%  [arXiv:1003.2030 [nucl-ex]].
  %%CITATION = doi:10.1126/science.1183980;%%
  %161 citations counted in INSPIRE as of 18 Sep 2018

\bibitem{starprc83}
M.~M.~Aggarwal {\it et al.} (STAR Collaboration),
  %``Strange and Multi-strange Particle Production in Au+Au Collisions at $\sqrt{s_{NN}}$ = 62.4 GeV,''
  Phys.\ Rev.\ C {\bf 83}, 024901 (2011).
 %[arXiv:1010.0142 [nucl-ex]].
 %%CITATION = PHRVA,C83,024901;%%

%\cite{Agakishiev:2011ar}
\bibitem{Agakishiev:2011ar} 
G.~Agakishiev {\it et al.} (STAR Collaboration),
  %``Strangeness Enhancement in Cu+Cu and Au+Au Collisions at $\sqrt{s_{NN}} = 200$ GeV,''
  Phys.\ Rev.\ Lett.\  {\bf 108}, 072301 (2012).
%  doi:10.1103/PhysRevLett.108.072301
%  [arXiv:1107.2955 [nucl-ex]].
  %%CITATION = doi:10.1103/PhysRevLett.108.072301;%%
  %68 citations counted in INSPIRE as of 14 Feb 2019

%%%%%%% PHENIX %%%%%%%%%%

\bibitem{phenix_lambda}
K.~Adcox {\it et al.} (PHENIX Collaboration),
  %``Measurement of the Lambda and anti-Lambda particles in Au+Au collisions at s(NN)**(1/2) = 130-GeV,''
  Phys.\ Rev.\ Lett.\  {\bf 89}, 092302 (2002).
 %[arXiv:nucl-ex/0204007].
 %%CITATION = PRLTA,89,092302;%%

%%%%%%% LHC/ALICE %%%%%%%%%%

\bibitem{lhc}
%\cite{Abelev:2013xaa}
%\bibitem{Abelev:2013xaa} 
B.~Abelev {\it et al.} (ALICE Collaboration),
  %``$K^0_S$ and $\Lambda$ production in Pb-Pb collisions at $\sqrt{s_{NN}}$ = 2.76 TeV,''
  Phys.\ Rev.\ Lett.\  {\bf 111}, 222301 (2013).
\bibitem{lhc2}
B.~Abelev {\it et al.} (ALICE Collaboration),
  %``Multi-strange baryon production at mid-rapidity in Pb-Pb collisions at $\sqrt{s_{NN}}$ = 2.76 TeV,''
  Phys.\ Lett.\ B {\bf 728}, 216 (2014),
  Erratum: Phys.\ Lett.\ B {\bf 734}, 409 (2014).
  %%CITATION = ARXIV:1307.5530;%%
  %5 citations counted in INSPIRE as of 26 Sep 2013

%\cite{Chen:2018tnh}
\bibitem{Chen:2018tnh} 
J.~Chen, D.~Keane, Y.~G.~Ma, A.~Tang and Z.~Xu,
  %``Antinuclei in Heavy-Ion Collisions,''
  Phys.\ Rept.\  {\bf 760}, 1 (2018).
%  doi:10.1016/j.physrep.2018.07.002
%  [arXiv:1808.09619 [nucl-ex]].
  %%CITATION = doi:10.1016/j.physrep.2018.07.002;%%
  %6 citations counted in INSPIRE as of 14 Feb 2019

%%%%%%%%%% Theory %%%%%%%%%%%

\bibitem{becattini_prc}
F.~Becattini, J.~Cleymans, A.~Keranen, E.~Suhonen and K.~Redlich,
  %``Features of particle multiplicities and strangeness production in central heavy ion collisions between 1.7A-GeV/c and 158A-GeV/c,''
  Phys.\ Rev.\ C {\bf 64}, 024901 (2001).
 %[arXiv:hep-ph/0002267].
 %%CITATION = PHRVA,C64,024901;%%

\bibitem{pbmnpa}
P.~Braun-Munzinger, J.~Cleymans, H.~Oeschler and K.~Redlich,
  %``Maximum relative strangeness content in heavy ion collisions around 30-GeV/A,''
  Nucl.\ Phys.\ A {\bf 697}, 902 (2002).
 %[arXiv:hep-ph/0106066].
 %%CITATION = NUPHA,A697,902;%%

\bibitem{pbm_shm}
A.~Andronic, P.~Braun-Munzinger and J.~Stachel,
  %``Hadron production in central nucleus-nucleus collisions at chemical freeze-out,''
  Nucl.\ Phys.\ A {\bf 772}, 167 (2006).
 %[arXiv:nucl-th/0511071].
 %%CITATION = NUPHA,A772,167;%%

%\cite{Redlich:2001kb}
\bibitem{Redlich:2001kb}
K.~Redlich and A.~Tounsi,
  %``Strangeness enhancement and energy dependence in heavy ion collisions,''
  Eur.\ Phys.\ J.\ C {\bf 24}, 589 (2002).
%  [arXiv:hep-ph/0111261].
  %%CITATION = EPHJA,C24,589;%%

%\cite{Adams:2003kv}
\bibitem{Adams:2003kv} 
J.~Adams {\it et al.} (STAR Collaboration),
  %``Transverse momentum and collision energy dependence of high p(T) hadron suppression in Au+Au collisions at ultrarelativistic energies,''
  Phys.\ Rev.\ Lett.\  {\bf 91}, 172302 (2003).
%  doi:10.1103/PhysRevLett.91.172302
%  [nucl-ex/0305015].
  %%CITATION = doi:10.1103/PhysRevLett.91.172302;%%
  %831 citations counted in INSPIRE as of 14 Feb 2019

%\cite{Adcox:2003nr}
\bibitem{Adcox:2003nr} 
K.~Adcox {\it et al.} (PHENIX Collaboration),
  %``Single identified hadron spectra from s(NN)**(1/2) = 130-GeV Au+Au collisions,''
  Phys.\ Rev.\ C {\bf 69}, 024904 (2004).
  %[nucl-ex/0307010].
  %%CITATION = NUCL-EX/0307010;%%

\bibitem{starb2m}
B.~I.~Abelev {\it et al.} (STAR Collaboration),
  %``Energy dependence of pi+-, p and anti-p transverse momentum spectra for Au+Au collisions at s(NN)**(1/2) = 62.4 and 200-GeV,''
  Phys.\ Lett.\ B {\bf 655}, 104 (2007).
%  doi:10.1016/j.physletb.2007.06.035
% [nucl-ex/0703040]. 

\bibitem{Abelev:2006jr} 
B.~I.~Abelev {\it et al.} (STAR Collaboration),
  %``Identified baryon and meson distributions at large transverse momenta from Au+Au collisions at s(NN)**(1/2) = 200-GeV,''
  Phys.\ Rev.\ Lett.\  {\bf 97}, 152301 (2006).
%doi:10.1103/PhysRevLett.97.152301
% [nucl-ex/0606003].

%\cite{Abelev:2014laa}
\bibitem{Abelev:2014laa} 
B.~B.~Abelev {\it et al.} (ALICE Collaboration),
  %``Production of charged pions, kaons and protons at large transverse momenta in pp and Pb–Pb collisions at $\sqrt{s_{\rm NN}}$ =2.76 TeV,''
  Phys.\ Lett.\ B {\bf 736}, 196 (2014).
%  doi:10.1016/j.physletb.2014.07.011
%  [arXiv:1401.1250 [nucl-ex]].
  %%CITATION = doi:10.1016/j.physletb.2014.07.011;%%
  %191 citations counted in INSPIRE as of 14 Feb 2019

\bibitem{coal}
%\cite{Hwa:2001ih}
%\bibitem{Hwa:2001ih} 
  R.~C.~Hwa and C.~B.~Yang,
  %``Centrality dependence of baryon and meson momentum distributions in proton nucleus collisions,''
  Phys.\ Rev.\ C {\bf 65}, 034905 (2002),
  Erratum: Phys.\ Rev.\ C {\bf 67}, 059902 (2003).
  %doi:10.1103/PhysRevC.67.059902, 10.1103/PhysRevC.65.034905
  %[nucl-th/0108043].
  %%CITATION = doi:10.1103/PhysRevC.67.059902, 10.1103/PhysRevC.65.034905;%%
  %21 citations counted in INSPIRE as of 07 Mar 2017
\bibitem{coal2}
%\cite{Hwa:2002zu}
%\bibitem{Hwa:2002zu} 
  R.~C.~Hwa and C.~B.~Yang,
  %``Inclusive distributions for hadronic collisions in the Valon recombination model,''
  Phys.\ Rev.\ C {\bf 66}, 025205 (2002).
  %doi:10.1103/PhysRevC.66.025205
  %[hep-ph/0204289].
  %%CITATION = doi:10.1103/PhysRevC.66.025205;%%
  %105 citations counted in INSPIRE as of 07 Mar 2017

\bibitem{reconbination0} 
D.~Molnar and S.~A.~Voloshin,
  %``Elliptic flow at large transverse momenta from quark coalescence,''
  Phys.\ Rev.\ Lett.\  {\bf 91}, 092301 (2003).

\bibitem{reconbination1} 
R.~C.~Hwa and C.~B.~Yang,
  %``Production of strange particles at intermediate pT in central Au+Au collisions at high energies,''
  Phys.\ Rev.\ C {\bf 75}, 054904 (2007).

\bibitem{reconbination2} 
R.~J.~Fries, B.~Muller, C.~Nonaka and S.~A.~Bass,
  %``Hadronization in heavy ion collisions: Recombination and fragmentation of partons,''
  Phys.\ Rev.\ Lett.\  {\bf 90}, 202303 (2003).

\bibitem{reconbination21} 
R.~J.~Fries, B.~Muller, C.~Nonaka and S.~A.~Bass,
  %``Hadron production in heavy ion collisions: Fragmentation and recombination from a dense parton phase,''
  Phys.\ Rev.\ C {\bf 68}, 044902 (2003).

\bibitem{reconbination3} 
V.~Greco, C.~M.~Ko and P.~Levai,
  %``Parton coalescence and anti-proton / pion anomaly at RHIC,''
  Phys.\ Rev.\ Lett.\  {\bf 90}, 202302 (2003).

\bibitem{reconbination31} 
V.~Greco, C.~M.~Ko and P.~Levai,
  %``Parton coalescence at RHIC,''
  Phys.\ Rev.\ C {\bf 68}, 034904 (2003).

%%%%%%PID%%%%%%%%%

\bibitem{STARNIM} 
K.~H.~Ackermann {\it et al.} (STAR Collaboration),
  %``STAR detector overview,''
  Nucl.\ Instrum.\ Meth.\ A {\bf 499}, 624 (2003).

\bibitem{STARTPC} 
  M.~Anderson  {\it et al.},
  %``The Star time projection chamber: A Unique tool for studying high multiplicity events at RHIC,''
  Nucl.\ Instrum.\ Meth.\ A {\bf 499}, 659 (2003).

%\cite{Llope:2003ti}
\bibitem{Llope:2003ti} 
W.~J.~Llope {\it et al.},
  %``The TOFp / pVPD time-of-flight system for STAR,''
  Nucl.\ Instrum.\ Meth.\ A {\bf 522}, 252 (2004).
%  doi:10.1016/j.nima.2003.11.414
%  [nucl-ex/0308022].
  %%CITATION = doi:10.1016/j.nima.2003.11.414;%%
  %74 citations counted in INSPIRE as of 15 Feb 2019

\bibitem{STARTRG}
%\cite{Bieser:2002ah}
%\bibitem{Bieser:2002ah} 
  F.~S.~Bieser {\it et al.},
  %``The STAR trigger,''
  Nucl.\ Instrum.\ Meth.\ A {\bf 499}, 766 (2003).
  %doi:10.1016/S0168-9002(02)01974-5
  %%CITATION = doi:10.1016/S0168-9002(02)01974-5;%%
  %106 citations counted in INSPIRE as of 07 Mar 2017

%\cite{Bonner:2003bv}
\bibitem{Bonner:2003bv} 
B.~Bonner {\it et al.},
  %``A single Time-of-Flight tray based on multigap resistive plate chambers for the STAR experiment at RHIC,''
  Nucl.\ Instrum.\ Meth.\ A {\bf 508}, 181 (2003).
%  doi:10.1016/S0168-9002(03)01347-0
  %%CITATION = doi:10.1016/S0168-9002(03)01347-0;%%
  %76 citations counted in INSPIRE as of 15 Feb 2019

%\cite{Miller:2007ri}
\bibitem{Miller:2007ri}
M.~L.~Miller, K.~Reygers, S.~J.~Sanders and P.~Steinberg,
  %``Glauber modeling in high energy nuclear collisions,''
  Ann.\ Rev.\ Nucl.\ Part.\ Sci.\  {\bf 57}, 205 (2007).

%\cite{Olive:2016xmw}
\bibitem{Olive:2016xmw} 
  C.~Patrignani {\it et al.} (Particle Data Group),
  %``Review of Particle Physics,''
  Chin.\ Phys.\ C {\bf 40}, 100001 (2016).
%  doi:10.1088/1674-1137/40/10/100001
  %%CITATION = doi:10.1088/1674-1137/40/10/100001;%%
  %460 citations counted in INSPIRE as of 07 Mar 2017

%\cite{Shao:2005iu}
\bibitem{Shao:2005iu} 
  M.~Shao, O.~Y.~Barannikova, X.~Dong, Y.~Fisyak, L.~Ruan, P.~Sorensen and Z.~Xu,
  %``Extensive particle identification with TPC and TOF at the STAR experiment,''
  Nucl.\ Instrum.\ Meth.\ A {\bf 558}, 419 (2006).
%  doi:10.1016/j.nima.2005.11.251
%  [nucl-ex/0505026].
  %%CITATION = doi:10.1016/j.nima.2005.11.251;%%
  %104 citations counted in INSPIRE as of 05 Nov 2016

\bibitem{bichsel} 
H.~Bichsel,
  %``A method to improve tracking and particle identification in TPCs and silicon detectors,''
  Nucl.\ Instrum.\ Meth.\ A {\bf 562}, 154 (2006).

%\cite{Adams:2004ux}
\bibitem{Adams:2004ux} 
  J.~Adams {\it et al.} (STAR Collaboration),
  %``phi meson production in Au + Au and p+p collisions at s(NN)**(1/2) = 200-GeV,''
  Phys.\ Lett.\ B {\bf 612}, 181 (2005).
%  doi:10.1016/j.physletb.2004.12.082
 % [nucl-ex/0406003].
  %%CITATION = doi:10.1016/j.physletb.2004.12.082;%%
  %178 citations counted in INSPIRE as of 05 Nov 2016

%\cite{Abelev:2007rw}
\bibitem{Abelev:2007rw} 
  B.~I.~Abelev {\it et al.} (STAR Collaboration),
  %``Partonic flow and phi-meson production in Au + Au collisions at s(NN)**(1/2) = 200-GeV,''
  Phys.\ Rev.\ Lett.\  {\bf 99}, 112301 (2007).
%  doi:10.1103/PhysRevLett.99.112301
  %[nucl-ex/0703033 [NUCL-EX]].
  %%CITATION = doi:10.1103/PhysRevLett.99.112301;%%
  %177 citations counted in INSPIRE as of 05 Nov 2016

%\cite{Abelev:2008aa}
\bibitem{Abelev:2008aa} 
  B.~I.~Abelev {\it et al.} (STAR Collaboration),
  %``Measurements of phi meson production in relativistic heavy-ion collisions at RHIC,''
  Phys.\ Rev.\ C {\bf 79}, 064903 (2009).
%  doi:10.1103/PhysRevC.79.064903
  %[arXiv:0809.4737 [nucl-ex]].
  %%CITATION = doi:10.1103/PhysRevC.79.064903;%%
  %99 citations counted in INSPIRE as of 05 Nov 2016

\bibitem{huilong}
H.~Long, PhD thesis, UCLA, (2002).

\bibitem{geant}
V.~Fine and P.~Nevski, Proc.\ CHEP {\bf 2000}, 143 (2000).

%\cite{Schnedermann:1993ws}
\bibitem{Schnedermann:1993ws} 
E.~Schnedermann, J.~Sollfrank and U.~W.~Heinz,
  %``Thermal phenomenology of hadrons from 200-A/GeV S+S collisions,''
  Phys.\ Rev.\ C {\bf 48}, 2462 (1993).
%  doi:10.1103/PhysRevC.48.2462
%  [nucl-th/9307020].
  %%CITATION = doi:10.1103/PhysRevC.48.2462;%%
  %840 citations counted in INSPIRE as of 25 May 2019

%\cite{Cleymans:2004bf}
\bibitem{Cleymans:2004bf} 
 J.~Cleymans, A.~Forster, H.~Oeschler, K.~Redlich and F.~Uhlig,
  %``On the chemical equilibration of strangeness-exchange reaction in heavy-ion collisions,''
Phys.\ Lett.\ B {\bf 603}, 146 (2004).
%  doi:10.1016/j.physletb.2004.09.078
%  [hep-ph/0406108].
  %%CITATION = doi:10.1016/j.physletb.2004.09.078;%%
  %19 citations counted in INSPIRE as of 18 May 2019

\bibitem{becattiniprc73}
F.~Becattini, J.~Manninen and M.~Gazdzicki,
  %``Energy and system size dependence of chemical freeze-out in relativistic nuclear collisions,''
  Phys.\ Rev.\ C {\bf 73}, 044905 (2006).

\bibitem{cleymans}
J.~Cleymans,
  %``Strangeness: Theoretical status,''
  arXiv:nucl-th/9704046.

\bibitem{seewagatoo}
A.~Bazavov {\it et al.},
  %``Additional Strange Hadrons from QCD Thermodynamics and Strangeness Freezeout in Heavy Ion Collisions,''
  Phys.\ Rev.\ Lett.\  {\bf 113}, 072001 (2014).

\bibitem{E802}
L.~Ahle {\it et al.} (E802 Collaboration),
  %``Particle production at high baryon density in central Au+Au reactions at 11.6A GeV/c,''
  Phys.\ Rev.\ C {\bf 57}, R466 (1998).
 %%CITATION = PHRVA,C57,466;%%

%\cite{Alt:2006dk}
\bibitem{Alt:2006dk}
C.~Alt {\it et al.}  (NA49 Collaboration),
  %``Energy and centrality dependence of anti-p and p production and the anti-Lambda/anti-p ratio in Pb+Pb collisions between 20/A-GeV and 158/A-Gev,''
Phys.\ Rev.\ C {\bf 73}, 044910 (2006).
  %%CITATION = PHRVA,C73,044910;%%
  %67 citations counted in INSPIRE as of 26 Sep 2013

\bibitem{phenix_pid}
K.~Adcox {\it et al.} (PHENIX Collaboration),
  %``Centrality dependence of pi+ / pi-, K+ / K-, p and anti-p production from s(NN)**(1/2) = 13-=GeV Au+Au collisions at RHIC,''
  Phys.\ Rev.\ Lett.\  {\bf 88}, 242301 (2002).
 %[arXiv:nucl-ex/0112006].
 %%CITATION = PRLTA,88,242301;%%

\bibitem{starpid_200}
J.~Adams {\it et al.} (STAR Collaboration),
  %``Identified particle distributions in pp and Au+Au collisions at s(NN)**(1/2) = 200 GeV,''
  Phys.\ Rev.\ Lett.\  {\bf 92}, 112301 (2004).
 %[arXiv:nucl-ex/0310004].
 %%CITATION = PRLTA,92,112301;%%

%\cite{Abelev:2008ab}
\bibitem{Abelev:2008ab} 
B.~I.~Abelev {\it et al.} (STAR Collaboration),
  %``Systematic Measurements of Identified Particle Spectra in $p p, d^+$ Au and Au+Au Collisions from STAR,''
  Phys.\ Rev.\ C {\bf 79}, 034909 (2009).
  %%CITATION = ARXIV:0808.2041;%%
  %297 citations counted in INSPIRE as of 26 Sep 2013

%\cite{Bass:1998ca}
\bibitem{Bass:1998ca} 
  S.~A.~Bass {\it et al.},
  %``Microscopic models for ultrarelativistic heavy ion collisions,''
  Prog.\ Part.\ Nucl.\ Phys.\  {\bf 41}, 255 (1998).
%  [Prog.\ Part.\ Nucl.\ Phys.\  {\bf 41}, 225 (1998)]
%  doi:10.1016/S0146-6410(98)00058-1
%  [nucl-th/9803035].
  %%CITATION = doi:10.1016/S0146-6410(98)00058-1;%%
  %1341 citations counted in INSPIRE as of 08 Sep 2018

%\cite{Bleicher:1999xi}
\bibitem{Bleicher:1999xi} 
M.~Bleicher {\it et al.},
  %``Relativistic hadron hadron collisions in the ultrarelativistic quantum molecular dynamics model,''
  J.\ Phys.\ G {\bf 25}, 1859 (1999).
%  doi:10.1088/0954-3899/25/9/308
%  [hep-ph/9909407].
  %%CITATION = doi:10.1088/0954-3899/25/9/308;%%
  %894 citations counted in INSPIRE as of 17 Dec 2017

%\cite{Cassing:1999es}
\bibitem{Cassing:1999es} 
  W.~Cassing and E.~L.~Bratkovskaya,
  %``Hadronic and electromagnetic probes of hot and dense nuclear matter,''
  Phys.\ Rept.\  {\bf 308}, 65 (1999).
%  doi:10.1016/S0370-1573(98)00028-3
  %%CITATION = doi:10.1016/S0370-1573(98)00028-3;%%
  %588 citations counted in INSPIRE as of 17 Dec 2017

\bibitem{hsdurqmd}
E.~L.~Bratkovskaya, M.~Bleicher, M.~Reiter, S.~Soff, H.~Stoecker, M.~van Leeuwen, S.~A.~Bass and W.~Cassing,
  %``Strangeness dynamics and transverse pressure in relativistic nucleus-nucleus collisions,''
  Phys.\ Rev.\ C {\bf 69}, 054907 (2004).
 %[arXiv:nucl-th/0402026].
 %%CITATION = PHRVA,C69,054907;%%

%\cite{Randrup:2006uz}
\bibitem{Randrup:2006uz} 
J.~Randrup and J.~Cleymans,
  %``Maximum freeze-out baryon density in nuclear collisions,''
  Phys.\ Rev.\ C {\bf 74}, 047901 (2006).
  %%CITATION = PHRVA,C74,047901;%%
  %31 citations counted in INSPIRE as of 26 Sep 2013

%\cite{Adams:2003am}
\bibitem{Adams:2003am} 
J.~Adams {\it et al.} (STAR Collaboration),
  %``Particle type dependence of azimuthal anisotropy and nuclear modification of particle production in Au + Au collisions at s(NN)**(1/2) = 200-GeV,''
  Phys.\ Rev.\ Lett.\  {\bf 92}, 052302 (2004).
%  doi:10.1103/PhysRevLett.92.052302
%  [nucl-ex/0306007].
  %%CITATION = doi:10.1103/PhysRevLett.92.052302;%%
  %617 citations counted in INSPIRE as of 18 May 2019

\bibitem{Cronin1} 
J.~W.~Cronin, H.~J.~Frisch, M.~J.~Shochet, J.~P.~Boymond, P.~A.~Piroue and R.~L.~Sumner,
  %``Production of hadrons with large transverse momentum at 200, 300, and 400 GeV,''
  Phys.\ Rev.\ D {\bf 11}, 3105 (1975).

\bibitem{jinhui1} 
J.~H.~Chen, F.~Jin, D.~Gangadharan, X.~Z.~Cai, H.~Z.~Huang and Y.~G.~Ma,
  %``Parton Distributions at Hadronization from Bulk Dense Matter Produced at RHIC,''
  Phys.\ Rev.\ C {\bf 78}, 034907 (2008).

\end{thebibliography}
\end{document}